\documentclass[11pt,a4paper]{article}
\pdfoutput=1
\usepackage{jcappub}

\usepackage{ifpdf}

\usepackage[T1]{fontenc}

\usepackage{afterpage}

\usepackage{subfig}
\usepackage{amssymb}
\usepackage{amsmath}
\usepackage{graphicx}
\usepackage{longtable}
\usepackage{verbatim}
\usepackage{amsfonts}
\usepackage{xcolor}
\usepackage{hyperref}
\usepackage[normalem]{ulem}
\usepackage{multirow}
\usepackage{enumitem}

\newcommand{\bear}{\begin{array}}
\newcommand{\ear}{\end{array}}

\newcommand{\beq}{\begin{eqnarray}}
\newcommand{\eeq}{\end{eqnarray}}
\newcommand{\beqa}{\begin{eqnarray}}
\newcommand{\eeqa}{\end{eqnarray}}

\def\OMIT#1{{}}
\newcommand{\lsim}{\mathrel{\rlap{\lower4pt\hbox{\hskip1pt$\sim$}}
    \raise1pt\hbox{$<$}}}         
\newcommand{\gsim}{\mathrel{\rlap{\lower4pt\hbox{\hskip1pt$\sim$}}
    \raise1pt\hbox{$>$}}}         

\newcommand{\be}{\begin{equation}}
\newcommand{\ee}{\end{equation}}
\newcommand{\ba}{\begin{eqnarray}}
\newcommand{\ea}{\end{eqnarray}}

\def\lsim{\mathrel{\rlap{\lower4pt\hbox{\hskip1pt$\sim$}}
    \raise1pt\hbox{$<$}}}         
\def\gsim{\mathrel{\rlap{\lower4pt\hbox{\hskip1pt$\sim$}}
    \raise1pt\hbox{$>$}}}         

\DeclareMathOperator{\sech}{sech}

\renewcommand{\tilde}{\widetilde}

\definecolor{darkgreen}{RGB}{50,120,10}

\usepackage[capitalise]{cleveref}

\newcommand{\crefrangeconjunction}{--}

\usepackage{xcolor}
\usepackage{bm}

\vspace*{-30mm}

\title{\boldmath Using {\it Gaia} DR2 to Constrain Local Dark Matter Density and Thin 
Dark Disk}
\author[a, 1]{Jatan Buch %
\note{ORCID: \url{https://orcid.org/0000-0001-6672-6750}}}
\emailAdd{jatan\textunderscore buch@brown.edu}
\author[a]{, Shing Chau (John) Leung %
\note{ORCID: \url{https://orcid.org/0000-0002-8326-9544}}}
\emailAdd{shing\textunderscore chau\textunderscore leung@brown.edu}
\author[a]{, and JiJi Fan %
\note{ORCID: \url{https://orcid.org/0000-0002-3774-5626}}} 
\emailAdd{jiji\textunderscore fan@brown.edu}
\affiliation[a]{Department of Physics, Brown University, Providence, RI, 02912, USA}
\vspace*{1cm}

\abstract{We use stellar kinematics from the latest {\it Gaia} data release (DR2) to measure the local dark matter (DM) density $\rho_{\rm DM}$ in a heliocentric cylinder of radius $R= 150 \ {\rm pc}$ and half-height $z= 200 \ {\rm pc}$. We also explore the prospect of using our analysis to estimate the DM density in local substructure by setting constraints on the surface density and scale height of a thin dark disk aligned with the baryonic disk and formed due to dissipative dark matter self-interactions. Performing the statistical analysis within a Bayesian framework for three types of tracers, we obtain ${\rho_{\rm DM}= 0.016 \pm 0.010}$ M$_\odot$/pc$^3$ for A stars; early G stars give a similar result, while F stars yield a significantly higher value. For a thin dark disk, A stars set the strongest constraint: excluding surface densities (5-12) M$_\odot$/pc$^2$ for scale heights below 100 pc with 95\% confidence. The upper bound of this constraint implies ${\lsim} \, 1\%$ of the Milky Way DM mass is present in a dissipative dark sector. Comparing our results with those derived using {\it Tycho-Gaia} Astrometric Solution (TGAS) data, we find that the uncertainty in our measurements of the local DM content is dominated by systematic errors that arise from assumptions of our dynamical analysis in the low $z$ region. Furthermore, there will only be a marginal reduction in these uncertainties with more data in the {\it Gaia} era. We comment on the robustness of our method and discuss potential improvements for future work.
}

\begin{document}

\maketitle

\section{Introduction}

The second release of data collected by the European Space Agency's {\it Gaia} telescope provides the positions and proper motions, with unprecedented precision, of more than one billion sources in the Milky Way (MW)~\cite{2016A&A...595A...1G,2018arXiv180409365G,2018arXiv180409366L,2018arXiv180409368E,2018arXiv180409367R,2018arXiv180409371S, 2018arXiv180409369C, 2018arXiv180409372K, 2018arXiv180409376L}. With the release of line-of-sight velocities for about seven million stars, DR2 also allows, for the first time, a dynamical analysis with a self-consistent measurement of the 6D phase space for a stellar population.

DR2 presents an exciting opportunity to use the vertical velocity and number density distributions of different populations of stars that trace the gravitational potential for precisely determining the total matter density, including baryons and dark matter (DM), in the local solar neighborhood. Significant progress has been made in modeling the local baryon budget (interstellar gas, stars, stellar remnants) and its uncertainties~\cite{Flynn:2006tm, Bovy:2013raa, Mckee:2015, Kramer:2016dew} since Oort's early estimate~\cite{1932BAN.....6..249O} of the baryon density. Meanwhile, kinematic methods for estimating the local DM density rely on constraining the total matter content using motions of tracers after assuming a model for the baryons and attributing any additional density, within uncertainty, to DM. These methods are based on: {\it a)} the Jeans analysis that reduces the collisionless Boltzmann equation for the phase space distribution function into a set of moment equations by integrating over all velocities, and {\it b)} the Poisson equation which uses the total matter density in all components to calculate the gravitational potential. In this work, we primarily focus on the 1D distribution function method developed by Refs.~\cite{1989MNRAS.239..571K,1989MNRAS.239..605K,1989MNRAS.239..651K, 1993AIPC..278..580F, 1994MNRAS.270..471F} and used by Refs.~\cite{Holmberg:1998xu, 2004MNRAS.352..440H} to constrain the local DM density with data from the {\it Hipparcos} satellite~\cite{vanLeeuwen:2005yx}. However, the approximations of isothermality and decoupling of radial and vertical motions in this method are only valid up to scale height $z \, {\sim} 1 \ {\rm kpc}$. Therefore, for using tracer data at high $z$, Refs.~\cite{garbari:2012, Bovy:2012tw} adopt the more general moment-based method to estimate the DM density. A non-parametric formulation of the moment-based method, described by Ref.~\cite{Silverwood2016} and implemented in Ref.~\cite{Silverwood2017}, uses SDSS/SEGUE G stars in a heliocentric cylinder with $ R \, {\sim} 1 \ {\rm kpc}$ and $0.5 \ {\rm kpc} \,  {\lsim} \, |z| \, {\lsim} \, 2.5  \ {\rm kpc}$, grouped by age, namely $\alpha$-young and $\alpha$-old stars, as tracers. Ref.~\cite{Bovy:2013raa} also uses SDSS/SEGUE G star data between $4 \ {\rm kpc} \, {\lsim} \, R \, {\lsim} \, 9 \ {\rm kpc}$ and $0.3 \ {\rm kpc} \, {\lsim} \, |z| \, {\lsim} \, 3 \ {\rm kpc}$ to constrain the stellar and DM density through action-based distribution function modeling \cite{Bovy2012a}. Their analysis incorporated the age information of tracers in a more sophisticated manner by constructing mono-abundance populations~\cite{Bovy2012d} that consist of stars with similar elemental abundances. The above discussion is by no means an exhaustive review of the different attempts at measuring the local DM density (most notably, it doesn't address dynamical measurements made by, for example, Refs.~\cite{Salucci:2010qr, Pato:2015dua}); instead, we refer interested readers to Refs.~\cite{2013PhR...531....1S, Read:2014qva}.

Besides determining the local DM density, one could apply the data of stellar kinematics to constrain more exotic DM distributions. A recent example is dissipative DM scenarios, in which self-interactions among a sub-dominant component of DM dissipates energy to form a (possibly thin) dark disk (DD), co-rotating with the baryonic disk~\cite{Fan:2013tia, Fan:2013yva}. Possible effects of such a DD and variants of dissipative DM scenarios have been studied further in Refs.~\cite{CyrRacine:2012fz, McCullough:2013jma, Fan:2013bea, Randall:2014lxa, Fischler:2014jda, Foot:2014uba, Randall:2014kta, Reece:2015lch, Foot:2016wvj, Shaviv:2016umn, Kramer:2016dqu, Kramer:2016dew, Agrawal:2017rvu, Agrawal:2017pnb, Buckley:2017ttd, DAmico:2017lqj, Caputo:2017zqh, Vattis:2018aen, Outmazgine:2018orx, Alexander:2018lno}. The formation of the disk from DM self-interactions is highly debatable~\cite{Ghalsasi:2017jna} and numerical simulations using a cooling prescription (as in Ref.~\cite{Rosenberg:2017qia}) are still absent. Inspite of these uncertainties regarding its formation, it is still worthwhile to use the stellar data to test the simplest possibility of a thin DD aligned with the baryonic disk and parametrized by only two parameters: the surface density, $\Sigma_{DD}$ and a scale height, $h_{DD}$. This has been carried out using {\it Hipparcos} data in Ref.~\cite{Kramer:2016dqu} and {\it Tycho--Gaia} Astrometric Solution (TGAS), a joint solution combining {\it Tycho}-2 catalog with early {\it Gaia} data, in Ref.~\cite{Schutz:2017tfp}.

\begin{figure*} 
	\begin{center}
\includegraphics[width=0.9\linewidth]{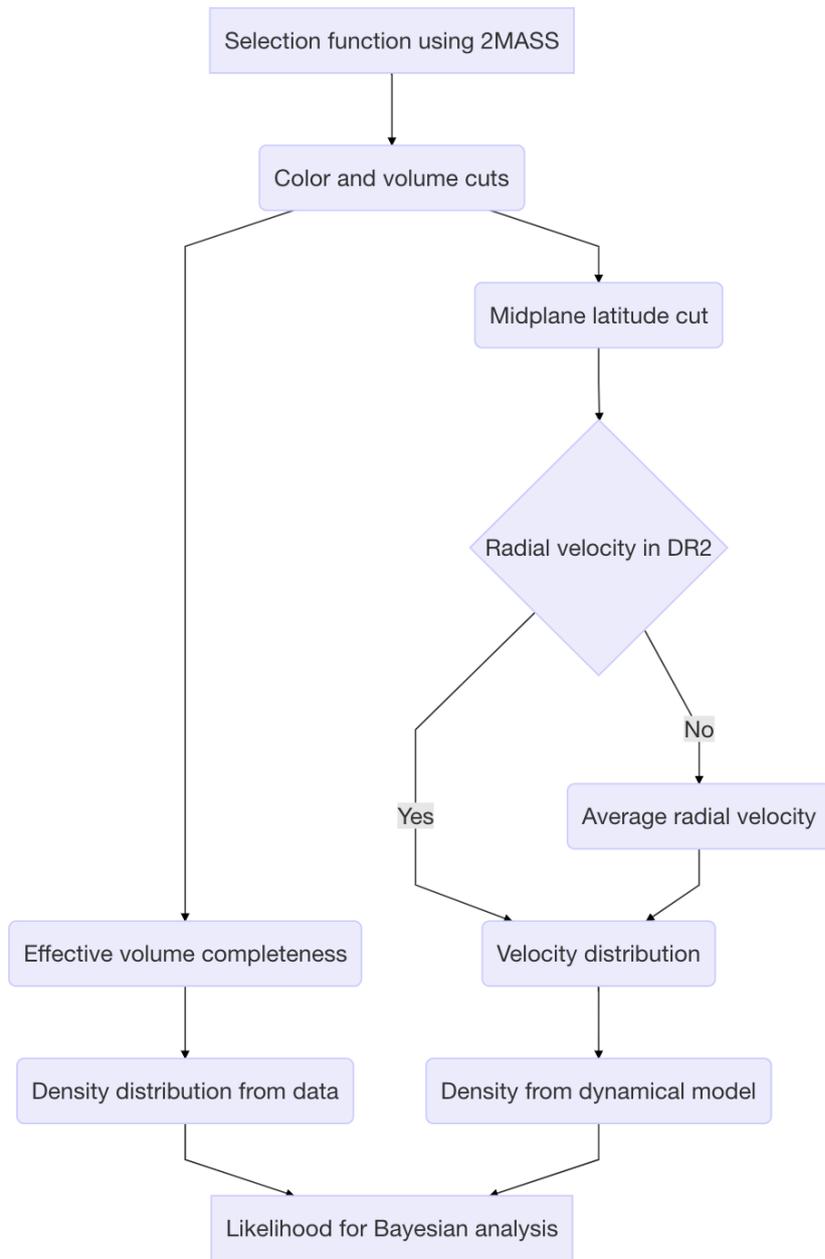}
	\end{center}
\caption{Flowchart of our analysis.}
\label{fig:flowchart}
\end{figure*}

In this article, we work with the second  {\it Gaia} data release (DR2)~\cite{2018arXiv180409365G} to estimate the local DM density as well as constrain thin DD models assuming that the DD is aligned with the baryonic disk. We follow the method in Refs.~\cite{Holmberg:1998xu, Kramer:2016dqu, Schutz:2017tfp} and use A, F and early G dwarf stars in the {\it Gaia} catalog as the tracers. In Section~\ref{sec:data}, we discuss the details of {\it Gaia} DR2 and the empirically determined survey selection function (Section~\ref{sec:Selection_Function}), which we use to construct the vertical number density profile and midplane velocity distribution for each tracer population in Sections~\ref{sec:numberdensity} and \ref{sec:midplane_velocity} respectively. Our fiducial analysis is described in Section~\ref{sec:analysis}. We use the 1D distribution function method summarized in Section~\ref{sec:PJ_Theory} to construct the equilibrium number density for the parameters of our mass model described in Section~\ref{sec:mass_model}. In Sections~\ref{sec:basic_setup} and \ref{sec:BM_discuss}, we introduce a Bayesian framework for comparing our predicted density with data for each tracer population while taking into account the uncertainties due to potential non-equilibrium effects. The important steps of our analysis are outlined as a flowchart in Fig.~\ref{fig:flowchart}. While our method is not new, we obtain interesting results, some of which are quite different from those based on TGAS. We present our results for the local DM content using {\it Gaia} DR2 in Section~\ref{sec:localdm_no_dd} and~\ref{sec:localdm_dd}, and list various sources of systematic uncertainties in the context of our method in Sections~\ref{sec:density_val}-\ref{sec:bary_dm}. Differences between constraints derived using DR2 and TGAS are discussed in Section~\ref{sec:localdm_tgas}. We conclude and comment on future directions in Section~\ref{sec:outlook}.

\section{Data Selection}
\label{sec:data}
The {\it Gaia} DR2 catalog contains ${\sim}1.7$ billion sources, among which ${\sim}1.3$ billion sources have a five-parameter astrometric solution: $(\alpha, \delta, \mu_{\tilde{\alpha}}, \mu_\delta, \varpi)$, representing positions and proper motions along the right ascension and declination, and parallax respectively. We emphasize that for DR2, the parallaxes and proper motions are based solely on {\it Gaia} measurements, unlike DR1 which depends on the {\it Tycho}-2 catalog~\cite{2018arXiv180409365G, 2018arXiv180409366L}. DR2 also provides photometric data in three passbands, $G$, $G_{\rm BP}$, and $G_{\rm RP}$, for a majority of the sources in the range $3 \, {\lesssim} \, G \, {\lesssim} \, 21$~\cite{2018arXiv180409368E, 2018arXiv180409367R}. Another new feature in DR2 is the line-of-sight radial velocities, $v_r$, for ${\sim} 7.2$ million stars brighter than $G_{\rm RVS} \, {\sim} \, 12$~\cite{2018arXiv180409371S, 2018arXiv180409369C, 2018arXiv180409372K}.

Despite a significant increase in statistics, the DR2 catalog is still incomplete for stars with $G<12$. We calculate the completeness fraction for the DR2 catalog by comparing it with the {\it Two Micron All-Sky Survey} ({\it 2MASS}\,) Point Source Catalog~\cite{2006AJ....131.1163S} that is $99\%$ complete down to its faint magnitude limit $J=15.8$ over almost the entire sky. The {\it 2MASS} catalog also provides the angular positions, $(\alpha, \delta)$, for each source along with its $J$ and $K_s$ magnitudes, which we use to categorize stars in DR2 by stellar type.

We query the {\it Gaia} archive\footnote{\url{https://gea.esac.esa.int/archive/}} for DR2 cross-matched with the {\it 2MASS} catalog, requiring the apparent magnitude $J < 13.5$, thereby cutting away stars that are either too dim for the main sequence, or too distant from the Sun.\footnote{In our selected volume, the apparent magnitude of all tracer stars satisfies $J < 12$.} As we discuss in the following section, the star counts in the {\it 2MASS} catalog with $J < 13.5$ are then used to determine the selection function, $S(J, J-K_s, \alpha, \delta)$, defined as the fraction of stars at a given $(J, J-K_s, \alpha, \delta)$ in the sky that are contained in the DR2 catalog.
 \subsection{Preliminaries: selection function, color and volume cuts}
 \label{sec:Selection_Function}
\begin{figure*}[!htb]
\subfloat{%
\includegraphics[scale=0.35]{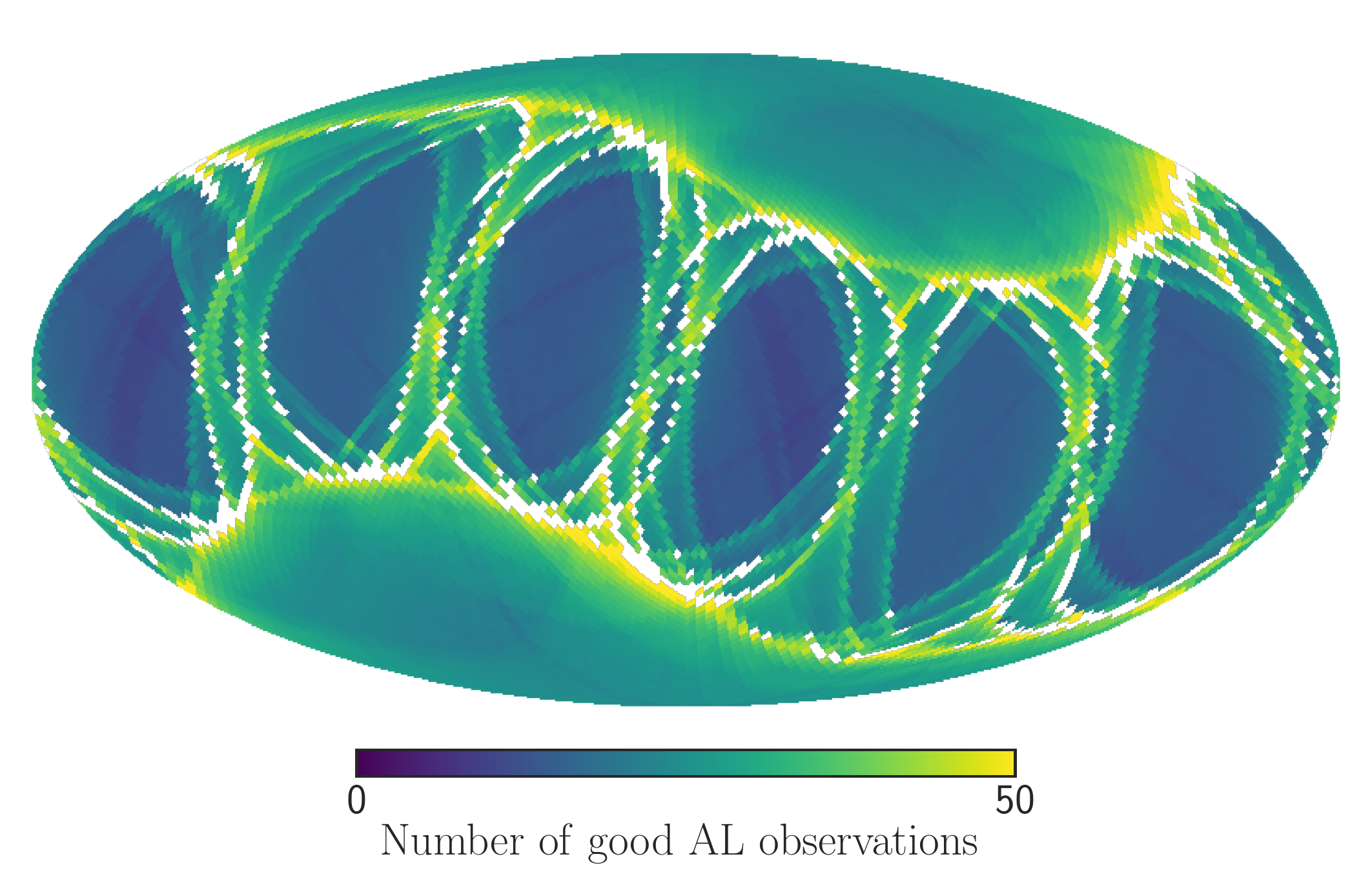}
}
\subfloat{%
\includegraphics[scale=0.35]{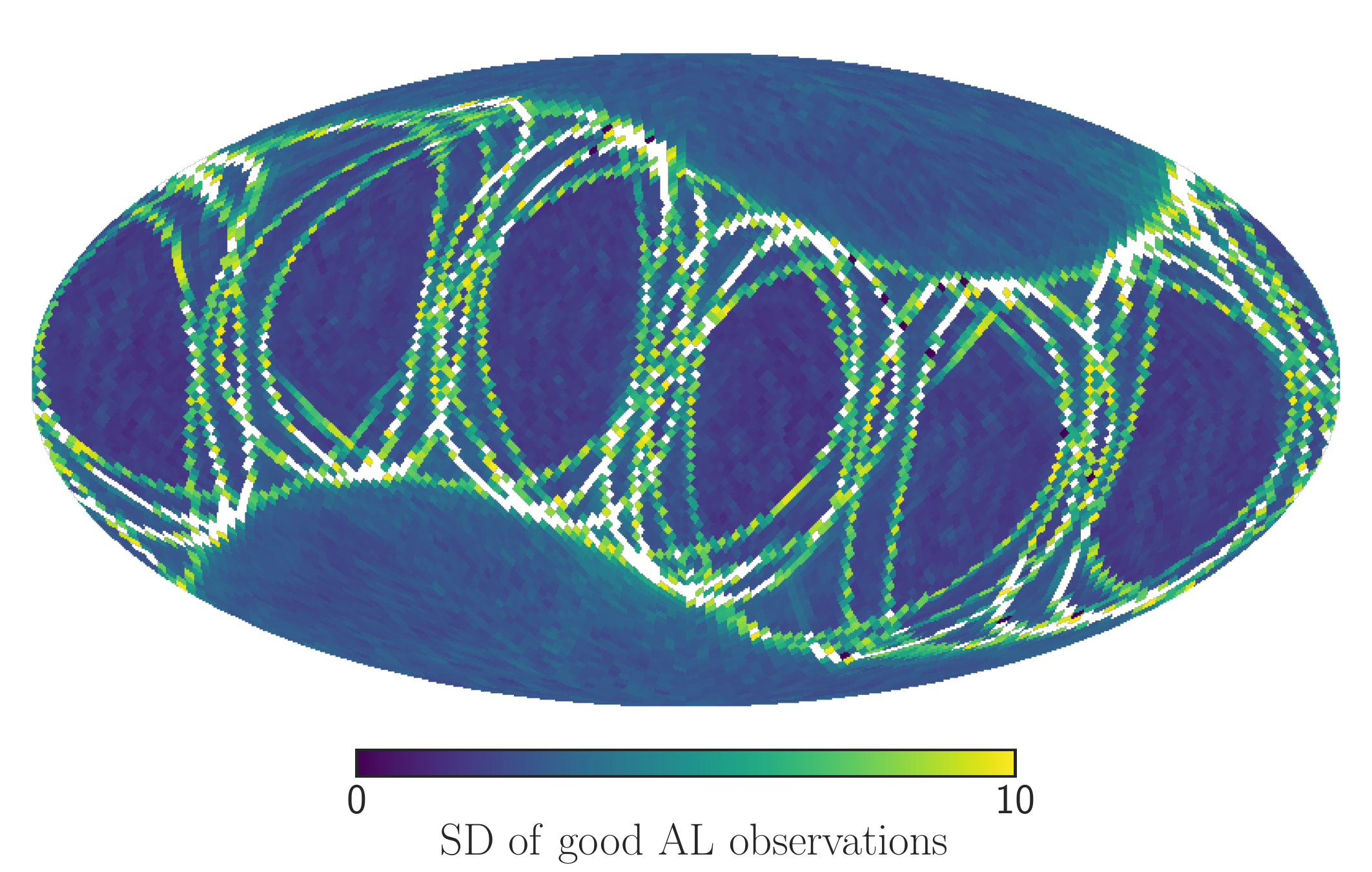}
}
\caption{Skymaps showing the number (left) and variance (right) of good AL observations in 
$3.36 \ {\rm deg}^2$ ($N_{\rm side}= 2^5$) HEALPix pixels. The white regions are the parts of the sky which do not pass our selection cuts defined in the main text. }
\label{sfig:skymap}
\end{figure*}

Since there has been no official release by the {\it Gaia} collaboration, we follow the procedure introduced in Ref.~\cite{Bovy:2017} to determine the selection function for the cross-matched data set. As a first step, using the \texttt{gaia\_tools} package,\footnote{\url{https://github.com/jobovy/gaia_tools}} we identify parts of the sky that satisfy,
\begin{enumerate}[label=(\roman*)]
\item Mean number of along-scan (AL) observations $\geq 8.5$;
\item Spread in the number of AL observations $\leq 10$.
\end{enumerate}
After these cuts, 95.6\% of the sky remains as shown in Fig.~\ref{sfig:skymap}. The overall completeness in these parts of the sky is largely isotropic, implying that our selection function only has a negligible dependence on sky position. Thus, for the rest of our analysis, we use the approximation~${S(J, J-K_s, \alpha, \delta) \approx S(J, J-K_s)}$. A similar approximation has been adopted for analyzing TGAS data in Ref.~\cite{Bovy:2017}, in which stronger cuts were used to identify the `good' parts of the sky.

Although the uncertainties of astrometric parameters in DR2 are an order of magnitude improvement over TGAS, there are still several systematic effects that we need to include in our data analysis. As suggested by Ref.~\cite{2018arXiv180502650Z}, we add 0.03 mas to the reported parallax to account for the global offset. Following Ref.~\cite{2018arXiv180409376L}, we also add in quadrature a systematic uncertainty of $\pm 0.1 \ {\rm mas}$ and $\pm 0.1 \ {\rm mas/yr}$ respectively to the reported values of parallax and proper motion uncertainties. We check that our analysis is not sensitive to the exact values of these numbers, only their order-of-magnitude estimate.

Next, we identify stellar populations that: a) are tracers of the local galactic potential (see Sec. 3.6 of Ref.~\cite{Read:2014qva} for a recent overview), and b) show a reasonable change in their number densities within the solar neighborhood (defined below). The most important factor in selecting tracer stars is their sensitivity to disequilibria, which could result in incompatible $\rho_{\rm DM}$ measurements~\cite{banik:2016}. While there is some disagreement in the literature about which stars, old~\cite{1992ApJ...389..234B, Holmberg:1998xu} or young~\cite{Silverwood2017}, are more suitable for an equilibrium analysis, we follow Ref.~\cite{Silverwood2017} in choosing younger A (A0-A9), F (F0-F9), and early G (G0-G3) dwarf stars (simply stars henceforth) which have lower scale heights and consequently shorter equilibration timescales, instead of older stars, in our analysis. 

We use the mean dwarf stellar locus from Ref.~\cite{2013ApJS..208....9P} to define each stellar type $t$ based on its absolute magnitude $\overline{M}_{J^t}$ and color $\left(\bar{J}^t -\bar{K}_s^t \right)$,
\beq
\begin{aligned}
J^t + 5  \log_{10} ({\varpi^t \, (\text{mas})}) - 10  \in \{\overline{M}_{J^t, \, \text{min}}, \overline{M}_{J^t, \, \text{max}} \} \\[0.5ex]
\left(J^t - K_s^t \right) \in \{\left(\bar{J}^t -\bar{K}_s^t \right)_{\text{min}}, \left(\bar{J}^t -\bar{K}_s^t \right)_{\text{max}} \}
\end{aligned}
\label{eq:cm_cuts}
\eeq
and include all A, F, and early G stars in a heliocentric cylinder with radius~$R = 150$ pc and half-height~$z= 200$ pc,
\beq
\left| \, \left(\frac{1}{\varpi}\right) \cos b \, \right| < 150 \, {\rm pc}, \quad \quad \left| \, \left(\frac{1}{\varpi}\right) \sin b \, \right| < 200 \, {\rm pc}
\label{eq:vol_cuts}
\eeq
where all the un-barred quantities are data from our cross-matched catalog. In Eq.~\ref{eq:cm_cuts} and~\ref{eq:vol_cuts}, we use $(1/\varpi)$ as an estimator of the distance, where $\varpi$ is the noisy parallax reported in the DR2 catalog. We treat it as an unbiased estimator since our tracer sample contains nearby stars within a distance, $(1/\varpi) < 250$ pc, and a low parallax uncertainty, $\sigma_\varpi {\lsim} \, 0.1$ mas. We discuss the uncertainties in our results due to the choice of cuts in Section~\ref{sec:density_val}.

 \subsection{Vertical number density distribution} 
 \label{sec:numberdensity}
To make any meaningful inferences about stellar dynamics, we require the volume complete number density of stars. Since {\it Gaia} DR2 is incomplete, the number density constructed from data must be corrected for the survey selection function (determined in Sec.~\ref{sec:Selection_Function}) to reflect the underlying {\it true} Poisson-distributed density. Ref.~\cite{Bovy:2017} accomplishes this by deriving a bin-by-bin normalization for each stellar population, defined as the {\it effective volume completeness}, under the assumption that all observed stars are samples from an inhomogeneous Poisson process. We follow the same procedure and summarize its important details in Appendix~\ref{sec:poisson}. 

\begin{figure*}[!ht]
	\begin{center}
\includegraphics[scale=.57]{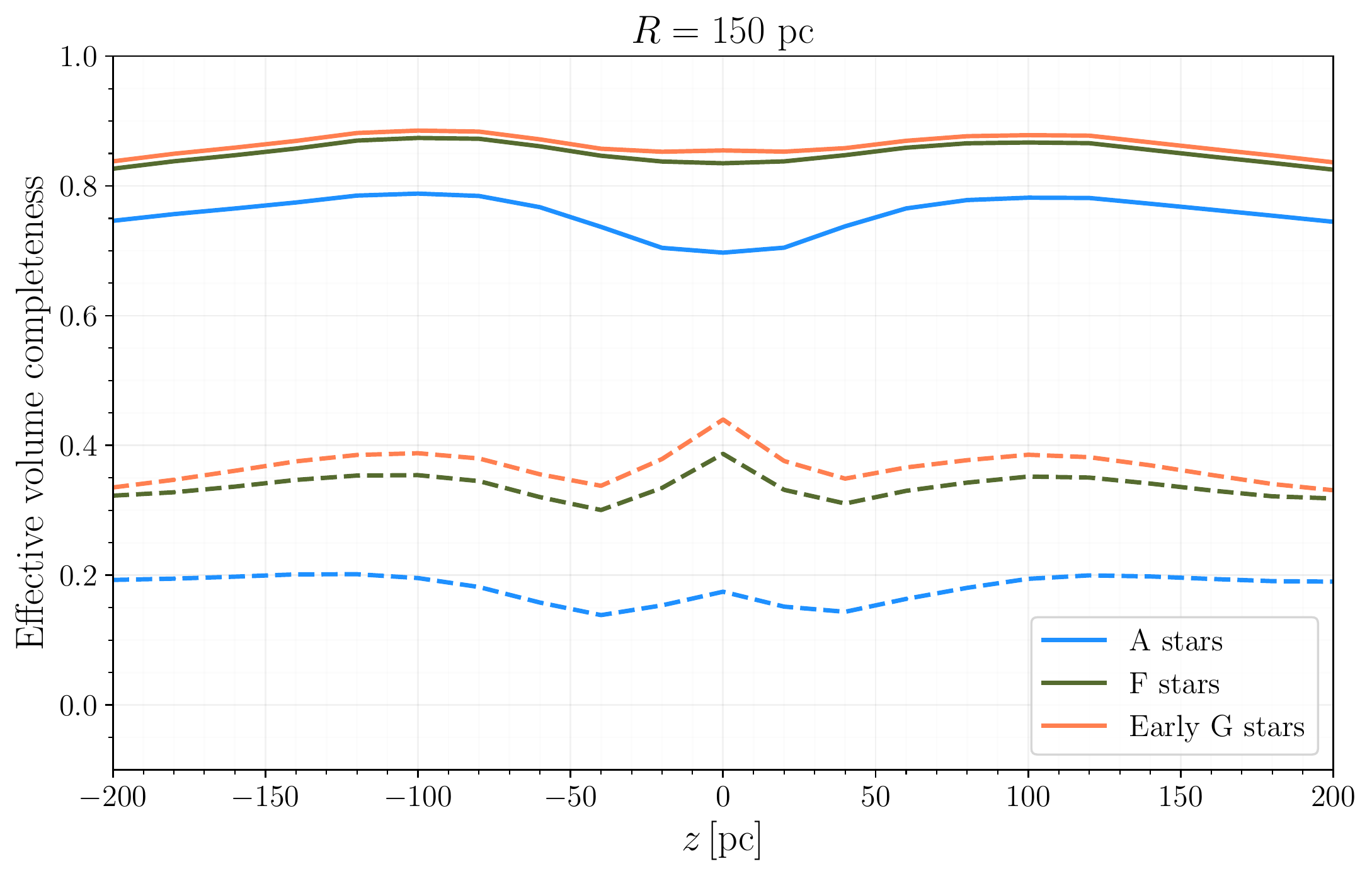}
\caption{Effective volume completeness for each stellar type. The completeness of DR2 (solid) is significantly improved as compared to TGAS (dashed) for A (blue), F (green), and early G (orange) stars.}
\label{fig:completeness}
	\end{center}
\end{figure*}

We calculate the effective volume completeness for our DR2 tracer populations using the \texttt{gaia\_tools} package and plot it as a function of scale height in Fig.~\ref{fig:completeness}. We also include the effective volume completeness for TGAS data as a reference, and note that the DR2 sample is significantly more complete.

 \begin{table}[!h]
\centering
\begin{tabular}{ | c  c | c  c | c  c |}  \hline
 \multicolumn{2}{|c}{Data set} &\multicolumn{2}{|c|}{{\it Gaia} DR2} & \multicolumn{2}{|c |}{TGAS} \\ \hline
  Type & Subtype & Total & Midplane & Total & Midplane\\ \hline\hline
  A & A0-A9 & 4445  & 321 & 1729  &  182 \\
  F & F0-F9 & 37707 & 2253 & 16789 & 1308 \\
  Early G & G0-G3 & 43332 & 2188 & 18653 & 1205 \\ \hline
\end{tabular}
\caption{Star counts in DR2 and TGAS catalogs for the heliocentric cylinder and the midplane 
region ($|b| < 5^{\text{o}}$) inside it.}
\label{table:counts}
\end{table}

\begin{figure*}[!ht]
	\begin{center}
\includegraphics[scale=.6]{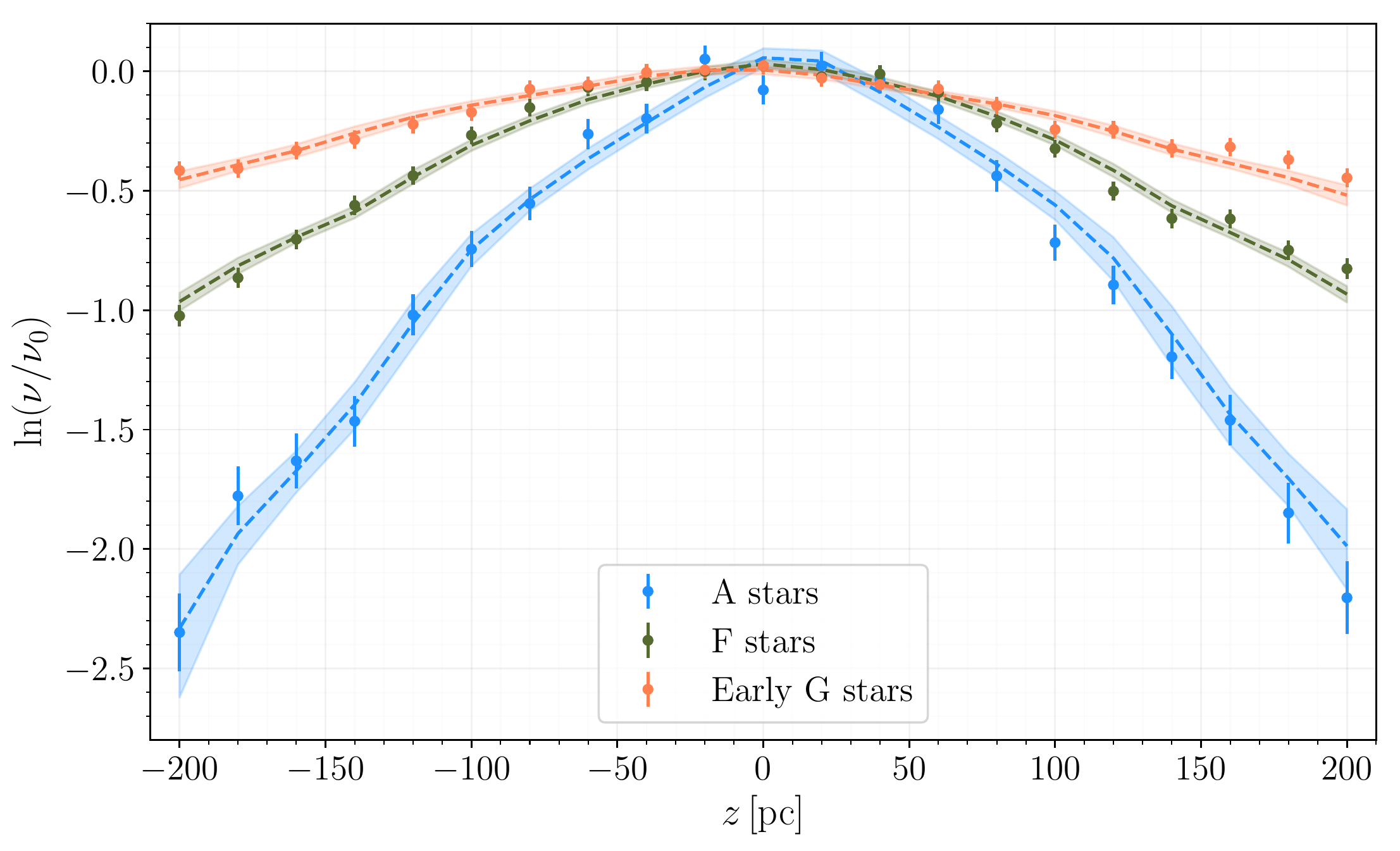}
\caption{Binned vertical number density profiles for A, F, and early G stars. Also shown is the median predicted density (dotted line) and the $68\%$ confidence interval (shaded region) for each tracer population obtained from our dynamical analysis described in Sec.~\ref{sec:analysis}.}
\label{fig:density}
	\end{center}
\end{figure*}

There are 4445 A, 37707 F, and 43332 early G stars in the solar neighborhood defined by our heliocentric cylinder. The volume complete vertical number density for the $i^{\rm th}$ tracer population, $\nu_ i^{\rm data}$  shown in Fig.~\ref{fig:density}, is obtained by dividing the number counts with the effective volume completeness in each $z$ bin. We choose 20 pc as the bin size based on parallax uncertainties as discussed in Appendix~\ref{sec:Uncertainty_analysis}. However, varying the bin size doesn't significantly affect the results of our analysis. We also present a comparison of star counts in the full volume and in the midplane (defined to be the region with $|b| < 5^{\text{o}}$ in the cylinder) between DR2 and TGAS in Table~\ref{table:counts}. 

The number density uncertainty for the $k^{\rm th}$ $z$ bin, $\left( \sigma^{2}_{\ln \nu_ i}\right)^{\rm data}$, is obtained by adding in quadrature the statistical uncertainty, $\sqrt{N_k}$ (see Eq.~\ref{eq:den_uncertain}), and a 3\% systematic uncertainty due to dust extinction. We expect the dust extinction to be important in the visible spectrum such as the $B$ and $V$ colors used in {\it Hipparcos} catalog, or the $G_{\rm BP} $ and $G_{\rm RP}$ used in DR2. However, colors in the infrared spectrum, {\it i.e.} the $J$ and $K_s$ colors used in our cross-matched DR2-{\it 2MASS} catalog, are associated with longer wavelengths and therefore much less affected by galactic dust. Ref.~\cite{Bovy:2017} finds that the effect of dust reddening on the number density of stars in the solar neighborhood defined using $J$ and $K_s$ is only $\lesssim 3$\% and mostly affects the overall normalization. Thus, instead of doing a full dust analysis, we conservatively adopt an overall $3\%$ correction for all three types of stars. We also find that this is only a sub-dominant uncertainty and doesn't significantly affect our constraints.

\subsection{Midplane velocity distribution}
\label{sec:midplane_velocity}
The last ingredient we need from the data is the vertical velocity distribution in the midplane, {\it i.e.} in the region near $z=0$. The vertical velocity of a star is approximated  by,
\beq
w = w_\odot + \frac{\kappa \mu_b}{\varpi} \cos b + v_r \sin b,
\label{eq:w}
\eeq
where $w_\odot$ is the Sun's vertical velocity that we determine by fitting a Gaussian distribution to the data, $\kappa = 4.74$ km yr s $^{-1}$ is a unit conversion constant, $\mu_b$ is the proper motion along the galactic latitude $b$ in mas/yr, $\varpi$ is the parallax in mas, and $v_r$ is the heliocentric radial velocity in km/s. Following the discussion below Eq.~\ref{eq:vol_cuts}, we note that for regions near the midplane, $(1/\varpi)$ is an unbiased estimator for distance in the formula for vertical velocity.

The `midplane region' can be defined by imposing a cut on either the: a) galactic latitude $|b|$, or b) height $|z|$.\footnote{At larger $b$ and consequently larger $z$, the kinematically hotter stars broaden the distribution~\cite{Holmberg:1998xu}. Meanwhile, simply choosing stars with $z=0$ yields poor statistics.} In our analysis, we follow Refs.~\cite{Kramer:2016dqu, Schutz:2017tfp} in choosing $|b| < 5^\circ$ as our midplane cut, and are left with 310, 2213 and 2166 A, F and early G stars respectively. We use radial velocities reported in DR2, when available, substituting $v_r$ by its mean value otherwise,
\beq
\langle v_R \rangle = - u_\odot \cos l \cos b - v_\odot \sin l \sin b - w_\odot  \sin b,
\label{eq:meanvr}
\eeq
where $u_{\odot} = 11.1 \pm 0.7^{\rm stat} \pm 1.0^{\rm sys}$ km/s and $v_{\odot} = 12.24 \pm 0.47^{\rm stat} \pm 2.0^{\rm sys}$ km/s as the $x$- and $y$-component of the Sun's peculiar velocity~\cite{2010MNRAS.403.1829S}. We also note that as $\sin b \ll 1$ (in radians) in our midplane region, $v_r$ only has a subdominant contribution to $w$ in Eq.~\ref{eq:w}, and the velocity uncertainties have a negligible effect on our results. We determine the Sun's vertical velocity, $w_\odot$, by considering the weighted mean of the best-fit Gaussian distribution to each tracer population's vertical velocity data in the midplane region. Our estimate, $\overline{w}_\odot=6.9\pm 0.2$ km/s, is consistent within $1\sigma$ of the value quoted in Ref.~\cite{2010MNRAS.403.1829S}. 

\begin{figure*}[!h]
\subfloat{%
\includegraphics[height= 7cm, width=.49\linewidth]{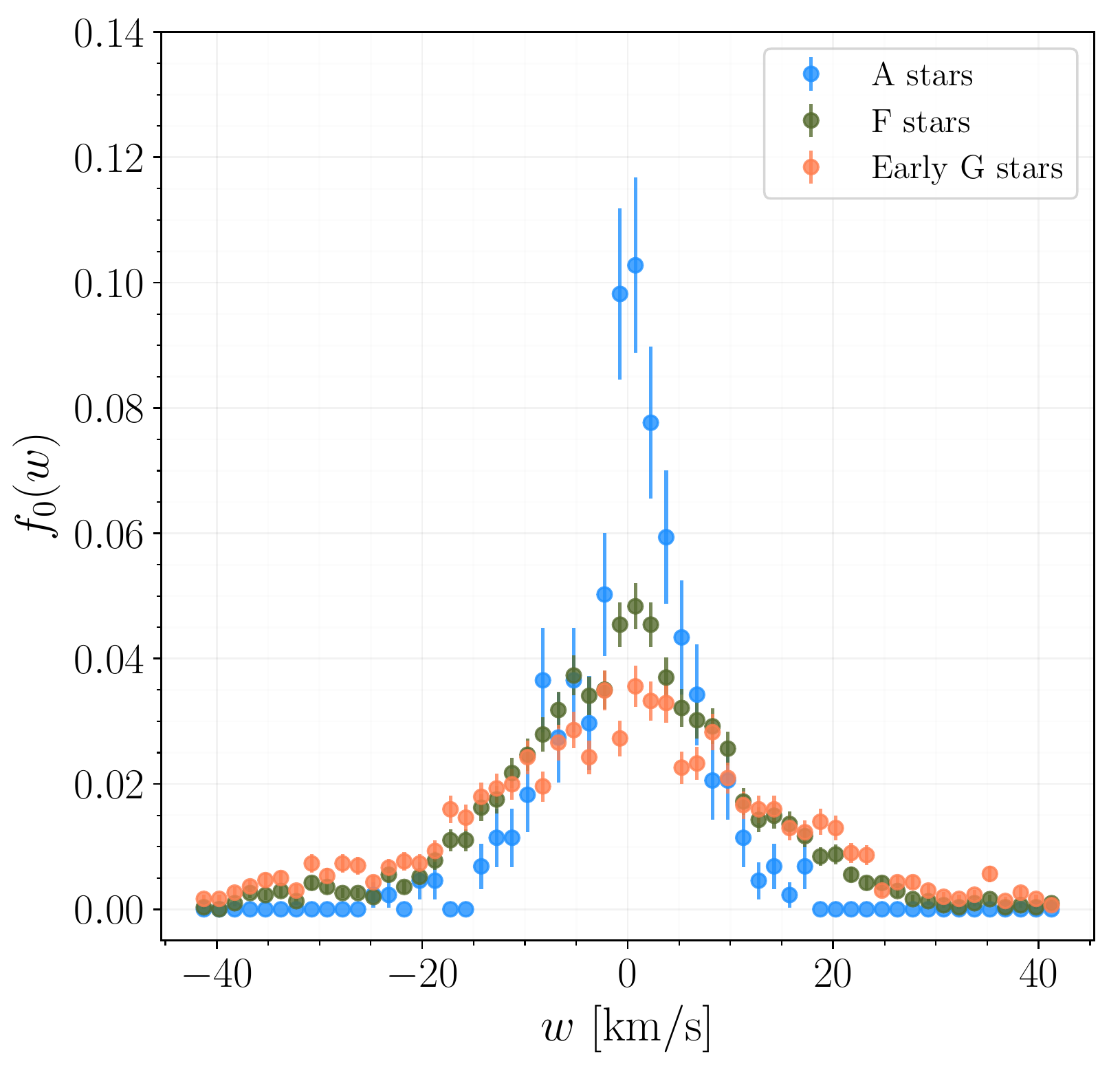}
}
\subfloat{%
\includegraphics[height= 7cm, width=.49\linewidth]{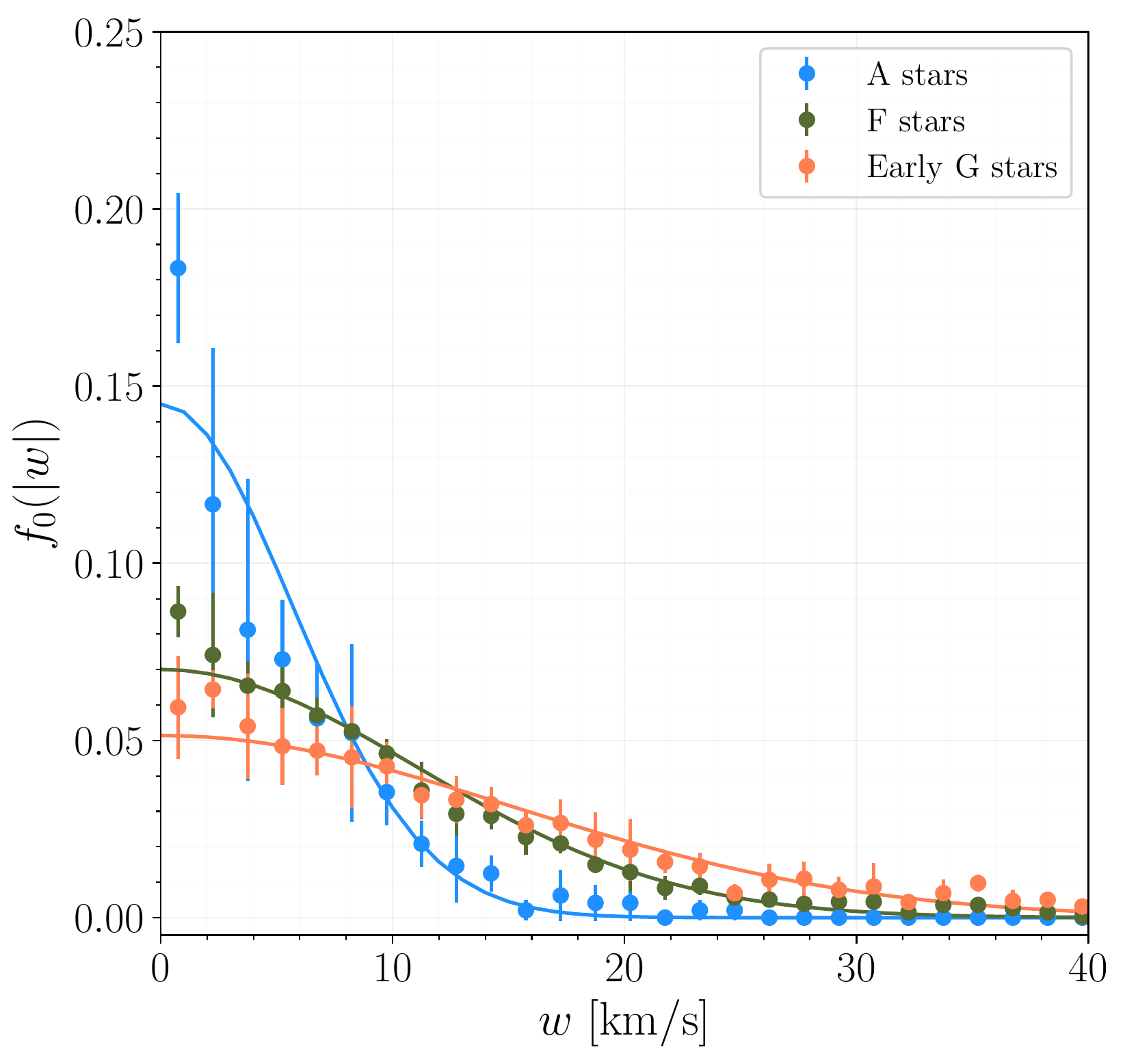}
}
\caption{Midplane velocity distributions of A, F, and early G stars after subtracting $w_\odot$  
(left). The best-fit Gaussian distribution to $f_0(|w|)$ with error bars that include contributions 
from the statistical uncertainty due to Poisson error and the asymmetry in $-|w|$ and $+|w|$ bins 
(right).}
\label{sfig:velocity}
\end{figure*}

Subtracting $\overline{w}_\odot$ from the stars' vertical velocity, we find the distributions are roughly symmetric about $w=0$. The resultant normalized midplane vertical velocity distribution $f_0(w)$ with a $w$-bin size of 1.5 km/s (see Appendix~\ref{sec:Uncertainty_analysis} for more details about this choice) is plotted in the left panel of Fig.~\ref{sfig:velocity}. We consider the asymmetry between star counts in $-|w|$ and $+|w|$ bins to be the systematic uncertainty, which may be attributed to non-equilibrium effects. We illustrate the magnitude of this uncertainty in the right panel of Fig.~\ref{sfig:velocity} by adding it in quadrature with the statistical error for every $w$ bin. In practice, however, we propagate these errors into the uncertainty of the prediction density as described in Sec.~\ref{sec:basic_setup}. 

We also explore the possibility of using the $z$-cut~\cite{garbari:2011} in Appendix~\ref{sec:midplane_bzcut} by including the radial velocities reported in DR2. Unfortunately, DR2 only contains radial velocities for approximately 2\% of A stars, 53\% of F stars, and 62\% of early G stars within $|z| < 20$ pc. We check that the percentage of tracers with radial velocity doesn't change significantly for values of $|z| \lesssim 50$ pc. In that case, only including stars with available $v_r$ could potentially introduce a selection bias, while approximating $v_r$ by its mean value might result in large errors at higher $b$ (even at low $z$). Thus, defining the midplane region using a $z$-cut isn't a viable option currently, but that could change with future data releases.

\section{Fiducial Analysis} \label{sec:analysis}
The main objective of our analysis is to infer the matter content of the solar neighborhood by fitting data observed by {\it Gaia} to a dynamical model. Sec.~\ref{sec:PJ_Theory} summarizes the 1D distribution function method~\cite{1989MNRAS.239..571K,1989MNRAS.239..605K,1989MNRAS.239..651K, 1993AIPC..278..580F, 1994MNRAS.270..471F} we adopt for constructing the equilibrium number density of a tracer population, while the parameters of our gravitational potential model are described in Sec.~\ref{sec:mass_model}. We outline the Bayesian statistical framework used for constraining these parameters in Secs.~\ref{sec:basic_setup} and~\ref{sec:BM_discuss}.

	\subsection{Equilibrium density modeling}\label{sec:PJ_Theory}

A self-gravitating stellar population with phase space distribution function (DF), $f(\mathbf{x}, \mathbf{v})$, satisfies the collisionless Boltzmann equation (CBE),
\beq
\frac{D f}{D t} \equiv \frac{\partial f}{\partial t} + (\nabla_{\mathbf{x}} f) \, .\, \mathbf{v}-  (\nabla_{\mathbf{x}} \Phi) \, . \,(\nabla_{\mathbf{v}} f) =0.
\label{eq:boltzmann_eq}
\eeq 
where $\mathbf{x}$ and $\mathbf{v}$ are the positions and velocities respectively, and $\Phi$ is the gravitational potential. Assuming axisymmetry for the local solar neighborhood, we use cylindrical polar coordinates for the rest of our analysis.

At this stage, we make two critical assumptions that allow us to drop the partial time derivative term in Eq.~\ref{eq:boltzmann_eq}: a) populations of all tracer stars are in equilibrium,\footnote{This assumption is central to {\it all} dynamical analyses of stars in the MW. We discuss its validity at some length in Sec.~\ref{sec:validation}.} , and b) the potential is time-independent, since the timescale for galactic processes is much longer than {\it Gaia}'s mission lifetime. Moreover, since we're interested in the dynamics of stars very close to the galactic plane ($|z| \lesssim 0.5$ kpc), we approximate the DF to be of the form,\footnote{In the language of Jeans modeling, this follows from the observed smallness of the so-called `tilt term' that couples the vertical and radial motions. }
\beq
f(\mathbf{x}, \mathbf{v}) \equiv f_{r, \phi} (r, v_r, \phi, v_\phi) \, f_{z} (z, v_z).
\label{eq:distfunc}
\eeq
Separability of the DF implies that the motion of a stellar population $i$ in the $z$-direction is independent of $\{R, \phi\}$, and follows the 1D CBE,
\beq
w \, \frac{\partial f_{z, i}}{ \partial z} - \frac{\partial \Phi}{\partial z} \, \frac{\partial f_{z, i}}{\partial w} =0.
\label{eq:boltzman_z}
\eeq 
which has a general solution of the form, $f_{z, i}(z, w)= F_z (E_z) \equiv F_z\left(w^2/2+\Phi(z)\right)$, where the vertical energy is defined as, $E_z = \frac{1}{2}w^2+\Phi(z)$.

We integrate the DF over vertical velocity (or equivalently, energy) to obtain the stellar number density~\cite{1993AIPC..278..580F},
\beq
\nu_i(z) &=&  \int_{-\infty}^\infty dw \, f_{z, i}(z,w) = \int_{-\infty}^\infty dw \, f_{z, i}(E_z)  \nonumber \\ 
&=& 2 \int_0^\infty dw f_{z=0,\, i}(0, \sqrt{w^2+2\Phi(z)}) \nonumber \\
&=& 2 \int_{\sqrt{2\Phi(z)}}^{\infty} \frac{f_{0, \, i}(|w|) \; w \; dw}{\sqrt{w^2 - 2 \Phi(z)}},
\label{eq:predict}
\eeq
where $f_{0, \, i}(|w|)$ is the midplane (absolute) velocity distribution of stars of a tracer population $i$, which we determine from {\it Gaia} data using the procedure described in Sec.~\ref{sec:midplane_velocity}.

	\subsection{Local matter content: baryons, halo DM, and a thin DD}\label{sec:mass_model}
We calculate the gravitational potential due to the total mass density, $\rho_{\rm tot}$, in the solar neighborhood through the Poisson equation,
\beq
\nabla^2 \Phi = \frac{\partial^2 \Phi}{\partial z^2} + \frac{1}{r} \frac{\partial}{\partial r} \left(r\frac{\partial \Phi}{\partial r} \right)  = 4 \pi G \rho_{\rm tot}(z),
\label{eq:poisson_eq}
\eeq
where the radial term, $\frac{1}{r} \frac{\partial}{\partial r} \left(r\frac{\partial \Phi}{\partial r} \right)$, effectively contributes a constant mass density\footnote{For an axisymmetric system, the radial term is related to Oort's constants. Strictly speaking, the Oort's constants and consequently the radial term also depend on $z$. However, since our tracers only explore a small volume close to the midplane, the variation is smaller than the measurement uncertainty~\cite{Bovy:2012tw}.} with a value $(3.4 \pm 0.6) \times 10^{-3}$ M$_\odot$/pc$^3$ determined from the TGAS data~\cite{2017MNRAS}. We have also assumed that the $R$ and $z$ components of the potential can be decoupled such that, $\Phi(R, z) = \Phi(R) + \Phi(z)$. 

\begin{table}[!ht]
\centering
	\begin{tabular}{| c  | c  | c  |}
	\hline
	 Baryonic components & $\rho(0)$ [M$_{\odot}$/pc$^3$] & $\sigma_z$ [km/s]	 \\ \hline \hline
         Molecular gas (H$_2$) & $0.0104 \pm 0.00312$ & $3.7 \pm 0.2$ \\
         Cold atomic gas (H$_{\rm I}$(1)) & $0.0277 \pm 0.00554$ & $7.1 \pm 0.5$ \\
         Warm atomic gas (H$_{\rm I}$(2)) & $0.0073 \pm 0.0007$ & $22.1 \pm 2.4$ \\
         Hot ionized gas (H$_{\rm II}$) & $0.0005 \pm 0.00003$ & $39.0 \pm 4.0$ \\
         Giant stars & $ 0.0006 \pm 0.00006$ & $15.5 \pm 1.6$ \\
         $M_V < 3$ & $ 0.0018 \pm 0.00018$ & $7.5 \pm 2.0$ \\
         $3<M_V<4$ & $ 0.0018 \pm 0.00018$ & $12.0 \pm 2.4$ \\
         $4<M_V<5$ & $0.0029 \pm 0.00029$ & $18.0 \pm 1.8$ \\
         $5<M_V<8$ & $ 0.0072\pm 0.00072$ & $18.5 \pm 1.9$ \\
         $M_V > 8$ (M dwarfs) & $0.0216 \pm 0.0028$ & $18.5 \pm 4.0$ \\
         White dwarfs & $0.0056 \pm 0.001$ & $20.0 \pm 5.0$ \\
         Brown dwarfs & $0.0015 \pm 0.0005$ & $20.0 \pm 5.0$ \\
         \hline
	\end{tabular}
\caption{Bahcall model consisting of midplane densities and velocity dispersions for $N_b$ baryonic components adapted from Ref.~\cite{Schutz:2017tfp}. The values and uncertainties, both observational and estimated, for all components have been compiled from Refs.~\cite{Flynn:2006tm, Read:2014qva, Mckee:2015, Kramer:2016dew}} \label{tab:baryons}
\end{table}

The total mass density contains contributions from $N_b$ baryon components, DM in the halo, and other gravitational sources such as a thin DD. The baryon mass density, $\rho_{b}$, is given by the Bahcall model that consists of a set of isothermal components for gas, stars, and star remnants~\cite{1984ApJ...276..169B, 1984ApJ...276..156B, 1984ApJ...287..926B},
\beq
\rho_b(z) = \sum_{i=1}^{N_b} \rho_i(0) e^{-\Phi(z)/\sigma_{z;i}^2} 
\label{eq:bahcall}
\eeq
where each isothermal component is characterized by its midplane density, $\rho(0)$, and vertical velocity dispersion, $\sigma_{z}$ as shown in Table~\ref{tab:baryons}.


We approximate the contribution of DM density from the smooth halo near the disk, $\rho_{\rm DM}$, to be constant. As shown by Eq. (28) in Ref.~\cite{Bovy:2012tw}, the DM density at or below $200$ pc is equal to that in the midplane up to a $2\%$ correction. 

\begin{figure*}[!ht]
	\begin{center}
\includegraphics[scale=.5]{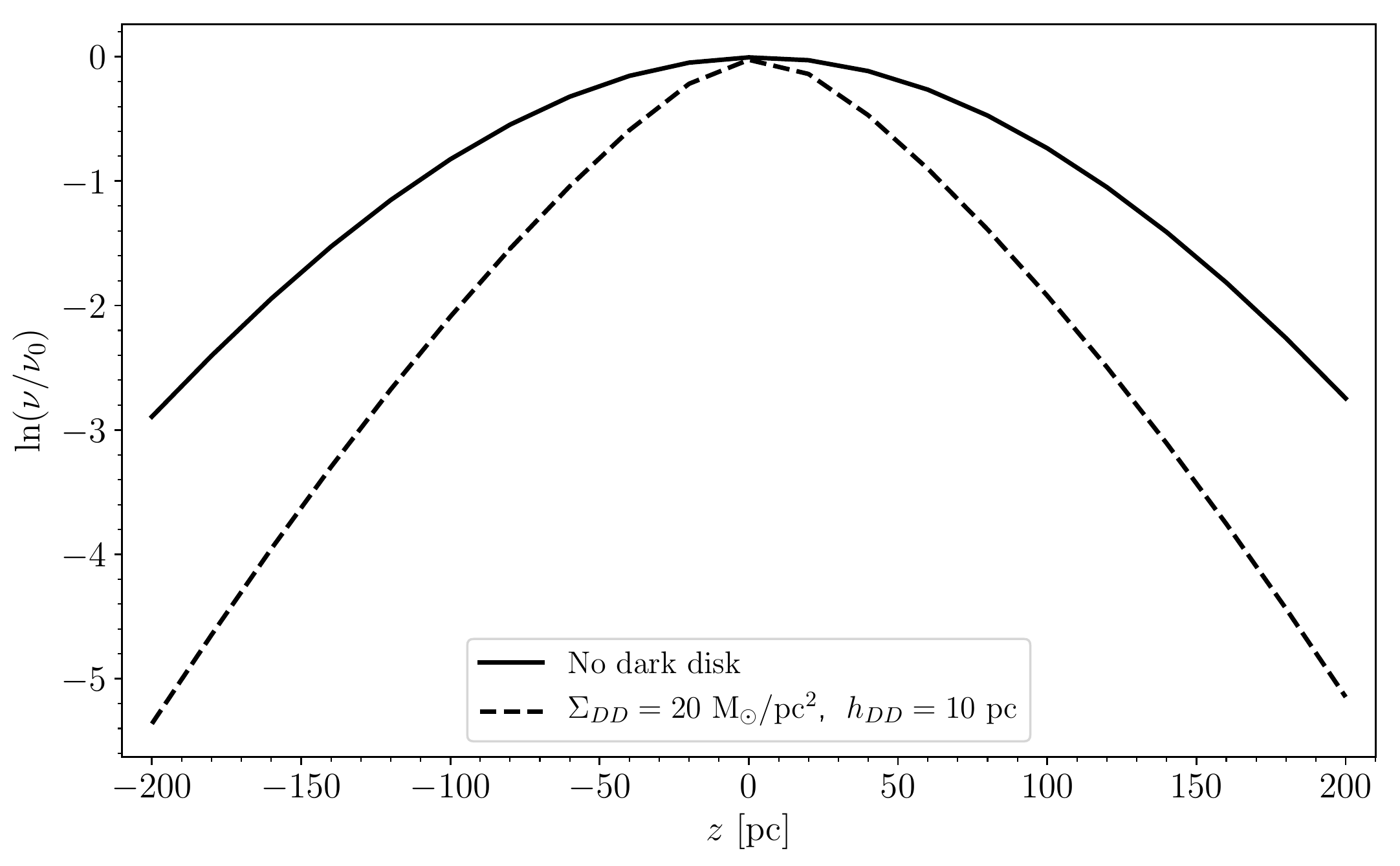}
\caption{The predicted number density of a tracer in a model containing a thin DD with surface density $\Sigma_{DD} = 20 \ {\rm M}_\odot/{\rm pc}^2$ and scale height $h_{DD}= 10 \ {\rm pc}$ (dashed). For comparison, we also plot the prediction of a model with the same matter content but without the thin DD (solid). }
\label{fig:thin_dd}
	\end{center}
\end{figure*}

In models with a thin DD, we assume that the DD is isothermal, axisymmetric, and perfectly aligned  with the baryonic disk. Following Ref.~\cite{1942ApJ....95..329S}, we choose the parametrization of the thin DD density to be, 
\beq
\rho_{DD}(z) = \frac{\Sigma_{DD}}{4h_{DD}} \sech^2 \left(\frac{z}{2 \, h_{DD}}\right),
\label{eq:DDSource}
\eeq 
where $\Sigma_{DD}$ is the surface density and $h_{DD}$ is the disk height. A thin DD aligned with the baryonic disk contributes an additional source of attractive potential, which pulls matter towards the midplane (see Section 2.2 of Ref.~\cite{Kramer:2016dqu} for an example with a toy model). This results in a narrowed {\it pinched} density profiles of tracers, as illustrated in Fig.~\ref{fig:thin_dd}. 

Thus, to obtain the local gravitational potential, we plug the total matter density, $\rho_{\rm tot}$, given by,
\beq
\rho_{\rm tot}(z)= \sum_{i=1}^{N_b} \rho_i(0) e^{-\Phi(z)/\sigma_{z;i}^2} +  \rho_{\rm DM} + \rho_{DD} (z).
\label{eq:rho_tot}
\eeq
into Eq.~\ref{eq:poisson_eq} and solve the resulting nonlinear, second-order differential equation numerically with \texttt{scipy.integrate.ODEint}. We also explicitly check that our results agree with those obtained by Refs.~\cite{Kramer:2016dqu, Schutz:2017tfp} using the iterative solver method.

To summarize, we calculate the equilibrium number density in the solar neighborhood for each tracer population, \textit{i.e.} A, F, and early G stars, by integrating its midplane velocity distribution (constructed from data) as a function of the parameters of our mass model using Eq.~\ref{eq:predict}. We also apply a Gaussian kernel smoothing to the result to approximate the effect of parallax uncertainties that smear the exact positions of stars. However, since the parallax uncertainties in DR2 are significantly reduced as compared to TGAS, this procedure only has a negligible effect on the predicted density.


\subsection{Likelihood, model uncertainties, and priors}  \label{sec:basic_setup}

Our model $\mathcal{M}$ is characterized by $\boldsymbol{\theta} = \{ \boldsymbol{\psi}, \boldsymbol{\xi} \} $, such that $\boldsymbol{\psi} = \{ \rho_{\rm DM},\Sigma_{DD}, h_{DD} \}$ are our parameters of interest, and $\boldsymbol{\xi}$ are the nuisance parameters which include: midplane densities, $\rho_k(0)$, and velocity dispersion, $\sigma_{z;k}$, for each baryonic component in the Bahcall model; overall normalization for each stellar population, $N_\nu$; height of sun above the midplane, $z_\odot$. 

For each tracer population, we use the likelihood, $p_{\nu}(d | \mathcal{M}, \boldsymbol{\theta})$, to fit the number density constructed from data with our model prediction in the presence of statistical and systematic uncertainties. Since each number density bin contains a large number of stars ($\mathcal{O}(1000$) for F and early G stars), the likelihood can be reasonably approximated by a Gaussian distribution,\footnote{We note that surveys like {\it Gaia} ({\it SDSS-SEGUE}) measure astrometric (spectrophotometric) parameters of individual stars. Thus, an ideal likelihood analysis should involve star-by-star predictions for these parameters drawn from a generative process that accounts for the survey selection function. The forward modeling approach of Refs.~\cite{Bovy2012d, Bovy:2013raa, Bovy:2016a}, for example, accomplishes this for several different scenarios. For DR2, such an analysis has been carried out by Ref.~\cite{Widmark:2018ylf} after this work appeared on arXiv, and their results agree with ours.}
\begin{equation}
p_{\nu}(d | \mathcal{M}, \boldsymbol{\theta}) = \prod_{i =1}^{N_z} \frac{1}{\sqrt{2 \pi \sigma^2_{\ln \nu_i}}} \exp \left( - \frac{(\ln(N_{\nu} \, \nu_{i}^{\rm mod}(\boldsymbol{\theta})) - \ln{\nu^{\rm data}_i})^2}{2 \, \sigma^2_{\ln \nu_i}(\boldsymbol{\theta})} \right),
\label{eq:p_nu}
\end{equation}
where $N_z$ is the number of $z$ bins, $\nu_i^{\rm mod}$ is the prediction of a model with parameters $\boldsymbol{\theta}$, and $\nu_i^{\rm data}$ is \textit{volume complete} number density constructed from data, as described in Sec.~\ref{sec:numberdensity}. Unlike Ref.~\cite{Schutz:2017tfp}, we do {\it not} multiply the likelihood functions for different stellar populations in our analysis since doing so assumes all populations are similar and trace the same galactic potential independently. This is a rather simplified assumption which ignores the evolution history of different stellar types. We comment more on this in Section~\ref{sec:localdm_no_dd}.

The total uncertainty, $\sigma^2_{\ln \nu_i}$, is obtained by adding in quadrature the data and the prediction uncertainties,
\begin{equation}
\sigma^2_{\ln \nu_i}(\boldsymbol{\theta}) = \left( \sigma^{2}_{\ln \nu_{i}}(\boldsymbol{\theta})\right)^{\rm mod} + \left( \sigma^{2}_{\ln \nu_ i}\right)^{\rm data}.
\label{eq:p_nu_error}
\end{equation}
The data uncertainty is discussed in Sec.~\ref{sec:numberdensity}, whereas the model uncertainty originates from uncertainties in the velocity profile $f_{ z=0}(|w|)$. It consists of two sources: {\it a)} the statistical uncertainty due to the finite sample size, and {\it b)} the systematic uncertainty due to possible non-equilibrium effects, which we characterize by the difference between $f_{z=0}(w > 0)$ and $f_{z=0}(w<0)$ following the treatment in Ref.~\cite{Schutz:2017tfp}.

Direct error propagation from uncertainties of $f_{z=0}(|w|)$ is difficult due to the large number of parameters and their correlations involved. Instead, we estimate the errors by bootstrap resampling. The bootstrap is a technique that extracts estimates for the mean and standard deviation of a given data set by repeated random sampling with replacement. For each stellar type, the raw midplane star data sets are bootstrapped multiple times to generate a suite of velocity distributions. For every velocity distribution, we use Eq.~\eqref{eq:predict} to derive a density distribution, and estimate the statistical uncertainty, $\left( \sigma^{2}_{\nu_ i}(z)\right)^{\rm mod,\,stat}$, as the bin-by-bin variance in the suite of density distributions. More details of the procedure are deferred to Appendix~\ref{sec:bootstrap}. 

We approximate the systematic uncertainty, $\left( \sigma^{2}_{\nu_ i}(z)\right)^{\rm mod,\,sys}$, by computing the difference between number densities predicted using the velocity distributions $f_{z=0}^{(w > 0)}$ and $f_{z=0}^{(w < 0)}$ for every unique value of the gravitational potential, 
\beq
\left( \sigma_{ \nu}\right)^{\rm mod,\,sys} \approx |\ln \nu^{(w > 0)}(z)-\ln\nu^{(w < 0)}(z)|.
\label{eq:predict_sys}
\eeq
The total uncertainty for the predicted number density, in every $z$ bin, is then given by,
\beq
\left( \sigma^{2}_{\nu_ i}(z)\right)^{\rm mod} = \left( \sigma^{2}_{\nu_ i}(z)\right)^{\rm mod,\,stat} + \left( \sigma^{2}_{\nu_ i}(z)\right)^{\rm mod,\,sys}.
\eeq
We find that the systematic uncertainties dominate over statistical ones in our analysis. The various sources of uncertainties in our analysis and their corresponding treatment are summarized in Table~\ref{tab:errors}. 

\begin{table}[h!]
\centering
	\begin{tabular}{|  c | c | c  |}
	\hline
	  Type & Source & Treatement \\ \hline \hline
	     \multirow{ 3}{*} {$\nu^{\text{data}}$} & Poisson & $\sqrt{N_k}$ in the $k$-th bin \\ \cline{2-3}
         & 3\% dust extinction & $0.03\times\nu^{\text{data}}$  \\ \cline{2-3}
         & {\it Gaia} systematic uncertainty & $\pm 0.1$ mas in $\varpi$; $\pm 0.1$ mas/yr in 
         $\mu_{\tilde{\alpha}}, \mu_\delta$ \\ \hline
         \multirow{ 3}{*}{$\nu^{\text{mod}}$} &  statistical errors of $f_{z=0}(|w|)$  &   bootstrap resampling \\ \cline{2-3}
         & $f_{z=0}(w>0) - f_{z=0}(w<0)$  & $|\ln \nu^{(+)}(z)-\ln\nu^{(-)}(z)|$ \\ \cline{2-3}
         & parallax uncertainty & Gaussian kernel smoothing \\ \hline
	\end{tabular}
\caption{Various sources of uncertainties and their treatment.} \label{tab:errors}
\end{table}

Our statistical analysis closely follows that of Ref.~\cite{Schutz:2017tfp} with one major 
difference: the treatment of velocity uncertainties. In Ref.~\cite{Schutz:2017tfp}, normalization of each velocity bin is also treated as a nuisance parameter, which adds an additional $20$-$30$ parameters to the analysis. In our approach, we propagate the velocity uncertainties, both statistical, estimated using bootstrap resampling, and systematic, into the prediction uncertainties. We check that these two methods yield similar results for TGAS data. 

Finally, to obtain the posterior distribution, we assume uniform prior distributions for all parameters except the baryonic ones; their priors follow a Gaussian distribution,
\begin{align}  \label{eq:baryon_prior}
p_b(\boldsymbol{\zeta}| \mathcal{M}) &=& \prod_{k =1}^{N_b} \left( \frac{1}{\sqrt{2 \pi \sigma^2_{\rho_k}}} \exp \left( - \frac{(\rho_k - \bar{\rho}_k)^2}{2 \, \sigma^2_{\rho_k}} \right) \right) \left( \frac{1}{\sqrt{2 \pi \sigma^2_{\sigma_{z;k}}}} \exp \left( - \frac{(\sigma_{z, k} - \bar{\sigma}_{z, k})^2}{2 \, \sigma^2_{\sigma_{z, k}}} \right) \right), 
\end{align}
where the mean and variance for each component are taken from Table~\ref{tab:baryons}. We summarize the details and ranges of assumed prior distributions for all parameters, $\boldsymbol{\theta}$, used in our analysis in Table~\ref{tab:prior_table}.

\begin{table}[h!]
\centering
	\begin{tabular}{| c |c|c| c |}
	\hline
	  Parameters & Prior type & Range & Total   \\ \hline \hline
	     $\rho_k(0)$, $\sigma_{z;k}$ & Gaussian & Eq.~(\ref{eq:baryon_prior}) & $24$  \\ \hline
		$N_{\nu}$ & Uniform & $[0.9, 2.0]$ & $3$ \\ \hline         
         $z_\odot$ & Uniform & $[-30.0, 30.0] \, \, \mathrm{pc}$ & $1$ \\ \hline
         $h_{DD}$ & Uniform & $[0.0, 100.0] \, \, \mathrm{pc}$  & $1$ \\ \hline
         $\rho_{\mathrm{DM}}$ & Uniform & $[0.0, 0.06] \, \, \mathrm{M}_\odot / \mathrm{pc^3}$  & $1$  \\ \hline
         $\Sigma_{DD}$ & Uniform & $[0.0, 30.0] \, \, \mathrm{M}_\odot / \mathrm{pc^2}$  & $1$ \\ \hline
       \end{tabular}
\caption{Prior distributions of model parameters. } \label{tab:prior_table}
\end{table}

\subsection{Sampling the posterior} \label{sec:BM_discuss}
The posterior probability density function (simply the posterior 
henceforth) of the parameters can be defined using Bayes' theorem,

\begin{equation}\label{eq:posterior}
p ( \boldsymbol{\theta} | \mathcal{M}, d) = \frac{p(d| \mathcal{M}, \boldsymbol{\theta}) p(\boldsymbol{\theta} | \mathcal{M})}{p(d | \mathcal{M})},
\end{equation} 
where the numerator is given by Eqs.~(\ref{eq:baryon_prior}) and (\ref{eq:p_nu}) and the 
denominator, referred to in the literature as `marginal likelihood' or `evidence', is defined as

\begin{equation}\label{eq:evidence}
p(d | \mathcal{M}) = \int \, p(d| \mathcal{M}, \boldsymbol{\theta}) p(\boldsymbol{\theta} | \mathcal{M}) \, d \boldsymbol{\theta}.
\end{equation}

We sample the posterior in Eq.~(\ref{eq:posterior}) with the Markov Chain Monte Carlo (MCMC) sampler \texttt{emcee}\footnote{\url{http://dfm.io/emcee/current/}} for estimating values of parameters and determining correlations between them. To draw samples from a $d$-dimensional parameter space, \texttt{emcee} implements the affine-invariant ensemble sampling algorithm of Ref.~\cite{GW2010} that is based on simultaneously evolving an ensemble of $N$ walkers. Since each walker in the ensemble independently samples the posterior, \texttt{emcee} is naturally suited for parallel computing on multicore systems (see Ref.~\cite{ForemanMackey:2012ig} for more details).

In our implementation, we let (100-300) walkers run for (15000-25000) steps depending on the stellar type and components ($\rho_{\rm DM}$ or $\rho_{\rm DM}$ + thin DD) of the local DM content. These numbers are chosen to achieve an acceptance fraction $a_f \approx 0.3$ \cite{optimal_scaling} for each walker. After accounting for the `warm-up' time, ${{\sim} 4000}$ steps, of the ensemble, we obtain ${\gsim} \, 2 \times 10^6$ samples on average for each iteration of our analysis. 

\section{Results}
\label{sec:BM_result}

	\subsection{Local DM density} \label{sec:localdm_no_dd}
We summarize the results from the posterior sampling for the analysis with baryons and a constant halo DM density $\rho_{\mathrm{DM}}$ in Table~\ref{tab:local_density_table}. The median value of $\rho_{\mathrm{DM}}$ obtained through our kinematic analysis of A and early G stars are similar to each other, while using F stars yields a significantly higher value. We also note that our value of $\rho_{\rm DM}$ determined using A and early G stars is consistent with previous measurements made using SDSS/SEGUE G star data~\cite{Yanny2009}, $\rho_{\rm DM}= 0.012^{+0.001}_{-0.002} \,\, \rm{M}_\odot / {\rm pc}^3$ (within $1 \sigma$) and $\rho_{\rm DM}= 0.008^{+0.025}_{-0.025} \,\, {\rm M}_\odot / {\rm pc}^3$ (within $2 \sigma$), by Refs.~\cite{Silverwood2017} and \cite{Bovy:2013raa} respectively.

While the $95\%$ credible region (CR) for measurements of $\rho_{\mathrm{DM}}$ with A, F, and early G stars in Fig.~\ref{sfig:exclusion} overlap and seem consistent with each other at the $2\sigma$ level, we emphasize that each tracer population doesn't necessarily probe the same galactic environment due to differences in age and star formation history.\footnote{For instance, each tracer type could have different sensitivity to non-equilibrium features of the MW~\cite{banik:2016}. Propagating these uncertainties to our estimates for the baryon and DM densities requires a detailed study using simulations, which is beyond the scope of this paper.} Consequently, without appropriate modeling of all prior information in a (hierarchical) Bayesian framework, results derived from different tracers should be compared with caution.

\begin{table}[!htb]
\centering
	\begin{tabular}{| c | c | c | c | c |}
	\hline
	  Stellar type &  $\rho_{\mathrm{DM}} \, \, [\mathrm{M}_\odot / \mathrm{pc^3}]$ &  $\rho_{\mathrm{DM}} \, \, [\mathrm{GeV}/ \mathrm{cm^3}]$ &  $\rho_{b} \, \, [\mathrm{M}_\odot / \mathrm{pc^3}]$ & $z_\odot \, \, [\mathrm{pc}]$ \\ \hline \hline
	    A stars & $0.016^{+0.010}_{-0.010}$ & $0.608^{+0.380}_{-0.380}$ & $0.088^{+0.007}_{-0.007}$ &  $8.80^{+3.74}_{-4.23}$ \\ \hline
		F stars & $0.039^{+0.008}_{-0.008}$ & $1.482^{+0.304}_{-0.304}$ & $0.089^{+0.007}_{-0.007}$ &  $2.04^{+2.84}_{-3.13}$ \\ \hline         
         G stars & $0.011^{+0.010}_{-0.009}$  & $0.418^{+0.380}_{-0.342}$ & $0.087^{+0.007}_{-0.007}$ &  $-8.82^{+5.32}_{-4.64}$\\ \hline
       \end{tabular}
\caption{Median posterior values with $1 \sigma$ errors for the local densities of baryons $\rho_b$ and halo DM $\rho_{\mathrm{DM}}$, and height of the sun above the midplane $z_\odot$. The 
halo DM density $\rho_{\mathrm{DM}}$ is expressed in both $\mathrm{M}_\odot / \mathrm{pc^3}$ (astronomical unit) and $\mathrm{GeV}/ \mathrm{cm^3}$ (particle physics unit), where 1 $\mathrm{M}_\odot / \mathrm{pc^3}$ $\approx$ 38 $\mathrm{GeV}/ \mathrm{cm^3}$.}
\label{tab:local_density_table}
\end{table}

	\subsection{Constraints on a thin DD} \label{sec:localdm_dd}
\begin{figure*}[!ht]
	\begin{center}
\includegraphics[scale=.4]{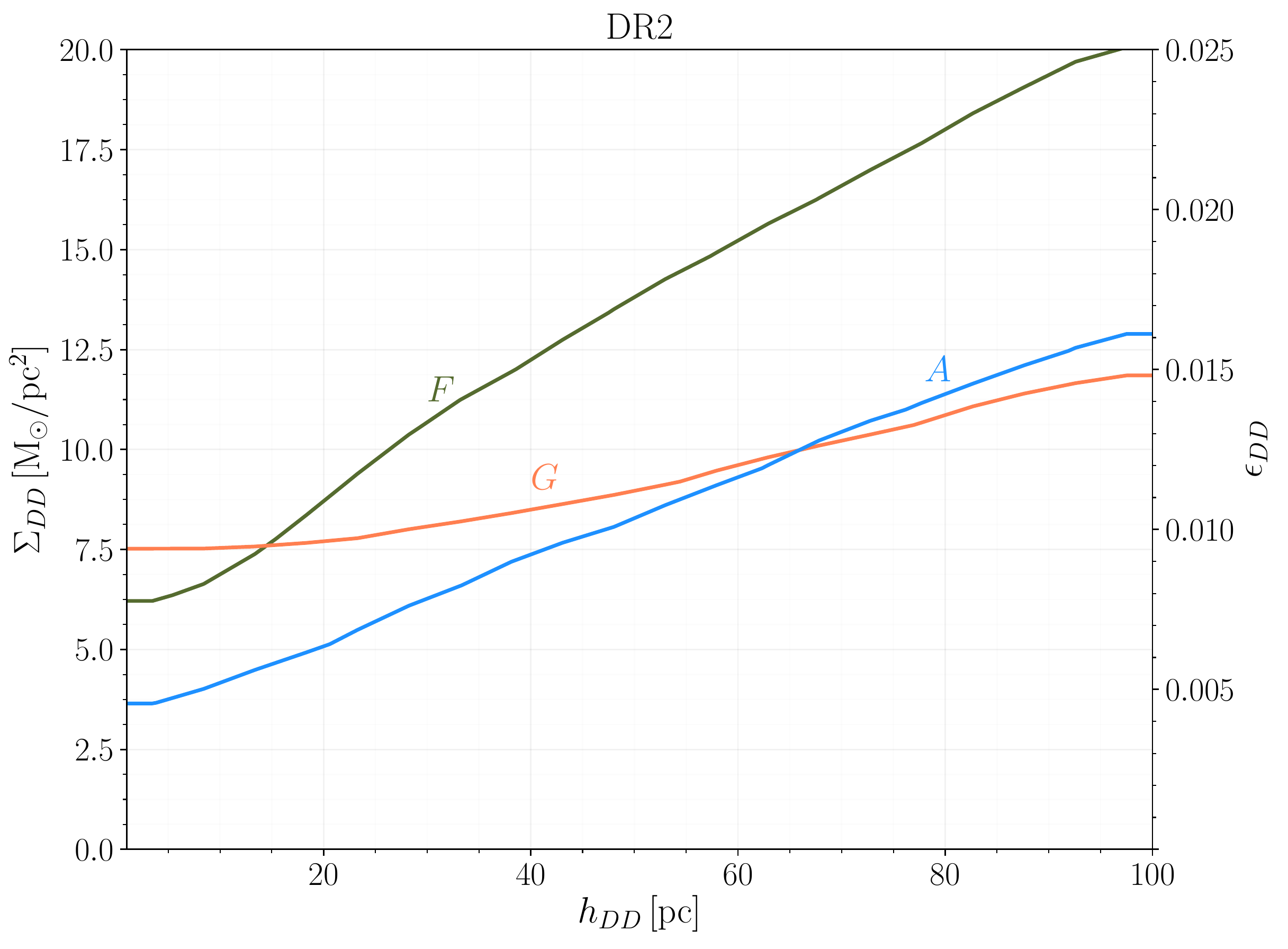}
\caption{$95\%$ CR upper limit contours for surface density $\Sigma_{DD}$ and scale height $h_{DD}$ of a thin DD for A (blue), F(green), and G (orange) stars using data from DR2 (left panel) and TGAS (right). The upper bound for the fraction of the total DM mass in the MW that could exist in a DD, $\epsilon_{DD}$, is also shown on the right side of each plot for reference.}
\label{fig:exclusion_dr2}
	\end{center}
\end{figure*}

We perform a full MCMC scan of the posterior after including a thin DD component along with local density of halo DM $\rho_{\rm DM}$, and plot the marginalized posteriors for thin DD parameters, $\rho_{\rm DM}$, and the total midplane baryon density $\rho_{b}$ in Figs.~\ref{fig:mcmc_A}\crefrangeconjunction \ref{fig:mcmc_G}. On the other hand, Fig.~\ref{fig:exclusion_dr2} gives the constraints on thin DD parameters after marginalizing over the uncertainties of the baryon mass model and asymmetries in velocity distribution. Given the exploratory nature of our analysis, this may be interpreted, at best, as {\it an approximate upper bound on the thin DD parameters}. 

We also use the constraint on $\Sigma_{DD}$ to estimate the amount of dissipative DM in the galaxy. Following Ref.~\cite{Fan:2013tia}, we relate the locally measured surface density of a thin DD to $\epsilon_{DD}$, the fraction of the total DM mass in the MW that has dissipative self-interactions and forms a DD, 
\beq
\Sigma_{DD} (R_\odot)= \frac{\epsilon_{DD} \, M_{\rm DM}^{\rm gal}}{2 \pi R_{DD}^2}\, \exp{(-R_\odot/R_{DD})} 
\eeq
where $M_{\rm DM}^{\rm gal} \, {\sim} \, 10^{12} M_\odot$ is the total DM mass in the MW, $R_{\odot} \, {\sim} \, 8.1$ kpc is the Sun's distance from the galactic center, and the scale radius of the thin DD is assumed to be equal to that of baryons, $R_{DD} = 2.15$ kpc \cite{Bovy:2013raa}. As indicated in Fig.~\ref{fig:exclusion_dr2}, only ${\sim} 1\%$ of the total DM mass could reside in a thin DD.

\section{Discussion}\label{sec:validation}
Our main results from the MCMC sampling of the posterior, e.g. for A stars, imply that the local DM content can accommodate a constant density $\rho_{\rm DM}= 0.016 \pm 0.010 \, \, {\rm M}_\odot/{\rm pc}^3$, or $\rho_{\rm DM}= 0.008^{+0.011}_{-0.008} \, \, {\rm M}_\odot/{\rm pc}^3$ and a thin DD with $\Sigma_{DD}= 2.99^{+3.75}_{-2.18} \,\, {\rm M}_\odot/{\rm pc}^2$, the precise value depending on $h_{DD}$. We observe that the $1\sigma$ uncertainties are fairly large in both cases and suggest high systematic noise in our determination. We discuss different sources of the uncertainties in Secs.~\ref{sec:density_val}-\ref{sec:bary_dm} and comment on the robustness of our dynamical analysis. Lastly, we cross-validate our statistical setup by repeating our analysis with TGAS data in the same galactic volume, and comparing the results with those from DR2 in Sec.~\ref{sec:localdm_tgas}.

\subsection{Effect of volume cuts} \label{sec:density_val}
We vary the cylinder radius $R$ and find that the tracers' vertical density distributions do not vary much for $R \lesssim 200$ pc. Increasing $R$ from 150 pc to 250 pc, though, results in an overall broadening of the density distributions. Ref.~\cite{Schutz:2017tfp} attributed a similar trend in TGAS data to the so-called `Eddington' bias, {\it i.e:} higher parallax uncertainties of distant stars could lead to a smearing of the density distributions at large $|z|$. However, as shown in Fig.~\ref{fig:parallax}, the parallax uncertainties are significantly reduced in DR2 and remain small at large $|z|$ even when $R$ is increased to 250 pc. Thus, it seems unlikely that the broadening of the density distributions is due to the `Eddington' bias. A more plausible option is the presence of local disequilibrium effects as we discuss in the following section. We note that our procedure would result in a lower local DM density estimate for a broader density distribution, since additional matter tends to pinch the predicted density profile as shown in Fig.~\ref{fig:thin_dd}.

\subsection{Disequilibria in the solar neighborhood?} \label{sec:disequilibria}
An implicit assumption in our modeling of the tracer density profile is that the local
neighborhood is axisymmetric and the stellar disk is in dynamic equilibrium. However, growing evidence in DR2 data for: asymmetry in the vertical number counts~\cite{Widrow2012, 2019MNRAS.482.1417B}; vertical waves in the disk at Sun's position~\cite{Gomez2013, Carlin2013, Widrow2014}; kinematic substructure~\cite{Antoja2018, Myeong:2018kfh, Necib:2018iwb}, warrants a closer look at sources of disequilibria in the solar neighborhood. We defer searches of local disequilibria and the corresponding revision of our traditional kinematic method outlined in Sec.~\ref{sec:PJ_Theory} to future work. Presently, we only approximate the effect of non-equilibrium behavior by propagating asymmetries in the midplane velocity distribution to the errors in the predicted density.

\subsection{Degeneracy between $\rho_{\rm DM}$ and $\rho_b$} \label{sec:bary_dm}
The marginalized posterior for each tracer in Fig.~\ref{sfig:exclusion} indicates a strong degeneracy between measurements of $\rho_b$ and $\rho_{\rm DM}$. As proposed by Ref.~\cite{1984ApJ...276..169B}, and recently implemented on simulated data by Ref.~\cite{garbari:2011}, this degeneracy can only be broken if any kinematic analysis includes the density falloff at larger $|z|$ away from the midplane. Since most of the baryonic matter is confined to the stellar disk with a scale height $\mathcal{O}({\rm kpc})$, any excess matter that causes the falloff can be attributed to (at least to leading order) to DM, allowing a more precise measurement of $\rho_{\rm DM}$ with smaller error bars. On the other hand, this introduces another layer of complexity as the coupling between the radial and vertical motions is no longer negligible at $|z| \, {\gsim} \,0.5 \ {\rm kpc}$ and must be modeled by simultaneously fitting to the velocity data~\cite{Budenbender:2014xra, Silverwood2016}. 

Meanwhile, the highly diagonal posterior in the $\rho_{\rm DM}$--$\Sigma_{DD}$ plane combined with identically flat posterior in the $\rho_{\rm DM}$--$\rho_{b}$ and $\Sigma_{DD}$--$\rho_{b}$ planes of Figs.~\ref{fig:mcmc_A}\crefrangeconjunction \ref{fig:mcmc_G} implies that introducing a thin DD in our analysis merely shifts some of the DM density from $\rho_{\rm DM}$ while increasing its relative error. Thus, to set realistic constraints on, or seek evidence for, DM density in the thin DD (or equivalently some form of extended substructure near the midplane) using our procedure, we would need more physical insight to break the degeneracy between different distributions of DM. 


As the discussion above indicates, our results are dominated by systematic errors stemming from an approximate modeling of non-equilibrium behavior and a strong degeneracy between different matter components near the midplane. {\it We note that these errors, in the context of the 1D distribution function method, may not be reduced significantly in future {\it Gaia} data releases.}
 
\subsection{Comparison of constraints between DR2 and TGAS}  \label{sec:localdm_tgas}
\begin{figure*}[!ht]
	\begin{center}
\includegraphics[scale=.5]{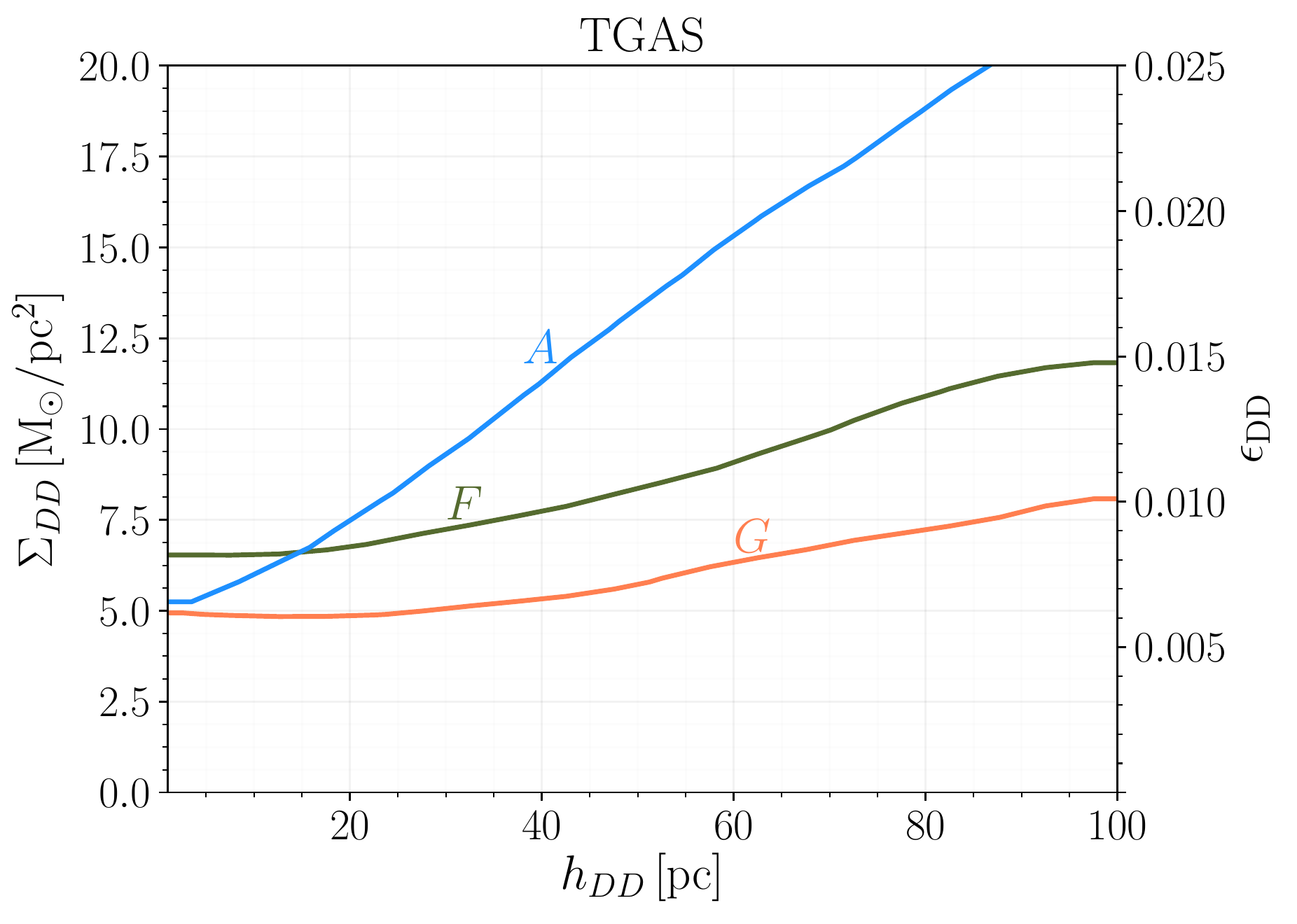}
\caption{Same as Fig.~\ref{fig:exclusion_dr2} but using TGAS data.}
\label{fig:exclusion_tgas}
	\end{center}
\end{figure*}

We plot the $95\%$ CR upper limit contours for the thin DD parameters using data from DR2 and TGAS in Fig.~\ref{fig:exclusion_dr2} and Fig.~\ref{fig:exclusion_tgas} respectively. Both sets of exclusion curves are significantly stronger than previous results based on the {\it Hipparcos} catalog~\cite{Kramer:2016dqu}. However, there are obvious differences between our results derived using DR2 and TGAS data.\footnote{Our TGAS results derived using a fully Bayesian analysis roughly agree with those of Ref.~\cite{Schutz:2017tfp}; see left panel of their Fig.~S17 in particular.}

Using TGAS data, early G stars exclude $\Sigma_{DD} \, {\approx} \, 5  \, \mathrm{M}_\odot / \mathrm{pc^3}$ depending on $h_{DD}$ while A stars set the weakest constraint. On the other hand, using DR2 data, A stars exclude $\Sigma_{DD} \, {\gsim} \, (5 - 12) \, \mathrm{M}_\odot / \mathrm{pc^3}$ while the weakest constraint is due to F stars.

Naively, we expect that there would be an (modest) improvement in the constraints from DR2 data compared to those from TGAS due to increased statistics (about a factor of ${\sim} 2.5$) and decreased parallax uncertainties (due to our choice of binning, these only affect the high $z$ bins). We check numerically that if we take central values from TGAS and uncertainties from DR2 to generate mock distributions for the tracers, the derived constraints on thin DD are indeed similar to those from TGAS data with minor improvements. Given this expectation, it seems counterintuitive that our DR2 constraints are different from the TGAS ones. 

Before discussing possible origins of the differences for each tracer population, we note that adding more matter pinches the density profile of tracer stars, such as the effect of thin DD discussed in Sec.~\ref{sec:PJ_Theory}. Thus, the narrower the profile from data or broader the predicted density is, the more matter that can be included, and weaker the constraint on local DM content.

\begin{figure*}[!ht]
\includegraphics[scale=0.465]{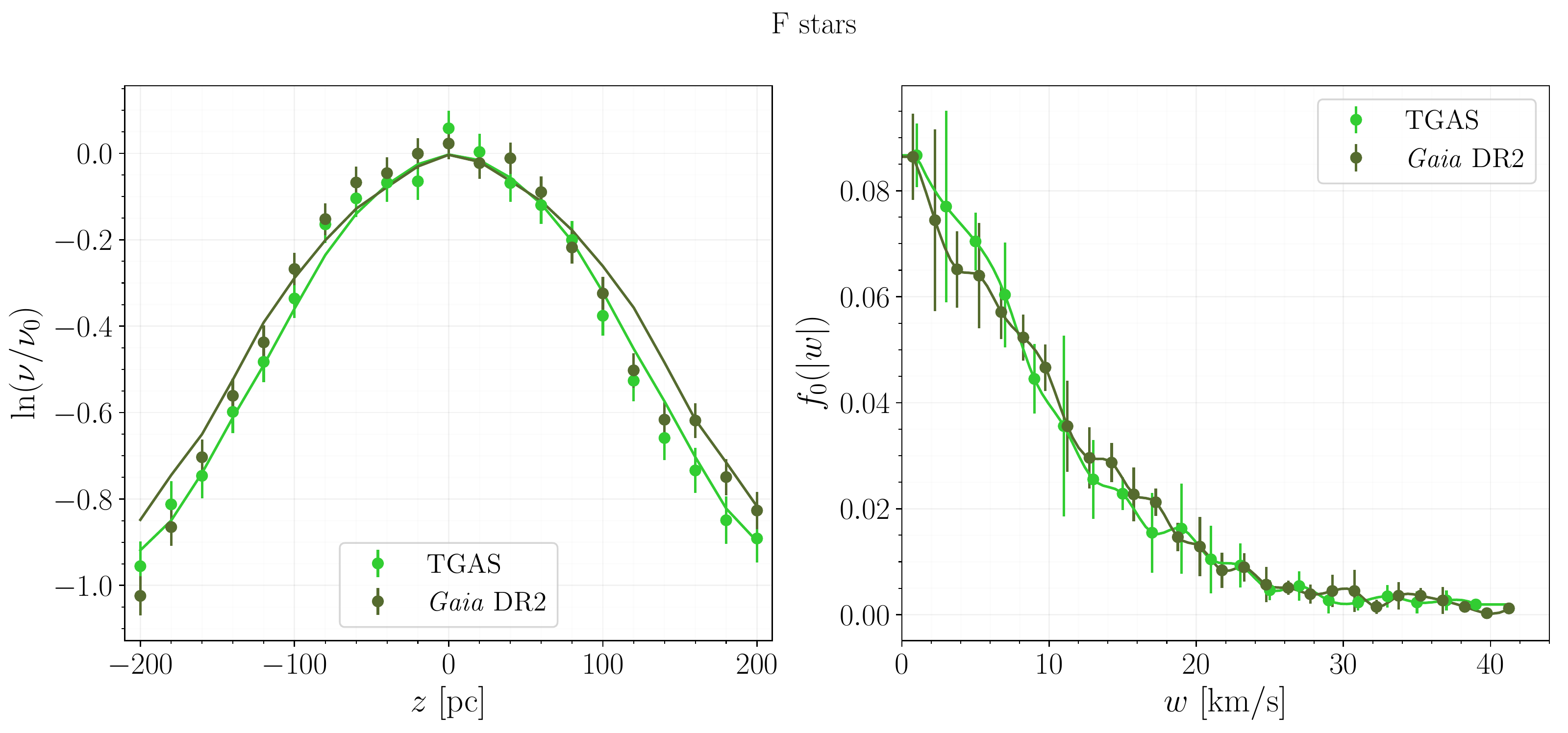}
\caption{F stars: (left) volume complete number density profiles overlaid with the predicted
density derived using the mean TGAS and DR2 velocity distributions assuming fiducial values for baryons and ${\rho_{\rm DM} = 0.02  \ \mathrm{M}_\odot / \mathrm{pc^3}}$; (right) midplane velocity distributions with interpolated fits to the data. Note that the TGAS velocity distribution has a bin size of 2 km/s while DR2 bin size is 1.5 km/s.}
\label{fig:f_star_density_vel}
\end{figure*}

The significant weakening of constraints for F stars stems from small differences in the midplane velocity distributions, as shown in the right panel of Fig.~\ref{fig:f_star_density_vel}. The DR2 velocity distribution is slightly broader. We verify that this trend in the velocity distribution is not an artifact of our choice of the midplane latitude cut or the binning of the velocity data. Although velocity (and vertical density) profiles from TGAS and DR2 are consistent with each other within uncertainties, the predicted density distribution with DR2 data is broader than that with TGAS data with fixed model parameters (one example is shown in the left panel of Fig.~\ref{fig:f_star_density_vel}). As a result, a higher density in DM components is required to fit the predicted density of F stars to the DR2 number density profile for a fiducial baryon mass model. 

\begin{figure*}[!h]
\includegraphics[scale=0.465]{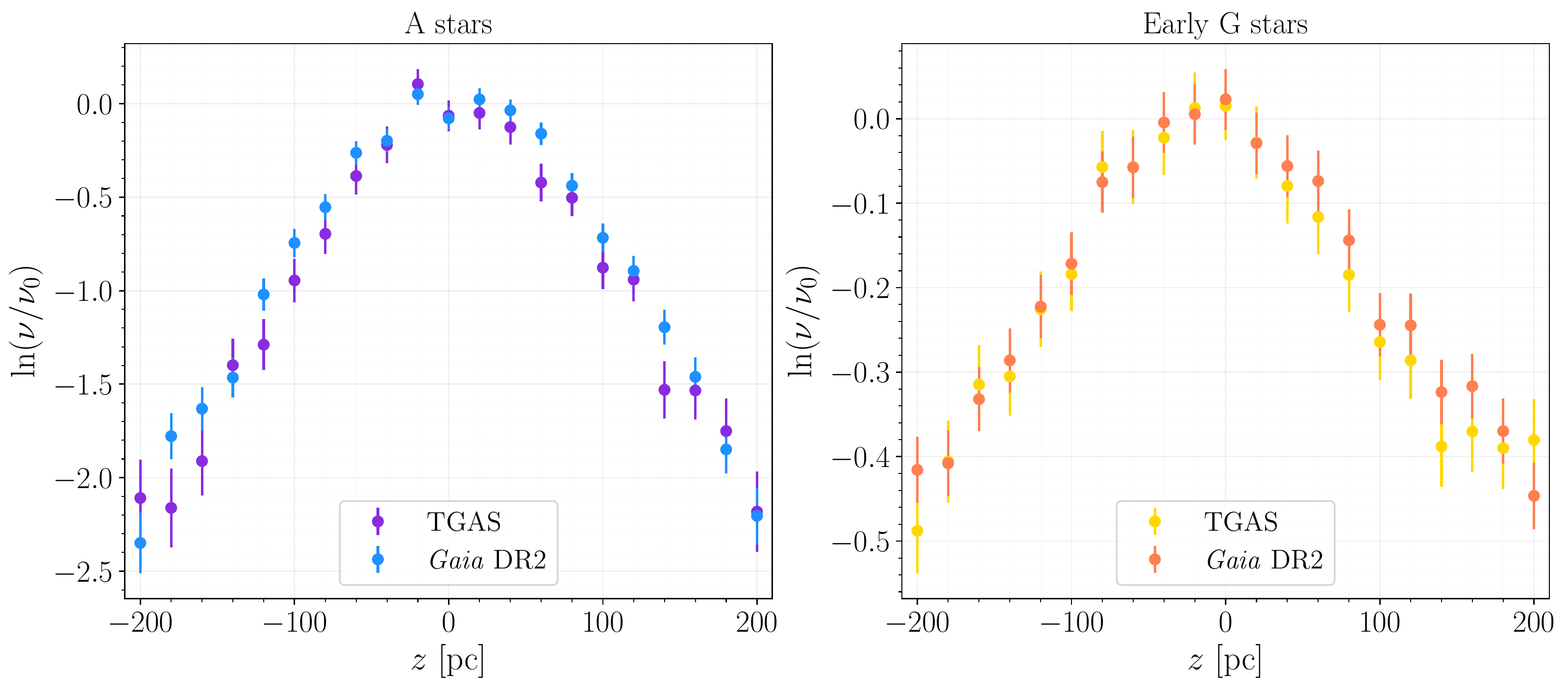}
\caption{Comparison of volume complete number density profiles in TGAS and DR2 data for A (left) and G (right) stars. }
\label{fig:a_g_star_density}
\end{figure*}

\begin{figure*}[!h]
\includegraphics[scale=0.465]{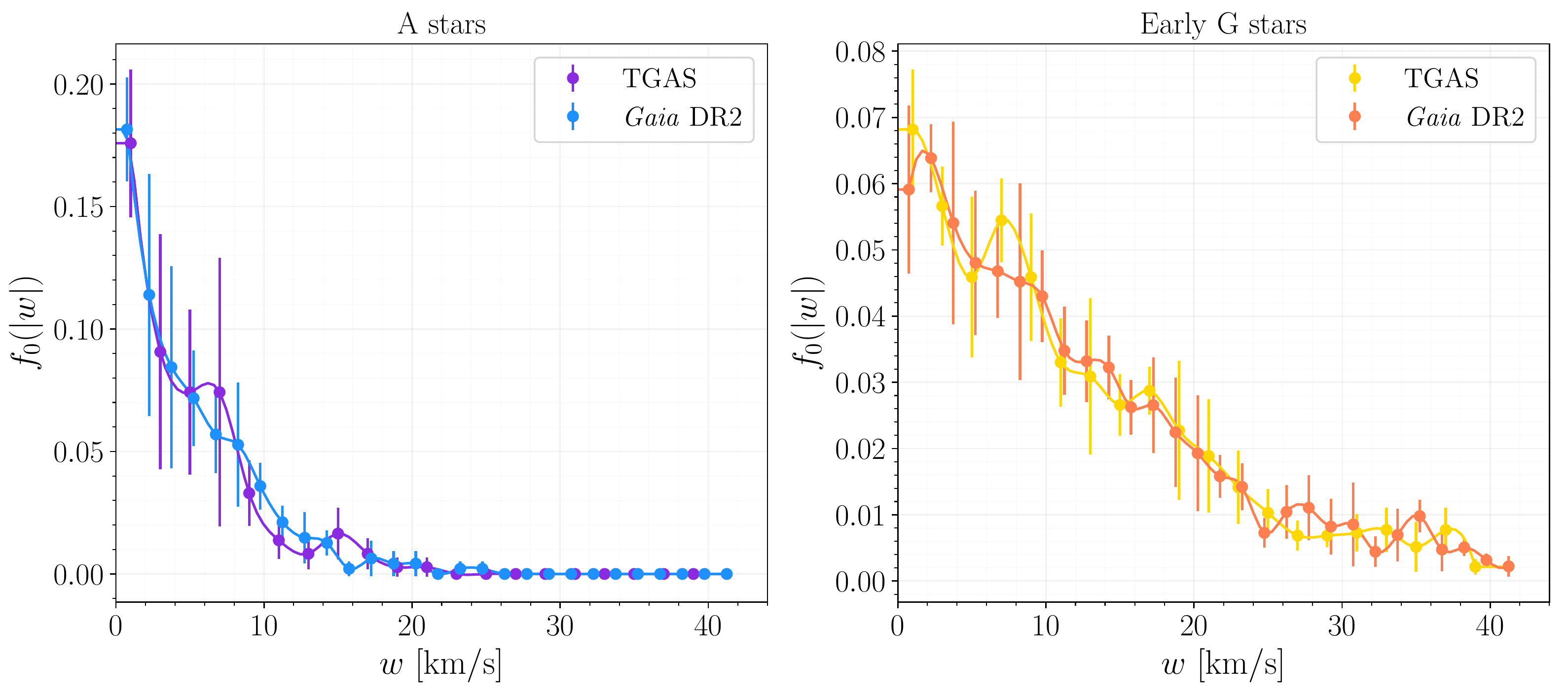}
\caption{Comparison of midplane velocity distributions in TGAS and DR2 data for A (left) and G
(right) stars. Note that the TGAS velocity distribution has a bin size of 2 km/s.}
\label{fig:a_g_star_velocity}
\end{figure*}

We also present the volume complete number density profiles and midplane velocity distributions for A and early G stars in Fig.~\ref{fig:a_g_star_density} and Fig.~\ref{fig:a_g_star_velocity}. From the plots, we note that all the distributions based on TGAS and DR2 data for both these tracers are also consistent within uncertainties, yet there are subtle differences. The velocity distributions using DR2 data are smoother compared to the TGAS ones with smaller systematic uncertainties from asymmetry between negative and positive velocity data.

The constraint from early G stars in the DR2 data set gets weaker due to both a slightly narrower density profile, and a slightly broader predicted density. However, in the case of A stars, the constraint gets considerably stronger at high $h_{DD}$ due to the reduction in the systematic errors from the asymmetry in the midplane velocity distribution. 

We reiterate that {\it Gaia} DR2 should be regarded as a different data catalog from TGAS, rather than just a statistical improvement over it \cite{2018arXiv180409365G}. DR1 incorporated positions from the Tycho-2 catalog to generate the five-parameter astrometric solution in the TGAS catalog, whereas, the DR2 catalog is independent from any other external catalogs with its own self-consistent astrometric solution. Any comparison between the constraints on local DM content from TGAS and DR2 should be made bearing this difference in mind. 

\section{Conclusions and Outlook}
\label{sec:outlook}
We apply the 1D distribution function method to {\it Gaia} DR2 and use stellar kinematics in the solar neighborhood to constrain the local DM density and properties of a thin DD aligned with the baryonic disk by performing our analysis within a Bayesian framework. We adopt young A, F, and early G stars as tracers as they have shorter equilibration timescales and consequently are expected not to be strongly affected by disequilibria. Using A stars gives an estimate of $\rho_{\rm DM}= 0.016 \pm 0.010$ M$_\odot$/pc$^3$ and sets the strongest constraint on the thin DD, excluding $\Sigma_{DD} \, {\gsim}$ (5-12) M$_\odot$/pc$^2$ depending on the scale height with 95\% confidence. This upper bound is used to constrain the amount of dissipative DM in the galaxy: a thin DD with $\Sigma_{DD} \, {\lsim} \, 12$ M$_\odot$/pc$^2$ and a scale radius ${\sim} 3 \, \rm{kpc}$ contains ${\lsim} \, 1\%$ of the total DM mass in the Milky Way \cite{Fan:2013tia}. While we obtain similar results from early G stars, F stars seem to prefer a higher value of the local DM content. Even though the distributions derived from DR2 are consistent with those from TGAS data within uncertainties, the allowed DM density and parameters of DD model are quite different for all tracers. In light of these results, we address the origins of the differences and discuss the robustness of our kinematic analysis. 

Our results also suggest that we need a better understanding of the physical origin of the systematic uncertainties, which we include in our analysis to account for the asymmetry in the midplane velocity distributions of tracers. One possibility is that with complete data for radial velocities, we could define the midplane region using the $z$-cut instead of the $b$-cut and obtain a more precise determination of the velocity distribution. Another possibility is to take a closer look at local disequilibria and their effects on traditional kinematic methods. Although we do not find any statistically significant evidence for non-equilibrium in the vertical density and velocity distributions in our samples, several analyses based on DR2 seem to suggest various sources of disequilibria at distances larger than the heliocentric cylinder we consider. In terms of baryon modeling, it could be useful to find a self-consistent, data-driven approach to determine the baryon distributions instead of assuming the isothermal Bahcall model. One way to achieve this would be to construct the mass density for stars directly from the data rather than treating it as an isothermal disk. 

For a more precise determination of the local DM density, a dynamical analysis could be performed using tracers at heights greater than the scale height of the stellar disk to minimize the latent degeneracy between baryons and DM. However, besides modeling effects of disequilibria, an analysis at larger scale height has to go beyond the 1D method and must include terms that couple the motions of tracers in different directions. We also see a degeneracy between parameters of ordinary DM and thin DD in the marginalized posteriors obtained through MCMC sampling. To break the degeneracy, we would need to distinguish between their effects on tracers by developing new observables and modeling priors that reflect these differences. 

\acknowledgments{We thank Ian Dell'Antonio, Eric Kramer, Tongyan Lin, Matt Reece, Ben Safdi, and Chih-Liang Wu for useful discussions. JB would like to thank Nicolas Garcia-Trillos and Alexander Fengler for extended conversations on MCMC sampling methods and Bayesian statistics. We are also grateful to the anonymous referee for their many insightful comments that greatly enriched the manuscript. This project was finished in part at the 2019 Santa Barbara Gaia Sprint, hosted by the Kavli Institute for Theoretical Physics (KITP) at the University of California, Santa Barbara.

This work has made use of data from the European Space Agency (ESA) mission {\it Gaia} (\url{https://www.cosmos.esa.int/Gaia}), processed by the {\it Gaia} Data Processing and Analysis Consortium (DPAC, \url{https://www.cosmos.esa.int/web/ Gaia/dpac/consortium}). Funding for the DPAC has been provided by national institutions, in particular the institutions participating in the {\it Gaia} Multilateral Agreement. It also makes use of data products from the Two Micron All Sky Survey ({\it 2MASS}), which is a joint project of the University of Massachusetts and the Infrared Processing and Analysis Center/California Institute of Technology, funded by the National Aeronautics and Space Administration (NASA) and the National Science Foundation (NSF). The results in this work were computed using the following open-source software: \texttt{IPython}~\cite{Perez:2007emg}, \texttt{matplotlib}~\cite{Hunter:2007ouj}, \texttt{scipy}~\cite{jones2001scipy}, \texttt{numpy}~\cite{vanderWalt:2011bqk}, \texttt{astropy}~\cite{astropy}, \texttt{gala}~\cite{gala}, \texttt{gaia\_tools}~\cite{Bovy:2017}, and \texttt{emcee}~\cite{ForemanMackey:2012ig}.}

This research was supported in part at KITP by the Heising-Simons Foundation and the National Science Foundation under Grant No. NSF PHY-1748958. JF is supported by the DOE grant DE-SC-0010010 and NASA grant 80NSSC18K1010.
\newpage
\appendix

\section{Constructing a Volume Complete Density} \label{sec:gaia_tools}
We use the \texttt{gaia\_tools} package to determine the: a) selection function by comparing the number counts in {\it Gaia} DR2 to those in {\it 2MASS}, and b) effective volume completeness in each $z$ bin using the Poisson likelihood approach introduced by Ref.~\cite{Bovy:2017}. We modify the default color-magnitude modeling in \texttt{gaia\_tools}, and discuss, in Appendix~\ref{sec:cm_model}, its effect on the completeness of the DR2 selection function. We highlight the important parts of the Poisson likelihood approach in Appendix~\ref{sec:poisson}.
 
\subsection{Color-magnitude modeling} \label{sec:cm_model}
The completeness for TGAS stars has a strong color dependence and drops off sharply at faint magnitudes, $J \gtrsim12$. To account for this effect, \texttt{gaia\_tools} calculates the completeness in each bin as a function of a color-dependent magnitude, $J_G$. However, as the faint end of DR2 extends well beyond $J {\sim} 12$, we use the $J$ magnitude instead of $J_G$ in our computation. 

\begin{figure}[!h]
	\centering$
	\begin{array}{cc}
       \includegraphics[width=0.4\linewidth]{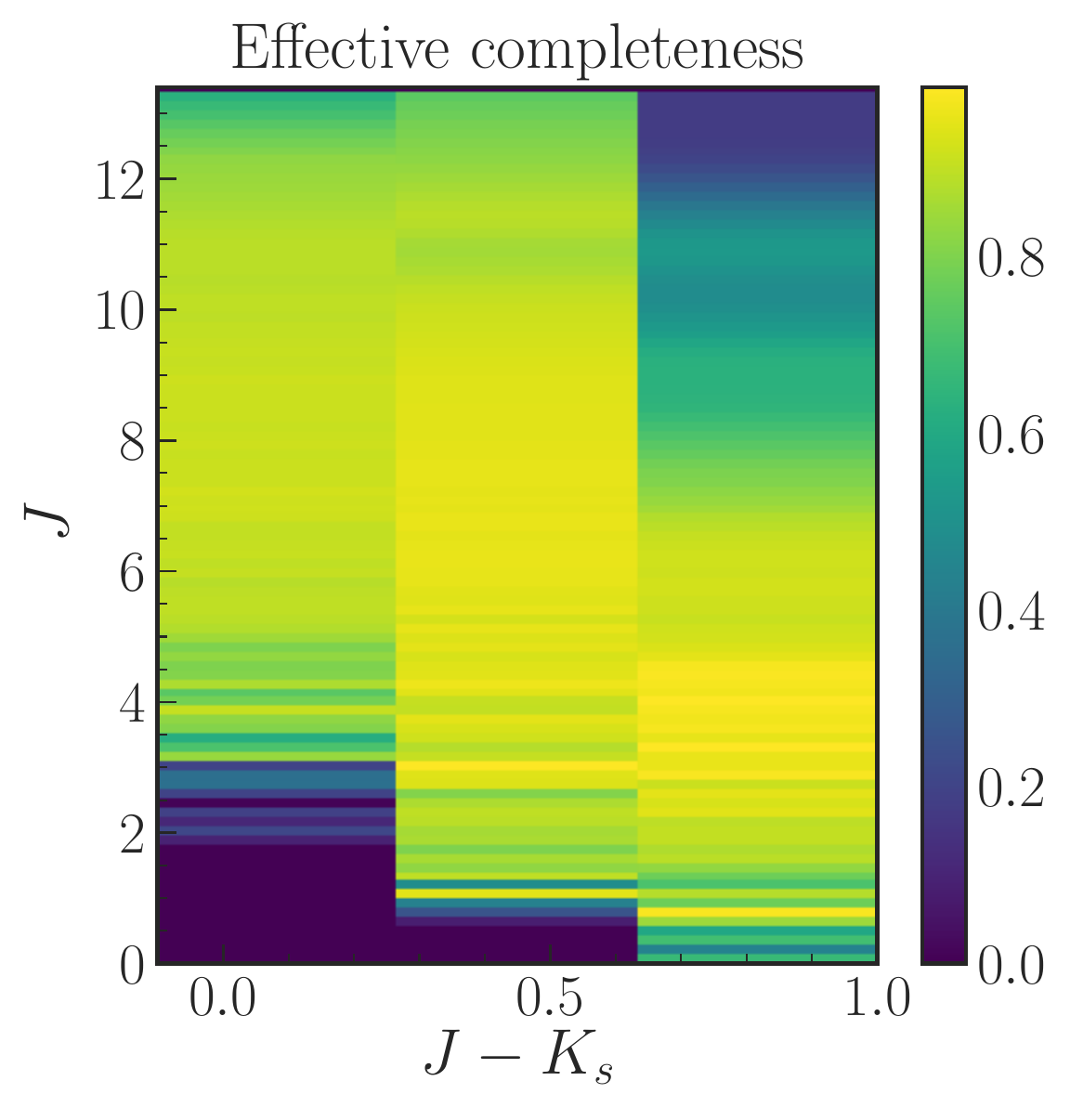}
       \includegraphics[width=0.4\linewidth]{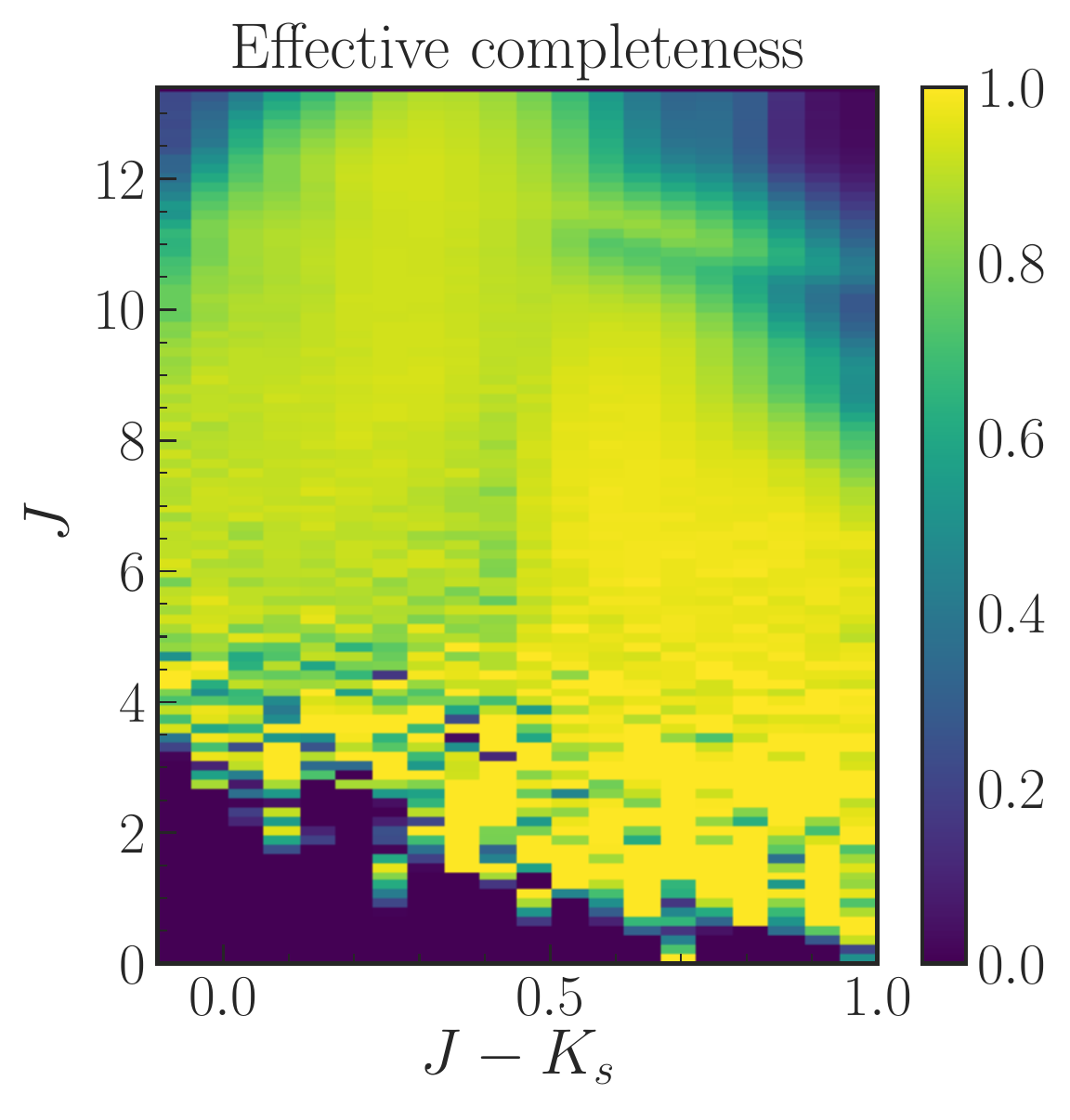}
	\end{array}$
    \caption{The effective completeness in color-magnitude space. Left: 3 $J-K_s$ bins. Right: 20 $J-K_s$ bins.}
    \label{fig:cmd_jk}
\end{figure}

As a consistency check, we also vary the $J-K_s$ color in the range $-0.05<J-K_s<1.05$ from the default 3 bins to 20 bins, as shown in Fig.~\ref{fig:cmd_jk}, and find that the variation on our number density profiles (through the effective volume completeness) is $\lesssim 2$\%. Thus, we conclude that the choice of binning has a negligible effect on the completeness of the selection function.

\subsection{Poisson process likelihood} \label{sec:poisson}
Given the the selection function of a survey, $S(J, J-K_s, \alpha, \delta)$, which indicates the fraction of stars observed at any $(J, J-K_s, \alpha, \delta)$, we are interested in determining the volume complete number density, $\nu_{\star} (X, Y, Z)$, for a particular stellar population. An important ingredient that relates the completeness in $(J, J-K_s, \alpha, \delta)$ space to real $(X, Y, Z)$ space is the (de-reddened) color-(absolute) magnitude density (CMD), $\rho_{\rm CMD} (M_J, [J-K_s]_0 | X, Y, Z)$. In case of {\it Gaia}, the CMD may be determined empirically using an external survey like {\it 2MASS} that is (almost) complete over the entire sky and a three-dimensional extinction map $(A_J, E(J - K_s))[X, Y, Z]$.

The number density, $\nu_{\star} (X, Y, Z)$, is determined by assuming that the observed stars are independent samples of an inhomogeneous Poisson process. This process is characterized by its rate function, $\lambda(O|\theta)$, that relates the observables, $O \equiv \{X, Y, Z, J, J-K_s\}$, measured by the survey to the model parameters $\theta$,
\beq
\begin{aligned}
\lambda(O|\theta)=& \, \nu_{\star} (X, Y, Z | \theta) \times | J(X, Y, Z; \alpha, \delta, D) | \\
& \rho_{\rm CMD} (M_J, [J-K_s]_0 | X, Y, Z) \, S(J, J-K_s, \alpha, \delta),
\end{aligned}
\label{eq:rate_param}
\eeq
where $| J(X, Y, Z; \alpha, \delta, D) | = D^2 \cos \delta$ is the Jacobian for the coordinate transformation. Dropping all terms independent of $\theta$, the likelihood of this process, $\mathcal{L}(\theta)$, can be written as,
\beq
\begin{aligned}
\ln \mathcal{L}(\theta) &= \sum_i \lambda (O_i | \theta) - \int d O \lambda (O | \theta) \\
&= \sum_i \nu_{\star} (X_i, Y_i, Z_i | \theta) - \int d D \, D^2 \, d \alpha \, d \delta  \, \cos \delta \nu_{\star} (X, Y, Z | \theta) \, \mathfrak{S}(\alpha, \delta, D),
\end{aligned}
\label{eq:ln_likhood}
\eeq
where $\mathfrak{S}(\alpha, \delta, D)$ is the effective selection function as defined by Ref.~\cite{Bovy:2017},
\beq
\mathfrak{S}(\alpha, \delta, D)=  \int d J \, d (J - K_s) \, \rho_{\rm CMD} (M_J, [J-K_s]_0 | X, Y, Z) \, S(J, J-K_s, \alpha, \delta).
\label{eq:sel_func}
\eeq
We can interpret the effective selection function as the fraction of stars of a stellar population at a distance $D$ and position $(\alpha, \delta)$ observed by the survey. 

With these ingredients in place, we can estimate the {\it true} underlying (binned) stellar density, $n_k$, from the observed number counts of stars, $N_k$, in non-overlapping bins, $\Pi_k (X, Y, Z)$. Thus, plugging in a parametric density law,
\beq
\nu_{\star} (X, Y, Z | \theta) = \sum_k n_k \, \Pi_k (X, Y, Z) 
\eeq
into the expression for the log-likelihood in Eq.~\ref{eq:ln_likhood}, we obtain,
\beq
\begin{aligned}
& \ln \mathcal{L}(\{n_k\}_k) \\
&= \sum_i \ln \sum_k n_k \Pi_k (X, Y, Z) - \int d D \, D^2 \, d \alpha \, d \delta  \, \cos \delta \, \sum_k n_k \, \Pi_k (X, Y, Z) \, \mathfrak{S}(\alpha, \delta, D) \\
&= \sum_k \left[ N_k \ln n_k - n_k \, \int d D \, D^2 \, d \alpha \, d \delta  \, \cos \delta \, \Pi_k (X, Y, Z) \, \mathfrak{S}(\alpha, \delta, D) \right], 
\end{aligned}
\label{eq:density}
\eeq
where the second equality follows from considering all possible combinations of $i$ stars distributed in $k$ identical bins. The maximum likelihood estimate (MLE) can be calculated analytically by differentiating the above equation with respect to $n_k$ and setting the $k$ derivatives to zero. We find that
\beq
\hat{n}_k = \frac{N_k}{\int d D \, D^2 \, d \alpha \, d \delta  \, \cos \delta \, \Pi_k (X, Y, Z) \, \mathfrak{S}(\alpha, \delta, D)}, 
\eeq
which can be written more compactly by defining $\Xi (\Pi_k)$ as the {\it effective volume completeness} per bin and $\Xi (\Pi_k) \, V(\Pi_k) = \int_{\Pi_k} d^3 x \, \mathfrak{S}(\alpha, \delta, D)$ as the effective volume,
\beq
\hat{n}_k = \frac{N_k}{\Xi (\Pi_k) \, V (\Pi_k)}. 
\eeq
The uncertainty in our estimate is easily evaluated by calculating the Fisher information,
\beq
 \sqrt{-\frac{\partial^2 \ln \mathcal{L}(\{n_k\}_k)}{\partial (\{n_k\}_k)^2 }} \equiv \sigma_{\hat{n}_k} = \frac{\hat{n}_k}{\sqrt{N_k}}. 
\label{eq:den_uncertain}
\eeq

\section{Uncertainty Analysis}
\label{sec:Uncertainty_analysis}
In this section, we discuss our choices of bin sizes in the vertical height $z$ and velocity $w$ for 
constructing the number density and midplane velocity distribution respectively. 

\begin{figure*}[!h]
	\begin{center}
\includegraphics[scale=0.395]{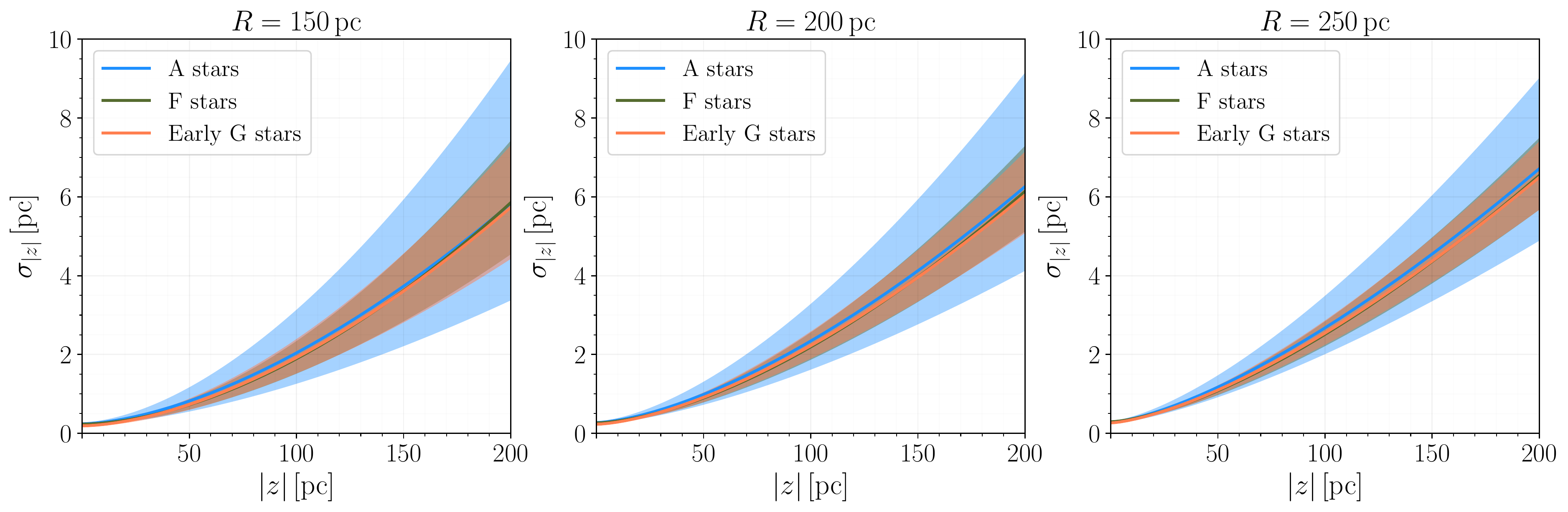}
\caption{$1\sigma$ spread in the uncertainty (at leading order) of $z$ as a function of $z$ for 
different radial cuts. }
\label{fig:parallax}
	\end{center}
\end{figure*}

The uncertainty in $z$ is given by,
\beq
\delta z^2 \ ({\rm kpc}^2)= \left(\frac{\sin b}{\varpi^2}\right)^2 \sigma_{\varpi}^2 + \Big(\frac{\cos b}{\varpi}\Big)^2 \sigma_{b}^2 + \Big(\frac{2\sin b \cos b}{\varpi^3} \Big) \, \sigma_{\varpi b}^2 
\label{eq:zbin_size}
\eeq
which is dominated by the parallax uncertainty due to the extra factor of $\varpi$ in unit of ${{\rm mas} \approx 10^{-9}}$ in the first term. We plot the uncertainty in $z$ (at leading 
order) as a function of $z$ for all tracers in Fig.~\ref{fig:parallax}. Although the maximum 
uncertainty is $\approx 10$ pc, we conservatively adopt 20 pc as the bin size to account for the underestimation of the reported uncertainties in DR2~\cite{2018arXiv180409366L}.

Similarly, the uncertainty in $w$ is
\beq
\left( \frac{\sigma_{w}}{w}\right)^2 = \left(\frac{\sigma_{\varpi}}{\varpi}\right)^2 +\Big(\frac{\sigma_{\mu_b}}{\mu_b}\Big)^2 + \ \text{subleading terms}.
\eeq
where the omitted terms are suppressed by $10^{-2}$ when $|b| < 5^\circ$. Around the 
midplane, $\sigma_{\mu_b}/\mu_b \, {\lsim} \, 0.2$, which translates to $\sigma_w \approx 1.5$ km/s. Therefore, we pick 1.5 km/s as the bin size for obtaining the $f_0(w)$ profile.

\section{Variation of Midplane Cut}
\label{sec:midplane_bzcut}
The midplane velocity profile is required in Eq.~\eqref{eq:predict} to predict the tracer density for a given mass model. With partial radial velocity measured by ${\it Gaia}$, we define the midplane in two ways: one is putting a cut on the galactic latitude $|b| < 5^{\text{o}}$ while the other is requiring $|z| < (20-50)$ pc~\cite{garbari:2011}. For both samples, we approximate $v_r$ by its mean value $\langle v_r \rangle$ in Eq.~\eqref{eq:meanvr} when there is no $v_r$ data available for a star. However, for the $z$-cut sample, we discard stars with $|b| > 5^{\text{o}}$ that do not have any $v_r$ data. 

\begin{figure}[!h]
	\centering$
	\begin{array}{cc}
       \includegraphics[width=0.32\linewidth]{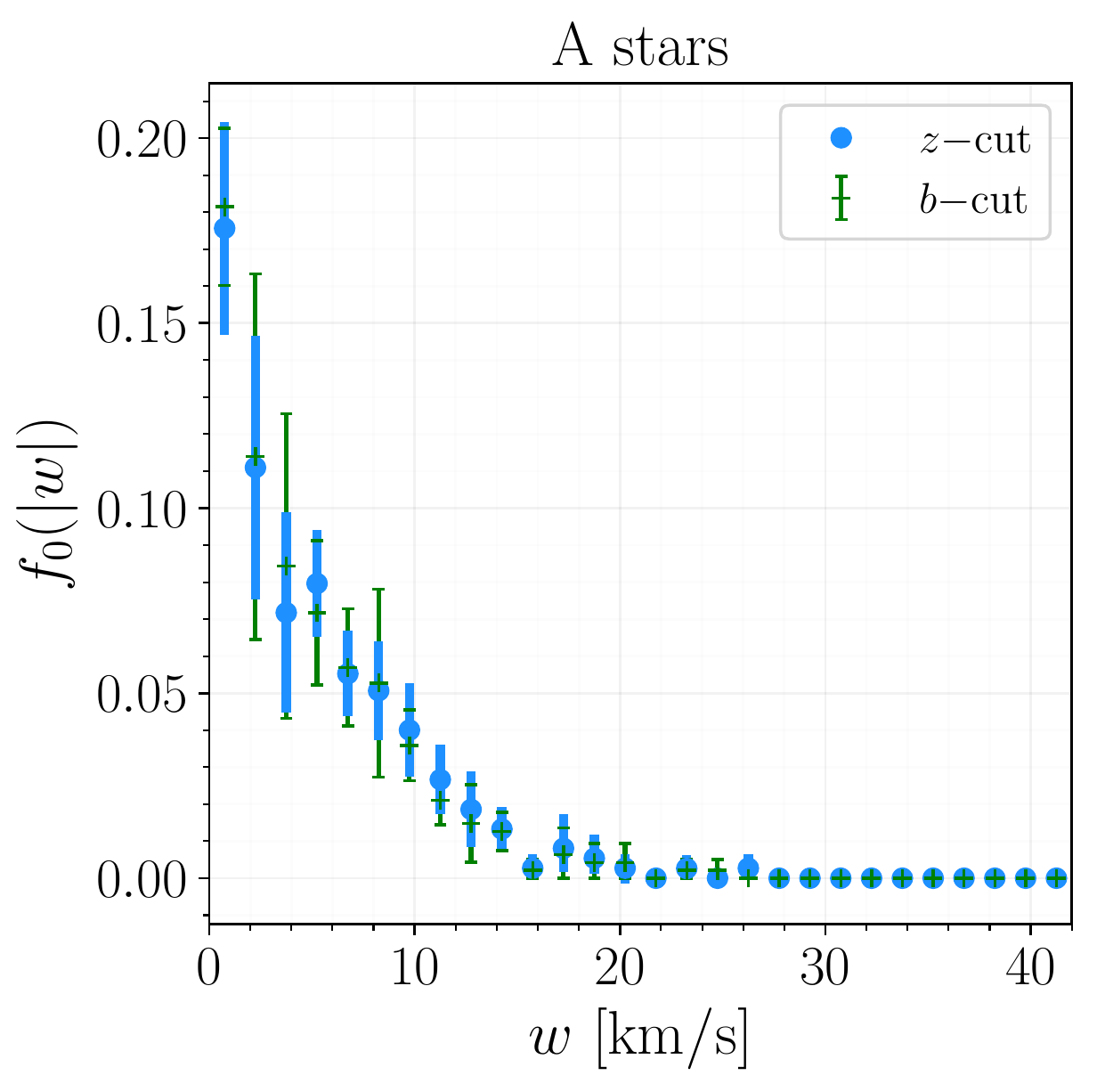}
       \includegraphics[width=0.32\linewidth]{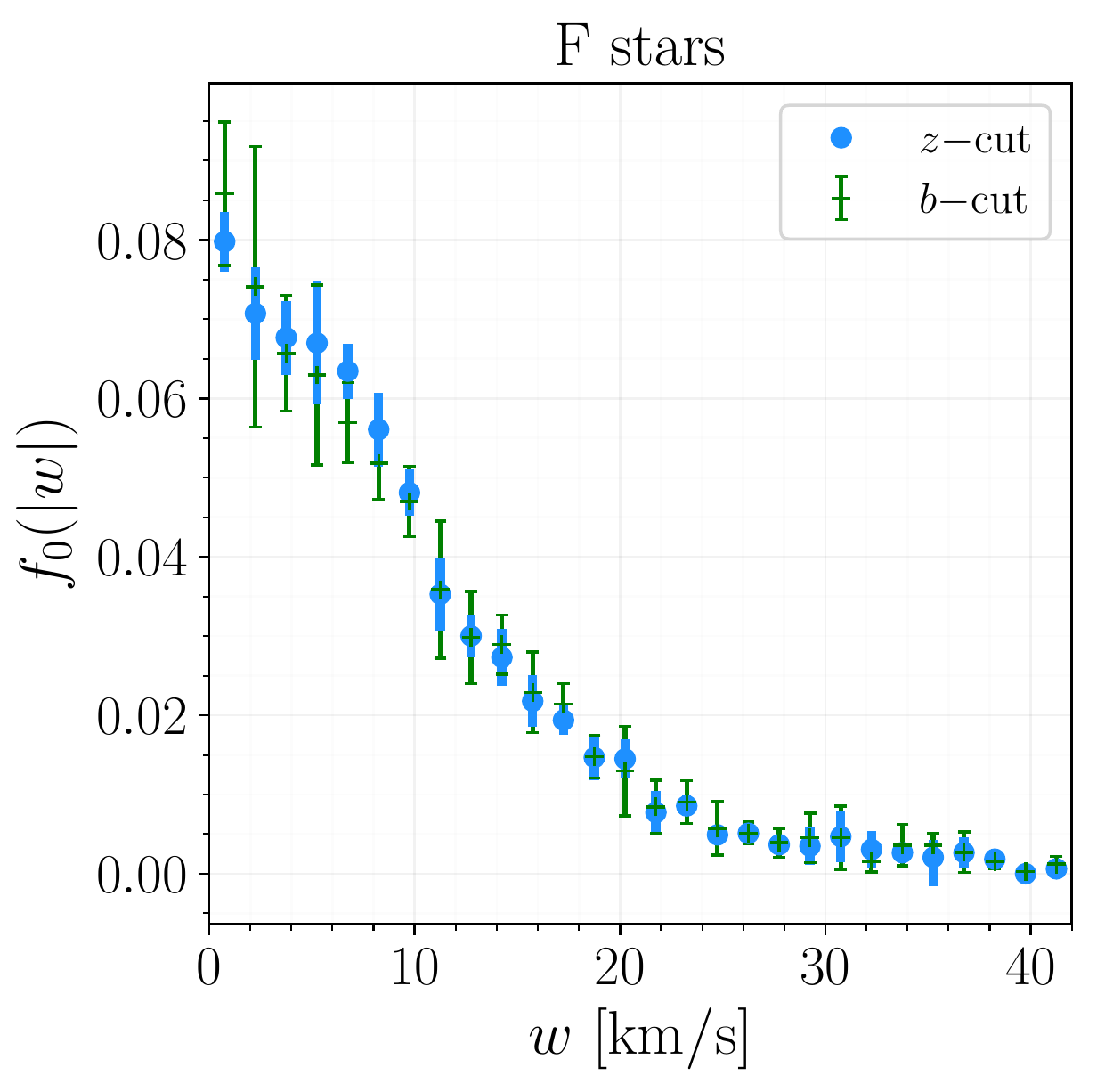}
       \includegraphics[width=0.32\linewidth]{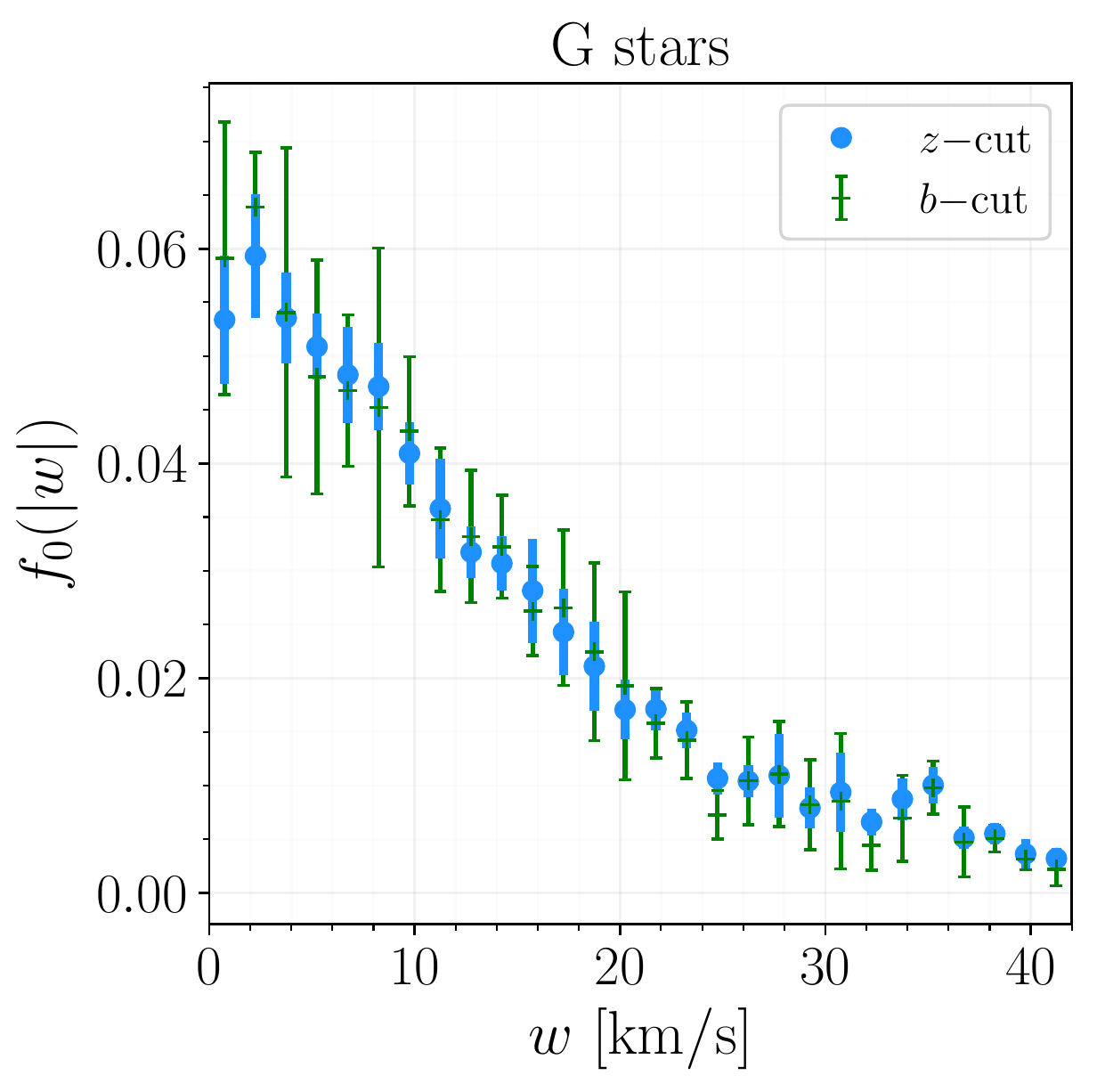}
	\end{array}$
    \caption{Midplane velocity distribution $f_0(|w|)$ for A (left), F (middle) and early G (right)
    stars. The distributions obtained using the $|b| < 5^{\text{o}}$ cut (green) and the $|z| < 20$
    pc cut (blue) are consistent within error bars. }
    \label{fig:fw_check}
\end{figure}

The midplane velocity distributions of the $z$- and $b$-cut samples are presented in Fig.~\ref{fig:fw_check} and agree with each other within 1$\sigma$ uncertainties. We note that the uncertainties in the midplane velocity data using $z$-cut are smaller than those using the $b$-cut. The uncertainties are dominated by systematics due to differences between $f(w>0)$ and $f(w<0)$. It turns out that the $z$-cut data is more symmetric about $z=0$ and thus has smaller uncertainties. In our analysis, we still use the $b$-cut sample, since there could be a potential selection bias in the $z$-cut sample, in which we discard a considerable fraction of stars with five-parameter astrometric solutions because we don't know their radial velocities. 

We also check the isothermality of the tracers by fitting the midplane data with Gaussian distributions. From the fits, we find that the velocity dispersions $\sigma_z$ are 5.7, 11.2, 15.0 km/s for A, F and early G stars respectively. The $\chi^2$ of the fits are 11.9, 20.3 and 35.4 for 16, 28, and 28 degrees of freedom respectively. The Gaussian (isothermal) distributions give reasonable fits for A and early G stars, but not as good a fit for F stars. In our analysis, we always use the distributions from data and never their Gaussian fits.

\section{Bootstrap Statistics}
\label{sec:bootstrap}
Bootstrap resampling is a standard statistical technique to acquire the mean and uncertainty when there is only one data set available and analytic propagation of uncertainty cannot be performed easily. The basic idea of the method is described below. 

Suppose we have a set of $N$ stars labelled as $S_N = \{X_1, X_2,\cdots,X_N\}$. 
Each star $X_k$ is associated with 6 dimensional phase space coordinates denoted by $\theta_k$. In bootstrap resampling, we make random draws \emph{with replacement} star-by-star from the original set of stars $S_N$. This generates a new data set $\widetilde{S}_N$ of the same size $N$, with each star labeled as $\widetilde{X}_k$. Since the draws are with replacement, we expect (many) duplicated coordinate values in the new data set, such as $\widetilde{X}_k = \theta_k$ and $\widetilde{X}_{k+1} = \theta_k$, for large $N$. Therefore, $\widetilde{S}_N \neq S_N$ in general.

We resample $B$ times the original data set $S_N$, labeling them as $\widetilde{S}^{(1)}_N$, $\widetilde{S}^{(2)}_N$, ..., $\widetilde{S}^{(B)}_N$. The variance of the underlying distribution in each $z$ bin can be estimated as follows,
\begin{equation}\label{eq:resample}
\widetilde{\sigma}^2_\nu(z) = \frac{1}{B}\sum^B_{k=1}(\widetilde{\nu}^{(k)}(z)-\overline{\widetilde{\nu}}(z))^2\,,\qquad\text{where \quad $\overline{\widetilde{\nu}}(z) = \frac{1}{B}\sum^B_{k=1}\widetilde{\nu}^{(k)}(z)$.}
\end{equation}
For sufficiently large $B$, it can be proven that $\widetilde{\sigma}^2_\nu\rightarrow\sigma^2_\nu$~\cite{Ivezic:2014:SDM:2578955}.

In propagating velocity uncertainties into prediction uncertainties via the bootstrap method, we choose $B=1000$ as a compromise between computational time and statistical precision. We take the bin-by-bin variance of all the predictions based on the 1000 resampled velocity sets as the estimator for the statistical uncertainty, $\left( \sigma^{2}_{\nu}\right)^{\rm mod,\,stat}$, of the predicted profile $\nu^{\rm mod}$.

\begin{figure*}[ph]
\subfloat{%
\includegraphics[height= 7cm, width=.49\linewidth]{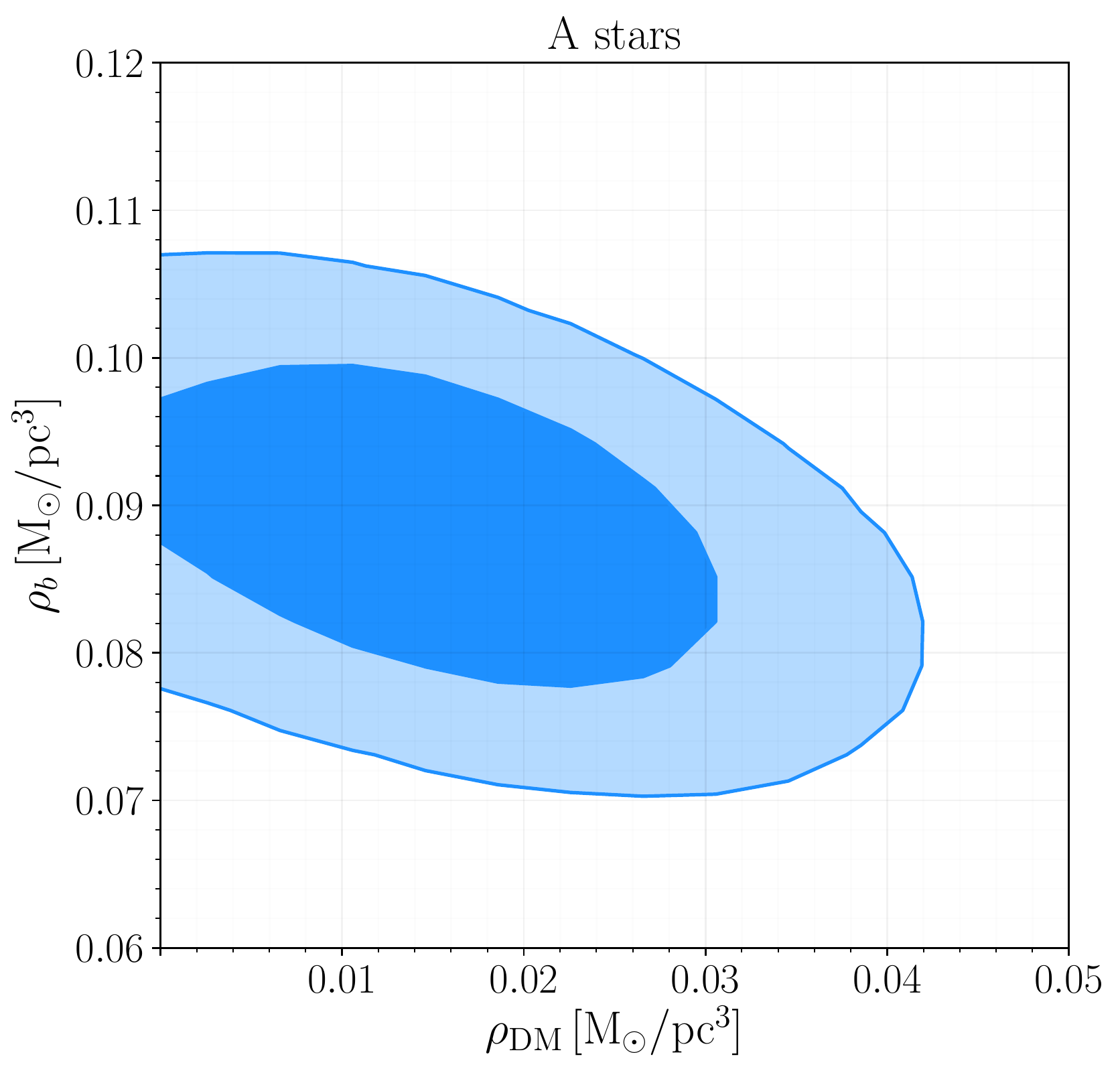}
}
\subfloat{%
\includegraphics[height= 7cm, width=.49\linewidth]{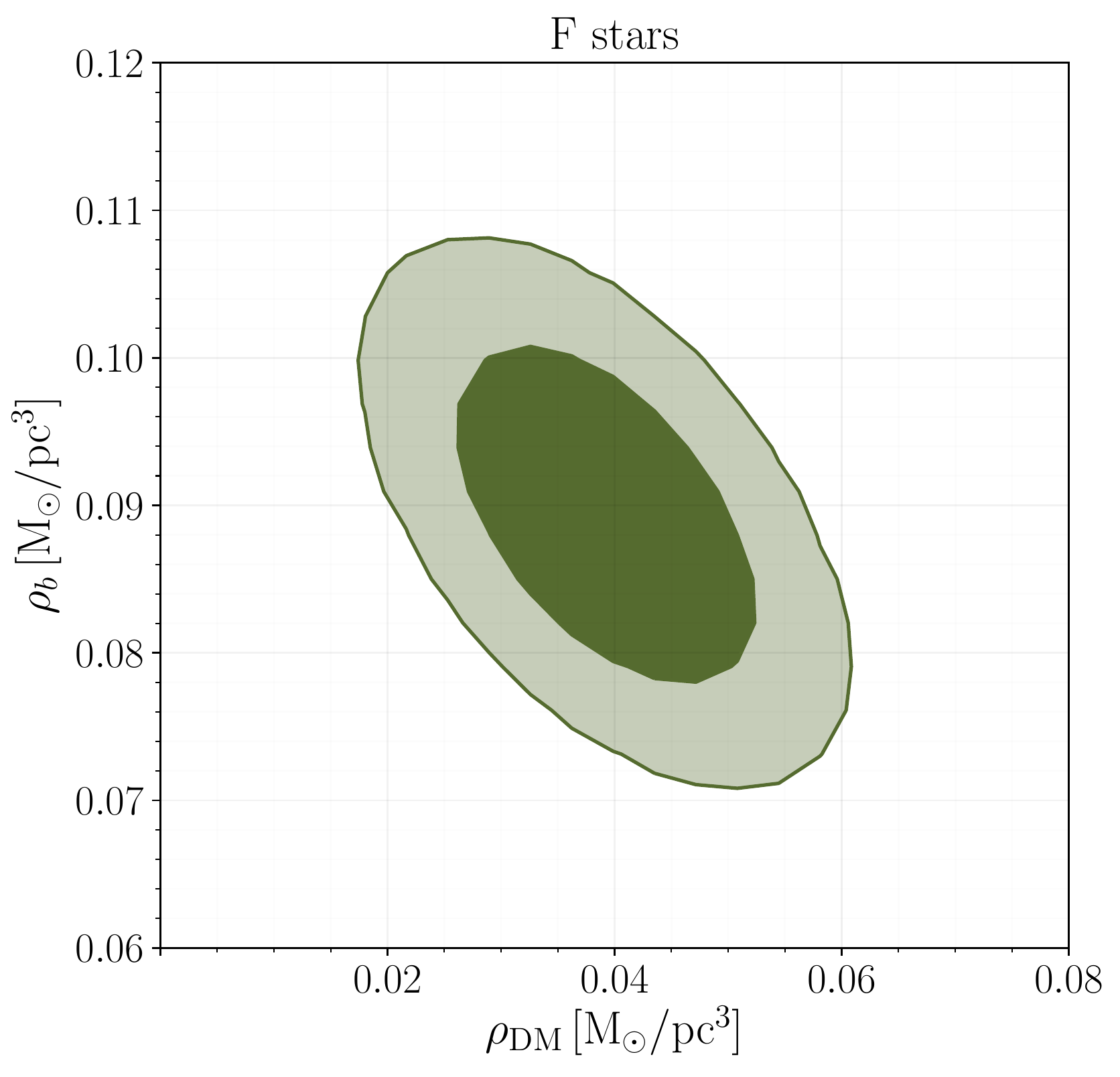}
}\\
\begin{center}
\subfloat{%
\includegraphics[height= 7cm, width=.49\linewidth]{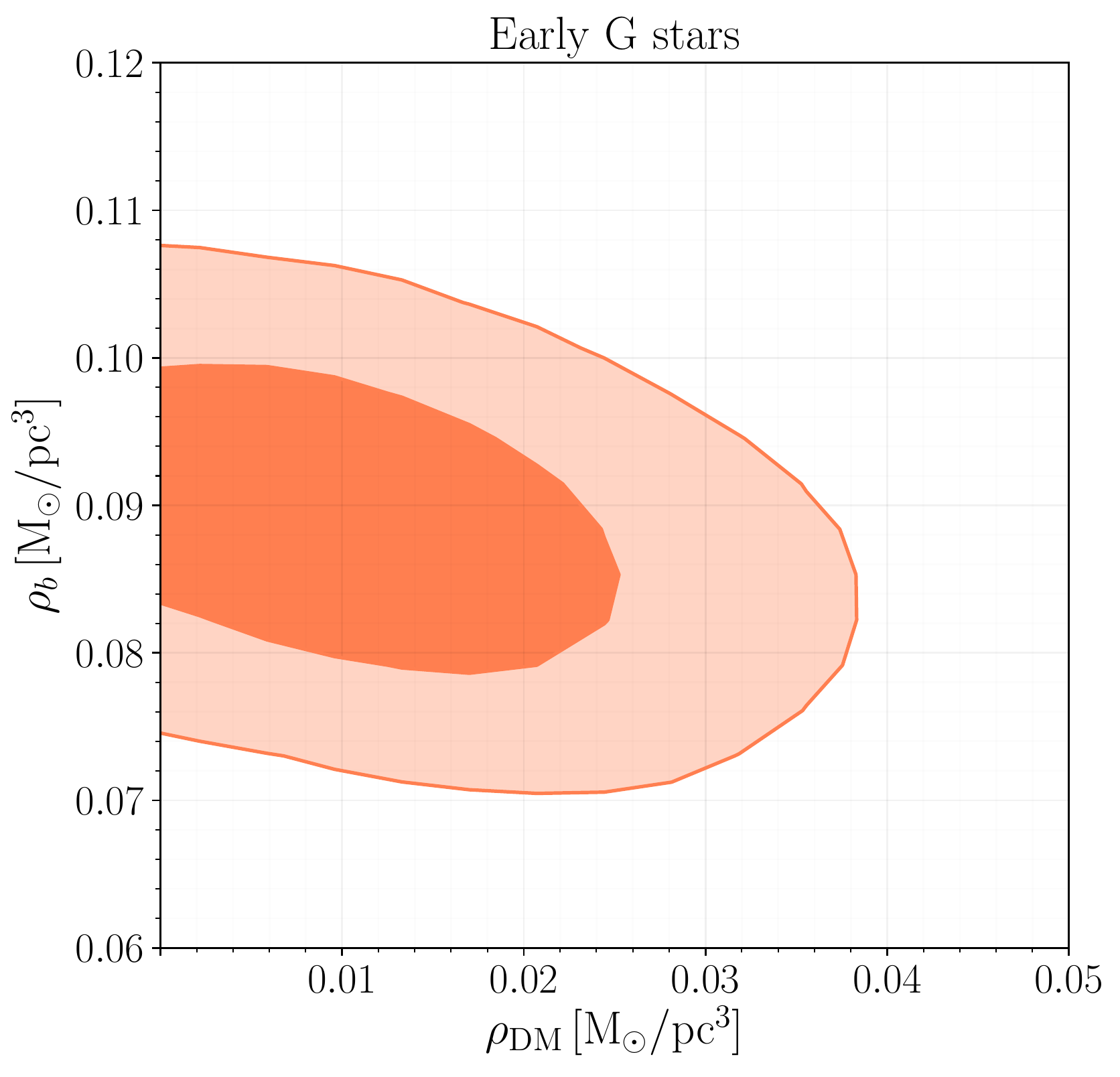}
}
\end{center}

\caption{Marginalized posteriors indicating the degeneracy between the local densities of 
baryons $\rho_b$ and halo DM $\rho_{\mathrm{DM}}$.}
\label{sfig:exclusion}
\end{figure*}

\begin{figure}[ph]
    \makebox[\linewidth]{
        \includegraphics[width=1.2\linewidth]{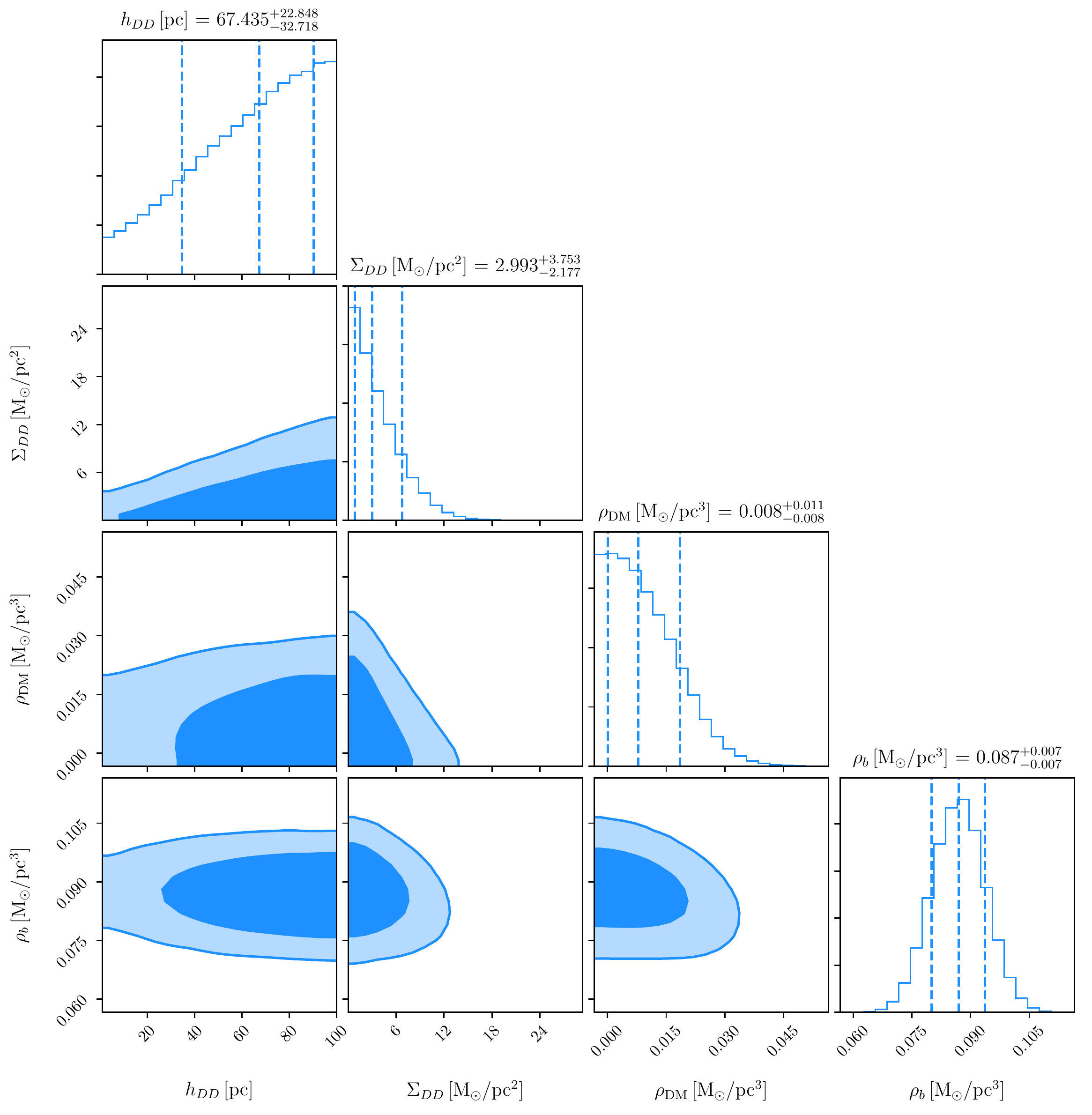}
    }
    \caption{Marginalized posterior distributions of thin DD parameters, local dark matter 	
    density $\rho_{DM}$, and the total baryon density in the midplane $\rho_b$ for A stars.
    The dark (light) shaded regions indicate the $68\%$ ($95\%$) credible regions,
    whereas the dashed lines represent the 16th, 50th, and 84th percent quantile values of
    the posterior distribution.}
    \label{fig:mcmc_A}
\end{figure}

\begin{figure}[ph]
    \makebox[\linewidth]{
        \includegraphics[width=1.2\linewidth]{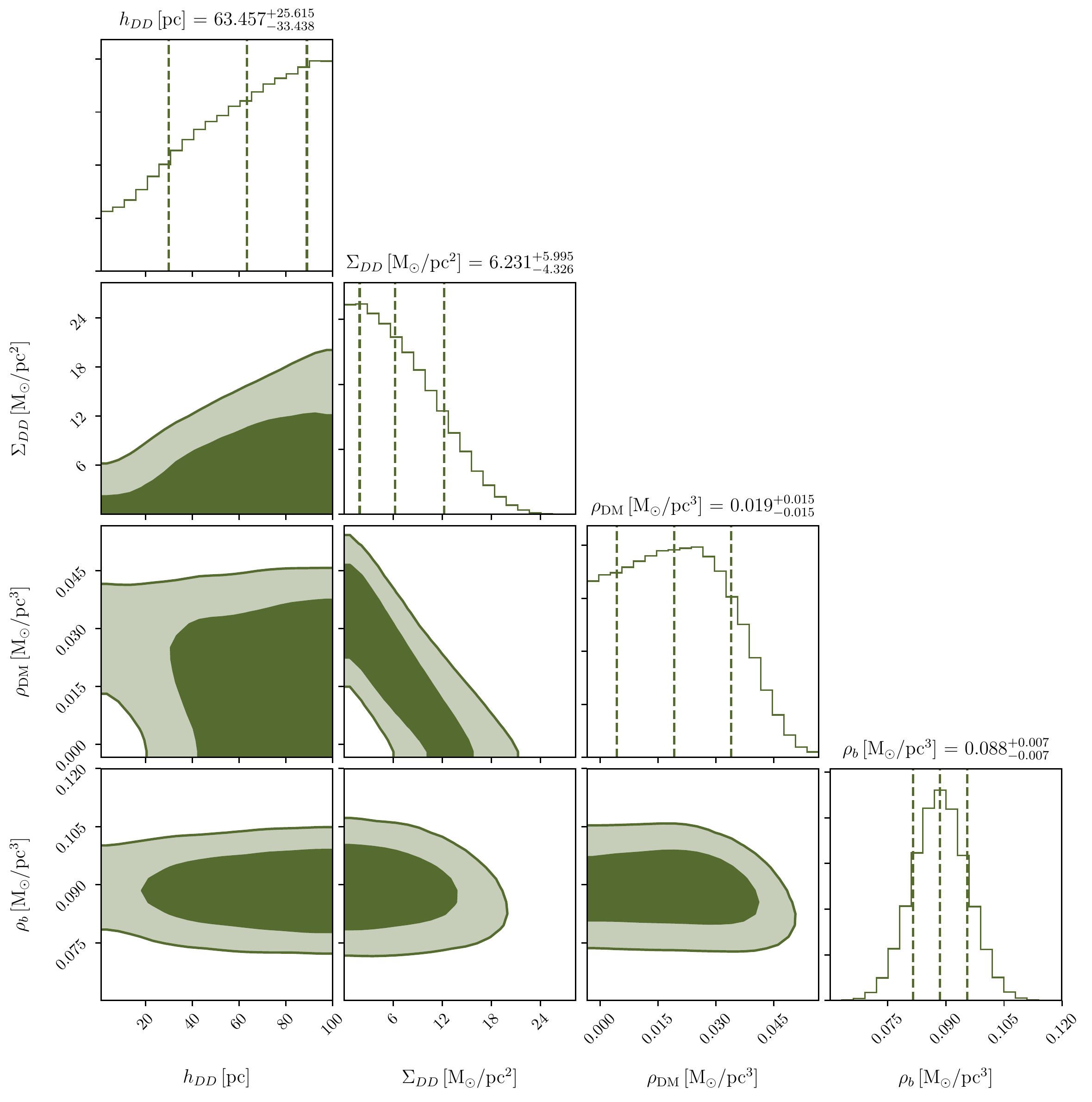}
    }
    \caption{Marginalized posterior distributions of thin DD parameters, local dark matter 	
    density $\rho_{DM}$, and the total baryon density in the midplane $\rho_b$ for F stars.
    The dark (light) shaded regions indicate the $68\%$ ($95\%$) credible regions,
    whereas the dashed lines represent the 16th, 50th, and 84th percent quantile values of
    the posterior distribution.}
    \label{fig:mcmc_F}
\end{figure}

\begin{figure}[ph]
    \makebox[\linewidth]{
        \includegraphics[width=1.2\linewidth]{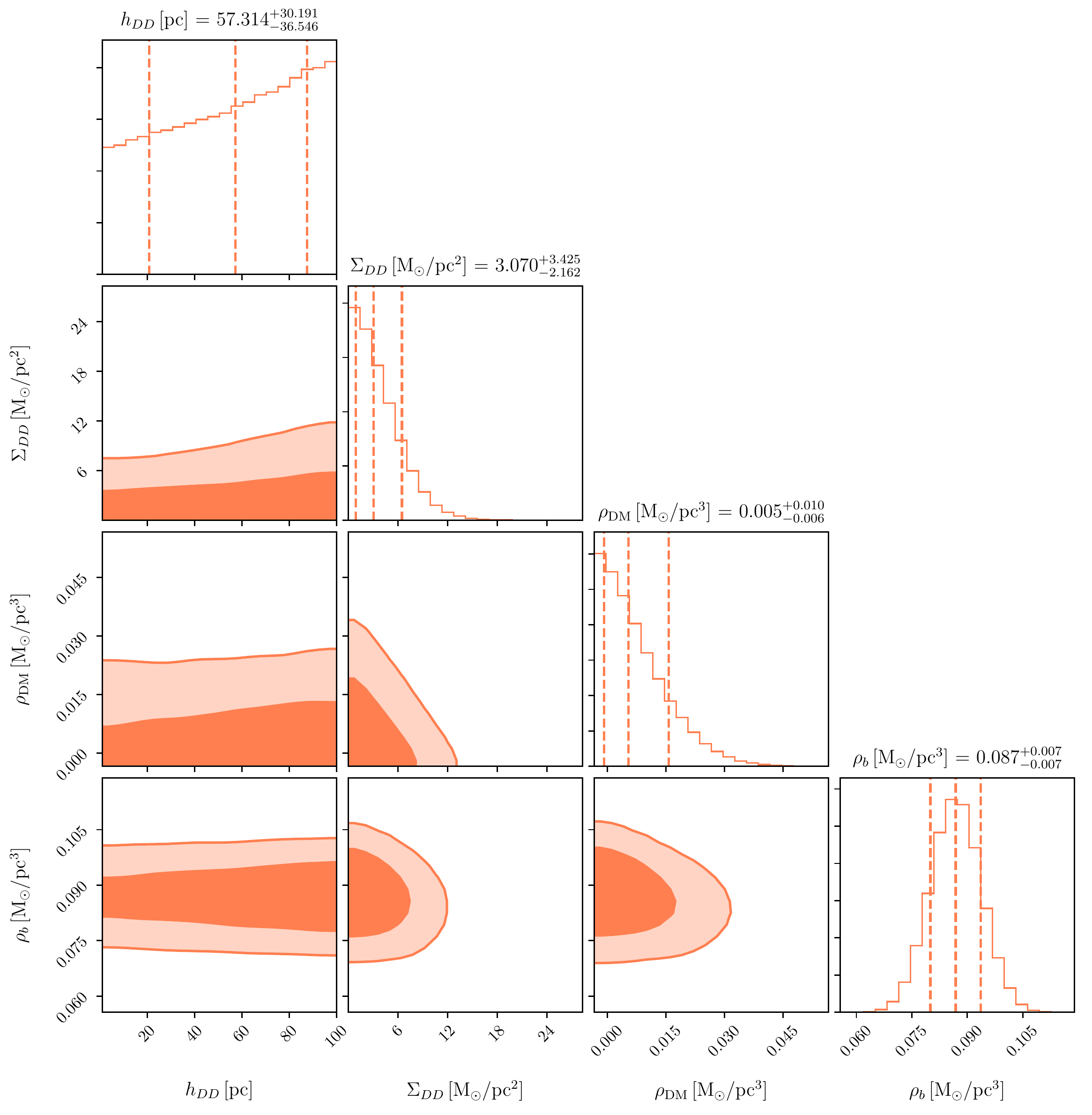}
    }
    \caption{Marginalized posterior distributions of thin DD parameters, local dark matter 	
    density $\rho_{DM}$, and the total baryon density in the midplane $\rho_b$ for G stars.
    The dark (light) shaded regions indicate the $68\%$ ($95\%$) credible regions,
    whereas the dashed lines represent the 16th, 50th, and 84th percent quantile values of
    the posterior distribution.}
    \label{fig:mcmc_G}
\end{figure}

\bibliography{ref}

\providecommand{\href}[2]{#2}\begingroup\raggedright\begin{thebibliography}{10}

\bibitem{2016A&A...595A...1G}
{Gaia Collaboration}, T.~{Prusti}, J.~H.~J. {de Bruijne}, A.~G.~A. {Brown},
  A.~{Vallenari}, C.~{Babusiaux} et~al., \emph{{The Gaia mission}},
  \href{https://doi.org/10.1051/0004-6361/201629272}{\emph{A\&A} {\bfseries
  595} (Nov., 2016) A1}, [\href{https://arxiv.org/abs/1609.04153}{{\ttfamily
  1609.04153}}].

\bibitem{2018arXiv180409365G}
{Gaia Collaboration}, A.~G.~A. {Brown}, A.~{Vallenari}, T.~{Prusti}, J.~H.~J.
  {de Bruijne}, C.~{Babusiaux} et~al., \emph{{Gaia Data Release 2. Summary of
  the contents and survey properties}}, {\emph{ArXiv e-prints} (Apr., 2018) },
  [\href{https://arxiv.org/abs/1804.09365}{{\ttfamily 1804.09365}}].

\bibitem{2018arXiv180409366L}
L.~{Lindegren}, J.~{Hernandez}, A.~{Bombrun}, S.~{Klioner}, U.~{Bastian},
  M.~{Ramos-Lerate} et~al., \emph{{Gaia Data Release 2: The astrometric
  solution}}, {\emph{ArXiv e-prints} (Apr., 2018) },
  [\href{https://arxiv.org/abs/1804.09366}{{\ttfamily 1804.09366}}].

\bibitem{2018arXiv180409368E}
D.~W. {Evans}, M.~{Riello}, F.~{De Angeli}, J.~M. {Carrasco}, P.~{Montegriffo},
  C.~{Fabricius} et~al., \emph{{Gaia Data Release 2: Photometric content and
  validation}}, {\emph{ArXiv e-prints} (Apr., 2018) },
  [\href{https://arxiv.org/abs/1804.09368}{{\ttfamily 1804.09368}}].

\bibitem{2018arXiv180409367R}
M.~{Riello}, F.~{De Angeli}, D.~W. {Evans}, G.~{Busso}, N.~C. {Hambly},
  M.~{Davidson} et~al., \emph{{Gaia Data Release 2: processing of the
  photometric data}}, {\emph{ArXiv e-prints} (Apr., 2018) },
  [\href{https://arxiv.org/abs/1804.09367}{{\ttfamily 1804.09367}}].

\bibitem{2018arXiv180409371S}
P.~{Sartoretti}, D.~{Katz}, M.~{Cropper}, P.~{Panuzzo}, G.~M. {Seabroke},
  Y.~{Viala} et~al., \emph{{Gaia Data Release 2: Processing the spectroscopic
  data}}, {\emph{ArXiv e-prints} (Apr., 2018) },
  [\href{https://arxiv.org/abs/1804.09371}{{\ttfamily 1804.09371}}].

\bibitem{2018arXiv180409369C}
M.~{Cropper}, D.~{Katz}, P.~{Sartoretti}, T.~{Prusti}, J.~H.~J. {de Bruijne},
  F.~{Chassat} et~al., \emph{{Gaia Radial Velocity Spectrometer}}, {\emph{ArXiv
  e-prints} (Apr., 2018) }, [\href{https://arxiv.org/abs/1804.09369}{{\ttfamily
  1804.09369}}].

\bibitem{2018arXiv180409372K}
D.~{Katz}, P.~{Sartoretti}, M.~{Cropper}, P.~{Panuzzo}, G.~M. {Seabroke},
  Y.~{Viala} et~al., \emph{{Gaia Data Release 2: Properties and validation of
  the radial velocities}}, {\emph{ArXiv e-prints} (Apr., 2018) },
  [\href{https://arxiv.org/abs/1804.09372}{{\ttfamily 1804.09372}}].

\bibitem{2018arXiv180409376L}
X.~{Luri}, A.~G.~A. {Brown}, L.~M. {Sarro}, F.~{Arenou}, C.~A.~L.
  {Bailer-Jones}, A.~{Castro-Ginard} et~al., \emph{{Gaia Data Release 2: using
  Gaia parallaxes}}, {\emph{ArXiv e-prints} (Apr., 2018) },
  [\href{https://arxiv.org/abs/1804.09376}{{\ttfamily 1804.09376}}].

\bibitem{Flynn:2006tm}
C.~Flynn, J.~Holmberg, L.~Portinari, B.~Fuchs and H.~Jahreiss, \emph{{On the
  mass-to-light ratio of the local Galactic disc and the optical luminosity of
  the Galaxy}},
  \href{https://doi.org/10.1111/j.1365-2966.2006.10911.x}{\emph{Mon. Not. Roy.
  Astron. Soc.} {\bfseries 372} (2006) 1149--1160},
  [\href{https://arxiv.org/abs/astro-ph/0608193}{{\ttfamily
  astro-ph/0608193}}].

\bibitem{Bovy:2013raa}
J.~Bovy and H.-W. Rix, \emph{{A Direct Dynamical Measurement of the Milky Way's
  Disk Surface Density Profile, Disk Scale Length, and Dark Matter Profile at 4
  kpc $\stackrel{<}{\sim}$ R $\stackrel{<}{\sim}$ 9 kpc}},
  \href{https://doi.org/10.1088/0004-637X/779/2/115}{\emph{Astrophys. J.}
  {\bfseries 779} (2013) 115},
  [\href{https://arxiv.org/abs/1309.0809}{{\ttfamily 1309.0809}}].

\bibitem{Mckee:2015}
C.~F. {McKee}, A.~{Parravano} and D.~J. {Hollenbach}, \emph{{Stars, Gas, and
  Dark Matter in the Solar Neighborhood}},
  \href{https://doi.org/10.1088/0004-637X/814/1/13}{\emph{Astrophys. J.}
  {\bfseries 814} (Nov., 2015) 13},
  [\href{https://arxiv.org/abs/1509.05334}{{\ttfamily 1509.05334}}].

\bibitem{Kramer:2016dew}
E.~D. Kramer and L.~Randall, \emph{{Interstellar Gas and a Dark Disk}},
  \href{https://doi.org/10.3847/0004-637X/829/2/126}{\emph{Astrophys. J.}
  {\bfseries 829} (2016) 126},
  [\href{https://arxiv.org/abs/1603.03058}{{\ttfamily 1603.03058}}].

\bibitem{1932BAN.....6..249O}
J.~H. {Oort}, \emph{{The force exerted by the stellar system in the direction
  perpendicular to the galactic plane and some related problems}}, {\emph{Bull.
  Astros. Inst. Netherlands} {\bfseries 6} (Aug., 1932) 249}.

\bibitem{1989MNRAS.239..571K}
K.~{Kuijken} and G.~{Gilmore}, \emph{{The mass distribution in the galactic
  disc. I - A technique to determine the integral surface mass density of the
  disc near the sun.}},
  \href{https://doi.org/10.1093/mnras/239.2.571}{\emph{Mon. Not. Roy. Astron.
  Soc.} {\bfseries 239} (Aug., 1989) 571--603}.

\bibitem{1989MNRAS.239..605K}
K.~{Kuijken} and G.~{Gilmore}, \emph{{The Mass Distribution in the Galactic
  Disc - II - Determination of the Surface Mass Density of the Galactic Disc
  Near the Sun}}, \href{https://doi.org/10.1093/mnras/239.2.605}{\emph{Mon.
  Not. Roy. Astron. Soc.} {\bfseries 239} (Aug., 1989) 605--649}.

\bibitem{1989MNRAS.239..651K}
K.~{Kuijken} and G.~{Gilmore}, \emph{{The Mass Distribution in the Galactic
  Disc - Part III - the Local Volume Mass Density}},
  \href{https://doi.org/10.1093/mnras/239.2.651}{\emph{Mon. Not. Roy. Astron.
  Soc.} {\bfseries 239} (Aug., 1989) 651--664}.

\bibitem{1993AIPC..278..580F}
B.~{Fuchs} and R.~{Wielen}, \emph{{Kinematical constraints on the dynamically
  determined local mass density of the Galaxy}},  in \emph{Back to the Galaxy}
  (S.~S. {Holt} and F.~{Verter}, eds.), vol.~278 of \emph{American Institute of
  Physics Conference Series}, pp.~580--583, 1993,
  \href{https://doi.org/10.1063/1.43942}{DOI}.

\bibitem{1994MNRAS.270..471F}
C.~{Flynn} and B.~{Fuchs}, \emph{{Density of dark matter in the Galactic
  disk}}, \href{https://doi.org/10.1093/mnras/270.3.471}{\emph{Mon. Not. Roy.
  Astron. Soc.} {\bfseries 270} (Oct., 1994) }.

\bibitem{Holmberg:1998xu}
J.~Holmberg and C.~Flynn, \emph{{The local density of matter mapped by
  hipparcos}},
  \href{https://doi.org/10.1046/j.1365-8711.2000.02905.x}{\emph{Mon. Not. Roy.
  Astron. Soc.} {\bfseries 313} (2000) 209--216},
  [\href{https://arxiv.org/abs/astro-ph/9812404}{{\ttfamily
  astro-ph/9812404}}].

\bibitem{2004MNRAS.352..440H}
J.~{Holmberg} and C.~{Flynn}, \emph{{The local surface density of disc matter
  mapped by Hipparcos}},
  \href{https://doi.org/10.1111/j.1365-2966.2004.07931.x}{\emph{Mon. Not. Roy.
  Astron. Soc.} {\bfseries 352} (Aug., 2004) 440--446},
  [\href{https://arxiv.org/abs/astro-ph/0405155}{{\ttfamily
  astro-ph/0405155}}].

\bibitem{vanLeeuwen:2005yx}
F.~van Leeuwen and E.~Fantino, \emph{{A New reduction of the raw Hipparcos
  data}}, \href{https://doi.org/10.1051/0004-6361:20053193}{\emph{Astron.
  Astrophys.} {\bfseries 439} (2005) 791--803},
  [\href{https://arxiv.org/abs/astro-ph/0505432}{{\ttfamily
  astro-ph/0505432}}].

\bibitem{garbari:2012}
S.~{Garbari}, C.~{Liu}, J.~I. {Read} and G.~{Lake}, \emph{{A new determination
  of the local dark matter density from the kinematics of K dwarfs}},
  \href{https://doi.org/10.1111/j.1365-2966.2012.21608.x}{\emph{Mon. Not. Roy.
  Astron. Soc.} {\bfseries 425} (Sept., 2012) 1445--1458},
  [\href{https://arxiv.org/abs/1206.0015}{{\ttfamily 1206.0015}}].

\bibitem{Bovy:2012tw}
J.~Bovy and S.~Tremaine, \emph{{On the local dark matter density}},
  \href{https://doi.org/10.1088/0004-637X/756/1/89}{\emph{Astrophys. J.}
  {\bfseries 756} (2012) 89},
  [\href{https://arxiv.org/abs/1205.4033}{{\ttfamily 1205.4033}}].

\bibitem{Silverwood2016}
H.~{Silverwood}, S.~{Sivertsson}, P.~{Steger}, J.~I. {Read} and G.~{Bertone},
  \emph{{A non-parametric method for measuring the local dark matter density}},
  \href{https://doi.org/10.1093/mnras/stw917}{\emph{Mon. Not. Roy. Astron.
  Soc.} {\bfseries 459} (July, 2016) 4191--4208},
  [\href{https://arxiv.org/abs/1507.08581}{{\ttfamily 1507.08581}}].

\bibitem{Silverwood2017}
S.~{Sivertsson}, H.~{Silverwood}, J.~I. {Read}, G.~{Bertone} and P.~{Steger},
  \emph{{The Local Dark Matter Density from SDSS-SEGUE G-dwarfs}},
  \href{https://doi.org/10.1093/mnras/sty977}{\emph{Mon. Not. Roy. Astron.
  Soc.} (Apr., 2018) }, [\href{https://arxiv.org/abs/1708.07836}{{\ttfamily
  1708.07836}}].

\bibitem{Bovy2012a}
Y.-S. {Ting}, H.-W. {Rix}, J.~{Bovy} and G.~{van de Ven}, \emph{{Constraining
  the Galactic potential via action-based distribution functions for
  mono-abundance stellar populations}},
  \href{https://doi.org/10.1093/mnras/stt1053}{\emph{"Mon. Not. Roy. Astron.
  Soc."} {\bfseries 434} (Sept., 2013) 652--660},
  [\href{https://arxiv.org/abs/1212.0006}{{\ttfamily 1212.0006}}].

\bibitem{Bovy2012d}
J.~{Bovy}, H.-W. {Rix}, C.~{Liu}, D.~W. {Hogg}, T.~C. {Beers} and Y.~S. {Lee},
  \emph{{The Spatial Structure of Mono-abundance Sub-populations of the Milky
  Way Disk}},
  \href{https://doi.org/10.1088/0004-637X/753/2/148}{\emph{Astrophys. J.}
  {\bfseries 753} (July, 2012) 148},
  [\href{https://arxiv.org/abs/1111.1724}{{\ttfamily 1111.1724}}].

\bibitem{Salucci:2010qr}
P.~Salucci, F.~Nesti, G.~Gentile and C.~F. Martins, \emph{{The dark matter
  density at the Sun's location}},
  \href{https://doi.org/10.1051/0004-6361/201014385}{\emph{Astron. Astrophys.}
  {\bfseries 523} (2010) A83},
  [\href{https://arxiv.org/abs/1003.3101}{{\ttfamily 1003.3101}}].

\bibitem{Pato:2015dua}
M.~Pato, F.~Iocco and G.~Bertone, \emph{{Dynamical constraints on the dark
  matter distribution in the Milky Way}},
  \href{https://doi.org/10.1088/1475-7516/2015/12/001}{\emph{JCAP} {\bfseries
  1512} (2015) 001}, [\href{https://arxiv.org/abs/1504.06324}{{\ttfamily
  1504.06324}}].

\bibitem{2013PhR...531....1S}
L.~E. {Strigari}, \emph{{Galactic searches for dark matter}},
  \href{https://doi.org/10.1016/j.physrep.2013.05.004}{\emph{Phys. Rep.}
  {\bfseries 531} (Oct., 2013) 1--88},
  [\href{https://arxiv.org/abs/1211.7090}{{\ttfamily 1211.7090}}].

\bibitem{Read:2014qva}
J.~I. Read, \emph{{The Local Dark Matter Density}},
  \href{https://doi.org/10.1088/0954-3899/41/6/063101}{\emph{J. Phys.}
  {\bfseries G41} (2014) 063101},
  [\href{https://arxiv.org/abs/1404.1938}{{\ttfamily 1404.1938}}].

\bibitem{Fan:2013tia}
J.~Fan, A.~Katz, L.~Randall and M.~Reece, \emph{{Dark-Disk Universe}},
  \href{https://doi.org/10.1103/PhysRevLett.110.211302}{\emph{Phys. Rev. Lett.}
  {\bfseries 110} (2013) 211302},
  [\href{https://arxiv.org/abs/1303.3271}{{\ttfamily 1303.3271}}].

\bibitem{Fan:2013yva}
J.~Fan, A.~Katz, L.~Randall and M.~Reece, \emph{{Double-Disk Dark Matter}},
  \href{https://doi.org/10.1016/j.dark.2013.07.001}{\emph{Phys. Dark Univ.}
  {\bfseries 2} (2013) 139--156},
  [\href{https://arxiv.org/abs/1303.1521}{{\ttfamily 1303.1521}}].

\bibitem{CyrRacine:2012fz}
F.-Y. Cyr-Racine and K.~Sigurdson, \emph{{Cosmology of atomic dark matter}},
  \href{https://doi.org/10.1103/PhysRevD.87.103515}{\emph{Phys. Rev.}
  {\bfseries D87} (2013) 103515},
  [\href{https://arxiv.org/abs/1209.5752}{{\ttfamily 1209.5752}}].

\bibitem{McCullough:2013jma}
M.~McCullough and L.~Randall, \emph{{Exothermic Double-Disk Dark Matter}},
  \href{https://doi.org/10.1088/1475-7516/2013/10/058}{\emph{JCAP} {\bfseries
  1310} (2013) 058}, [\href{https://arxiv.org/abs/1307.4095}{{\ttfamily
  1307.4095}}].

\bibitem{Fan:2013bea}
J.~Fan, A.~Katz and J.~Shelton, \emph{{Direct and indirect detection of
  dissipative dark matter}},
  \href{https://doi.org/10.1088/1475-7516/2014/06/059}{\emph{JCAP} {\bfseries
  1406} (2014) 059}, [\href{https://arxiv.org/abs/1312.1336}{{\ttfamily
  1312.1336}}].

\bibitem{Randall:2014lxa}
L.~Randall and M.~Reece, \emph{{Dark Matter as a Trigger for Periodic Comet
  Impacts}}, \href{https://doi.org/10.1103/PhysRevLett.112.161301}{\emph{Phys.
  Rev. Lett.} {\bfseries 112} (2014) 161301},
  [\href{https://arxiv.org/abs/1403.0576}{{\ttfamily 1403.0576}}].

\bibitem{Fischler:2014jda}
W.~Fischler, D.~Lorshbough and W.~Tangarife, \emph{{Supersymmetric Partially
  Interacting Dark Matter}},
  \href{https://doi.org/10.1103/PhysRevD.91.025010}{\emph{Phys. Rev.}
  {\bfseries D91} (2015) 025010},
  [\href{https://arxiv.org/abs/1405.7708}{{\ttfamily 1405.7708}}].

\bibitem{Foot:2014uba}
R.~Foot and S.~Vagnozzi, \emph{{Dissipative hidden sector dark matter}},
  \href{https://doi.org/10.1103/PhysRevD.91.023512}{\emph{Phys. Rev.}
  {\bfseries D91} (2015) 023512},
  [\href{https://arxiv.org/abs/1409.7174}{{\ttfamily 1409.7174}}].

\bibitem{Randall:2014kta}
L.~Randall and J.~Scholtz, \emph{{Dissipative Dark Matter and the Andromeda
  Plane of Satellites}},
  \href{https://doi.org/10.1088/1475-7516/2015/09/057}{\emph{JCAP} {\bfseries
  1509} (2015) 057}, [\href{https://arxiv.org/abs/1412.1839}{{\ttfamily
  1412.1839}}].

\bibitem{Reece:2015lch}
M.~Reece and T.~Roxlo, \emph{{Nonthermal production of dark radiation and dark
  matter}}, \href{https://doi.org/10.1007/JHEP09(2016)096}{\emph{JHEP}
  {\bfseries 09} (2016) 096},
  [\href{https://arxiv.org/abs/1511.06768}{{\ttfamily 1511.06768}}].

\bibitem{Foot:2016wvj}
R.~Foot and S.~Vagnozzi, \emph{{Solving the small-scale structure puzzles with
  dissipative dark matter}},
  \href{https://doi.org/10.1088/1475-7516/2016/07/013}{\emph{JCAP} {\bfseries
  1607} (2016) 013}, [\href{https://arxiv.org/abs/1602.02467}{{\ttfamily
  1602.02467}}].

\bibitem{Shaviv:2016umn}
N.~J. Shaviv, \emph{{The Paleoclimatic evidence for Strongly Interacting Dark
  Matter Present in the Galactic Disk}},
  \href{https://arxiv.org/abs/1606.02851}{{\ttfamily 1606.02851}}.

\bibitem{Kramer:2016dqu}
E.~D. Kramer and L.~Randall, \emph{{Updated Kinematic Constraints on a Dark
  Disk}}, \href{https://doi.org/10.3847/0004-637X/824/2/116}{\emph{Astrophys.
  J.} {\bfseries 824} (2016) 116},
  [\href{https://arxiv.org/abs/1604.01407}{{\ttfamily 1604.01407}}].

\bibitem{Agrawal:2017rvu}
P.~Agrawal, F.-Y. Cyr-Racine, L.~Randall and J.~Scholtz, \emph{{Dark
  Catalysis}}, \href{https://doi.org/10.1088/1475-7516/2017/08/021}{\emph{JCAP}
  {\bfseries 1708} (2017) 021},
  [\href{https://arxiv.org/abs/1702.05482}{{\ttfamily 1702.05482}}].

\bibitem{Agrawal:2017pnb}
P.~Agrawal and L.~Randall, \emph{{Point Sources from Dissipative Dark Matter}},
  \href{https://doi.org/10.1088/1475-7516/2017/12/019}{\emph{JCAP} {\bfseries
  1712} (2017) 019}, [\href{https://arxiv.org/abs/1706.04195}{{\ttfamily
  1706.04195}}].

\bibitem{Buckley:2017ttd}
M.~R. Buckley and A.~DiFranzo, \emph{{Collapsed Dark Matter Structures}},
  \href{https://doi.org/10.1103/PhysRevLett.120.051102}{\emph{Phys. Rev. Lett.}
  {\bfseries 120} (2018) 051102},
  [\href{https://arxiv.org/abs/1707.03829}{{\ttfamily 1707.03829}}].

\bibitem{DAmico:2017lqj}
G.~D'Amico, P.~Panci, A.~Lupi, S.~Bovino and J.~Silk, \emph{{Massive Black
  Holes from Dissipative Dark Matter}},
  \href{https://doi.org/10.1093/mnras/stx2419}{\emph{Mon. Not. Roy. Astron.
  Soc.} {\bfseries 473} (2018) 328--335},
  [\href{https://arxiv.org/abs/1707.03419}{{\ttfamily 1707.03419}}].

\bibitem{Caputo:2017zqh}
A.~Caputo, J.~Zavala and D.~Blas, \emph{{Binary pulsars as probes of a Galactic
  dark matter disk}},
  \href{https://doi.org/10.1016/j.dark.2017.10.005}{\emph{Phys. Dark Univ.}
  {\bfseries 19} (2018) 1--11},
  [\href{https://arxiv.org/abs/1709.03991}{{\ttfamily 1709.03991}}].

\bibitem{Vattis:2018aen}
K.~Vattis and S.~M. Koushiappas, \emph{{Self-interacting dark matter
  constraints in a thick dark disk scenario}},
  \href{https://arxiv.org/abs/1801.06556}{{\ttfamily 1801.06556}}.

\bibitem{Outmazgine:2018orx}
N.~J. Outmazgine, O.~Slone, W.~Tangarife, L.~Ubaldi and T.~Volansky,
  \emph{{Accretion of Dissipative Dark Matter onto Active Galactic Nuclei}},
  \href{https://arxiv.org/abs/1807.04750}{{\ttfamily 1807.04750}}.

\bibitem{Alexander:2018lno}
S.~Alexander and L.~Smolin, \emph{{The Equivalence Principle and the Emergence
  of Flat Rotation Curves}},
  \href{https://arxiv.org/abs/1804.09573}{{\ttfamily 1804.09573}}.

\bibitem{Ghalsasi:2017jna}
A.~Ghalsasi and M.~McQuinn, \emph{{Exploring the astrophysics of dark atoms}},
  \href{https://arxiv.org/abs/1712.04779}{{\ttfamily 1712.04779}}.

\bibitem{Rosenberg:2017qia}
E.~Rosenberg and J.~Fan, \emph{{Cooling in a Dissipative Dark Sector}},
  \href{https://doi.org/10.1103/PhysRevD.96.123001}{\emph{Phys. Rev.}
  {\bfseries D96} (2017) 123001},
  [\href{https://arxiv.org/abs/1705.10341}{{\ttfamily 1705.10341}}].

\bibitem{Schutz:2017tfp}
K.~Schutz, T.~Lin, B.~R. Safdi and C.-L. Wu, \emph{{Constraining a Thin Dark
  Matter Disk with Gaia}},  \href{https://arxiv.org/abs/1711.03103}{{\ttfamily
  1711.03103}}.

\bibitem{2006AJ....131.1163S}
M.~F. {Skrutskie}, R.~M. {Cutri}, R.~{Stiening}, M.~D. {Weinberg},
  S.~{Schneider}, J.~M. {Carpenter} et~al., \emph{{The Two Micron All Sky
  Survey (2MASS)}}, \href{https://doi.org/10.1086/498708}{\emph{AJ} {\bfseries
  131} (Feb., 2006) 1163--1183}.

\bibitem{Bovy:2017}
J.~{Bovy}, \emph{{Stellar inventory of the solar neighbourhood using Gaia
  DR1}}, \href{https://doi.org/10.1093/mnras/stx1277}{\emph{"Mon. Not. Roy.
  Astron. Soc."} {\bfseries 470} (Sept., 2017) 1360--1387},
  [\href{https://arxiv.org/abs/1704.05063}{{\ttfamily 1704.05063}}].

\bibitem{2018arXiv180502650Z}
J.~C. {Zinn}, M.~H. {Pinsonneault}, D.~{Huber} and D.~{Stello},
  \emph{{Confirmation of the zero-point offset in Gaia Data Release 2
  parallaxes using asteroseismology and APOGEE spectroscopy in the Kepler
  field}}, {\emph{ArXiv e-prints} (May, 2018) },
  [\href{https://arxiv.org/abs/1805.02650}{{\ttfamily 1805.02650}}].

\bibitem{banik:2016}
N.~{Banik}, L.~M. {Widrow} and S.~{Dodelson}, \emph{{Galactoseismology and the
  local density of dark matter}},
  \href{https://doi.org/10.1093/mnras/stw2603}{\emph{"Mon. Not. Roy. Astron.
  Soc."} {\bfseries 464} (Feb., 2017) 3775--3783},
  [\href{https://arxiv.org/abs/1608.03338}{{\ttfamily 1608.03338}}].

\bibitem{1992ApJ...389..234B}
J.~N. {Bahcall}, C.~{Flynn} and A.~{Gould}, \emph{{Local dark matter from a
  carefully selected sample}},
  \href{https://doi.org/10.1086/171201}{\emph{Astrophys. J.} {\bfseries 389}
  (Apr., 1992) 234--250}.

\bibitem{2013ApJS..208....9P}
M.~J. {Pecaut} and E.~E. {Mamajek}, \emph{{Intrinsic Colors, Temperatures, and
  Bolometric Corrections of Pre-main-sequence Stars}},
  \href{https://doi.org/10.1088/0067-0049/208/1/9}{\emph{ApJS} {\bfseries 208}
  (Sept., 2013) 9}, [\href{https://arxiv.org/abs/1307.2657}{{\ttfamily
  1307.2657}}].

\bibitem{2010MNRAS.403.1829S}
R.~{Sch{\"o}nrich}, J.~{Binney} and W.~{Dehnen}, \emph{{Local kinematics and
  the local standard of rest}},
  \href{https://doi.org/10.1111/j.1365-2966.2010.16253.x}{\emph{Mon. Not. Roy.
  Astron. Soc.} {\bfseries 403} (Apr., 2010) 1829--1833},
  [\href{https://arxiv.org/abs/0912.3693}{{\ttfamily 0912.3693}}].

\bibitem{garbari:2011}
S.~{Garbari}, J.~I. {Read} and G.~{Lake}, \emph{{Limits on the local dark
  matter density}},
  \href{https://doi.org/10.1111/j.1365-2966.2011.19206.x}{\emph{"Mon. Not. Roy.
  Astron. Soc."} {\bfseries 416} (Sept., 2011) 2318--2340},
  [\href{https://arxiv.org/abs/1105.6339}{{\ttfamily 1105.6339}}].

\bibitem{2017MNRAS}
J.~{Bovy}, \emph{{Galactic rotation in Gaia DR1}},
  \href{https://doi.org/10.1093/mnrasl/slx027}{\emph{Mon. Not. Roy. Astron.
  Soc.} {\bfseries 468} (June, 2017) L63--L67},
  [\href{https://arxiv.org/abs/1610.07610}{{\ttfamily 1610.07610}}].

\bibitem{1984ApJ...276..169B}
J.~N. {Bahcall}, \emph{{Self-consistent determinations of the total amount of
  matter near the sun}}, \href{https://doi.org/10.1086/161601}{\emph{Astrophys.
  J.} {\bfseries 276} (Jan., 1984) 169--181}.

\bibitem{1984ApJ...276..156B}
J.~N. {Bahcall}, \emph{{The distribution of stars perpendicular to galactic
  disk}}, \href{https://doi.org/10.1086/161600}{\emph{Astrophys. J.} {\bfseries
  276} (Jan., 1984) 156--168}.

\bibitem{1984ApJ...287..926B}
J.~N. {Bahcall}, \emph{{K giants and the total amount of matter near the sun}},
  \href{https://doi.org/10.1086/162750}{\emph{Astrophys. J.} {\bfseries 287}
  (Dec., 1984) 926--944}.

\bibitem{1942ApJ....95..329S}
L.~{Spitzer}, Jr., \emph{{The Dynamics of the Interstellar Medium. III.
  Galactic Distribution.}},
  \href{https://doi.org/10.1086/144407}{\emph{Astrophys. J.} {\bfseries 95}
  (May, 1942) 329}.

\bibitem{Bovy:2016a}
J.~{Bovy}, H.-W. {Rix}, G.~M. {Green}, E.~F. {Schlafly} and D.~P. {Finkbeiner},
  \emph{{On Galactic Density Modeling in the Presence of Dust Extinction}},
  \href{https://doi.org/10.3847/0004-637X/818/2/130}{\emph{ApJ} {\bfseries 818}
  (Feb., 2016) 130}, [\href{https://arxiv.org/abs/1509.06751}{{\ttfamily
  1509.06751}}].

\bibitem{Widmark:2018ylf}
A.~Widmark, \emph{{Measuring the local matter density using Gaia DR2}},
  \href{https://arxiv.org/abs/1811.07911}{{\ttfamily 1811.07911}}.

\bibitem{GW2010}
J.~{Goodman} and J.~{Weare}, \emph{{Ensemble samplers with affine invariance}},
  \href{https://doi.org/10.2140/camcos.2010.5.65}{\emph{Communications in
  Applied Mathematics and Computational Science, Vol.~5, No.~1, p.~65-80, 2010}
  {\bfseries 5} (2010) 65--80}.

\bibitem{ForemanMackey:2012ig}
D.~Foreman-Mackey, D.~W. Hogg, D.~Lang and J.~Goodman, \emph{{emcee: The MCMC
  Hammer}}, \href{https://doi.org/10.1086/670067}{\emph{Publ. Astron. Soc.
  Pac.} {\bfseries 125} (2013) 306--312},
  [\href{https://arxiv.org/abs/1202.3665}{{\ttfamily 1202.3665}}].

\bibitem{optimal_scaling}
G.~O.~Roberts, A.~Gelman and W.~R.~Gilks, \emph{Weak convergence and optimal
  scaling of random walk metropolis algorithms}, {\emph{The Annals of Applied
  Probability} {\bfseries 7} (04, 1997) }.

\bibitem{Yanny2009}
B.~{Yanny}, C.~{Rockosi}, H.~J. {Newberg}, G.~R. {Knapp}, J.~K.
  {Adelman-McCarthy}, B.~{Alcorn} et~al., \emph{{SEGUE: A Spectroscopic Survey
  of 240,000 Stars with g = 14-20}},
  \href{https://doi.org/10.1088/0004-6256/137/5/4377}{\emph{Astronomical. J.}
  {\bfseries 137} (May, 2009) 4377--4399},
  [\href{https://arxiv.org/abs/0902.1781}{{\ttfamily 0902.1781}}].

\bibitem{Widrow2012}
L.~M. {Widrow}, S.~{Gardner}, B.~{Yanny}, S.~{Dodelson} and H.-Y. {Chen},
  \emph{{Galactoseismology: Discovery of Vertical Waves in the Galactic Disk}},
  \href{https://doi.org/10.1088/2041-8205/750/2/L41}{\emph{Astrophys. J. Lett.}
  {\bfseries 750} (May, 2012) L41},
  [\href{https://arxiv.org/abs/1203.6861}{{\ttfamily 1203.6861}}].

\bibitem{2019MNRAS.482.1417B}
M.~{Bennett} and J.~{Bovy}, \emph{{Vertical waves in the solar neighbourhood in
  Gaia DR2}}, \href{https://doi.org/10.1093/mnras/sty2813}{\emph{Mon. Not. Roy.
  Astron. Soc.} {\bfseries 482} (Jan, 2019) 1417--1425},
  [\href{https://arxiv.org/abs/1809.03507}{{\ttfamily 1809.03507}}].

\bibitem{Gomez2013}
F.~A. {G{\'o}mez}, I.~{Minchev}, B.~W. {O'Shea}, T.~C. {Beers}, J.~S. {Bullock}
  and C.~W. {Purcell}, \emph{{Vertical density waves in the Milky Way disc
  induced by the Sagittarius dwarf galaxy}},
  \href{https://doi.org/10.1093/mnras/sts327}{\emph{"Mon. Not. Roy. Astron.
  Soc."} {\bfseries 429} (Feb., 2013) 159--164},
  [\href{https://arxiv.org/abs/1207.3083}{{\ttfamily 1207.3083}}].

\bibitem{Carlin2013}
J.~L. {Carlin}, J.~{DeLaunay}, H.~J. {Newberg}, L.~{Deng}, D.~{Gole},
  K.~{Grabowski} et~al., \emph{{Substructure in Bulk Velocities of Milky Way
  Disk Stars}},
  \href{https://doi.org/10.1088/2041-8205/777/1/L5}{\emph{Astrophys. J. Lett.}
  {\bfseries 777} (Nov., 2013) L5},
  [\href{https://arxiv.org/abs/1309.6314}{{\ttfamily 1309.6314}}].

\bibitem{Widrow2014}
L.~M. {Widrow}, J.~{Barber}, M.~H. {Chequers} and E.~{Cheng}, \emph{{Bending
  and breathing modes of the Galactic disc}},
  \href{https://doi.org/10.1093/mnras/stu396}{\emph{"Mon. Not. Roy. Astron.
  Soc."} {\bfseries 440} (May, 2014) 1971--1981},
  [\href{https://arxiv.org/abs/1404.4069}{{\ttfamily 1404.4069}}].

\bibitem{Antoja2018}
T.~{Antoja}, A.~{Helmi}, M.~{Romero-Gomez}, D.~{Katz}, C.~{Babusiaux},
  R.~{Drimmel} et~al., \emph{{Wrinkles in the Gaia data unveil a dynamically
  young and perturbed Milky Way disk}}, {\emph{ArXiv e-prints} (Apr., 2018) },
  [\href{https://arxiv.org/abs/1804.10196}{{\ttfamily 1804.10196}}].

\bibitem{Myeong:2018kfh}
G.~C. Myeong, N.~W. Evans, V.~Belokurov, J.~L. Sanders and S.~E. Koposov,
  \emph{{The Sausage Globular Clusters}},
  \href{https://arxiv.org/abs/1805.00453}{{\ttfamily 1805.00453}}.

\bibitem{Necib:2018iwb}
L.~Necib, M.~Lisanti and V.~Belokurov, \emph{{Dark Matter in Disequilibrium:
  The Local Velocity Distribution from SDSS-Gaia}},
  \href{https://arxiv.org/abs/1807.02519}{{\ttfamily 1807.02519}}.

\bibitem{Budenbender:2014xra}
A.~Budenbender, G.~van~de Ven and L.~L. Watkins, \emph{{The tilt of the
  velocity ellipsoid in the Milky Way disc}},
  \href{https://doi.org/10.1093/mnras/stv1314}{\emph{Mon. Not. Roy. Astron.
  Soc.} {\bfseries 452} (2015) 956--968},
  [\href{https://arxiv.org/abs/1407.4808}{{\ttfamily 1407.4808}}].

\bibitem{Perez:2007emg}
F.~Perez and B.~E. Granger, \emph{{IPython: A System for Interactive Scientific
  Computing}}, \href{https://doi.org/10.1109/MCSE.2007.53}{\emph{Comput. Sci.
  Eng.} {\bfseries 9} (2007) 21--29}.

\bibitem{Hunter:2007ouj}
J.~D. Hunter, \emph{{Matplotlib: A 2D Graphics Environment}},
  \href{https://doi.org/10.1109/MCSE.2007.55}{\emph{Comput. Sci. Eng.}
  {\bfseries 9} (2007) 90--95}.

\bibitem{jones2001scipy}
E.~Jones, T.~Oliphant, P.~Peterson et~al., \emph{{SciPy}: Open source
  scientific tools for {Python}},  2001--.

\bibitem{vanderWalt:2011bqk}
S.~van~der Walt, S.~C. Colbert and G.~Varoquaux, \emph{{The NumPy Array: A
  Structure for Efficient Numerical Computation}},
  \href{https://doi.org/10.1109/MCSE.2011.37}{\emph{Comput. Sci. Eng.}
  {\bfseries 13} (2011) 22--30},
  [\href{https://arxiv.org/abs/1102.1523}{{\ttfamily 1102.1523}}].

\bibitem{astropy}
{Astropy Collaboration}, T.~P. {Robitaille}, E.~J. {Tollerud}, P.~{Greenfield},
  M.~{Droettboom}, E.~{Bray} et~al., \emph{{Astropy: A community Python package
  for astronomy}}, \href{https://doi.org/10.1051/0004-6361/201322068}{\emph{A
  \& A} {\bfseries 558} (Oct., 2013) A33},
  [\href{https://arxiv.org/abs/1307.6212}{{\ttfamily 1307.6212}}].

\bibitem{gala}
A.~M. Price-Whelan, \emph{Gala: A python package for galactic dynamics},
  \href{https://doi.org/10.21105/joss.00388}{\emph{The Journal of Open Source
  Software} {\bfseries 2} (oct, 2017) }.

\bibitem{Ivezic:2014:SDM:2578955}
Z.~Ivezic, A.~J. Connolly, J.~T. VanderPlas and A.~Gray, \emph{Statistics, Data
  Mining, and Machine Learning in Astronomy: A Practical Python Guide for the
  Analysis of Survey Data}.
\newblock Princeton University Press, Princeton, NJ, USA, 2014.

\end{thebibliography}\endgroup
\bibliographystyle{jhep}
\end{document}